\documentclass[longauth,pdfpagelabels=false]{aa}  

\usepackage{float}
\usepackage{graphicx}
\usepackage{amsmath}
\usepackage{txfonts}
\usepackage{comment}
\usepackage{enumitem}
\usepackage{pdfpages}
\usepackage{rotating}
\usepackage{lscape}
\usepackage{placeins}
\usepackage{orcidlink}
\usepackage{hyperref}	
\hypersetup{colorlinks=true, linkcolor=blue, citecolor=blue, filecolor=blue, urlcolor=blue}

\newcommand{\msun}        {\mbox{\rm M$_\odot$}}
\newcommand{\msunperpcsq} {\mbox{\rm M$_\odot$~pc$^{-2}$}}

\newcommand{\msunperyr}   {\mbox{\rm M$_\odot$~yr$^{-1}$}}

\newcommand{\Mmol}        {\mbox{${\rm M}_{\rm mol}$}}

\newcommand{\Lco}         {\mbox{$L_{\rm CO}$}}

\newcommand{\htwo}{\ensuremath{\mathrm{H_2}}}

\newcommand{\Reff}{\mbox{\rm R$_{\rm eq}$}}
\newcommand{\qpah}{\ensuremath{q_\mathrm{PAH}}}
\newcommand{\aco}{\ensuremath{\alpha_\mathrm{CO}}}

\newcommand{\ipah}{\ensuremath{I^\mathrm{PAH}_{\rm F770W}}}

\newcommand{\norm}{\ensuremath{C_\mathrm{F770W}^\mathrm{PAH}}}
\newcommand{\mjysr}{MJy~sr$^{-1}$}

\newcommand{\sigmol}{\mbox{$\Sigma_{\rm mol}$}}
\newcommand{\sigsfr}{\mbox{$\Sigma_{\rm SFR}$}}

\defcitealias{Colombo_2015}{C15}
\defcitealias{Rosolowsky_2006}{R06}

\begin{document}

    \title{PHANGS-JWST: The largest extragalactic molecular cloud catalog traced by polycyclic aromatic hydrocarbon emission}
    \subtitle{}

    \author{
    Z.~Bazzi \inst{\ref{ubonn}}\orcidlink{0009-0001-1221-0975}\thanks{zbazzi@uni-bonn.de} \and
    D.~Colombo \inst{\ref{ubonn}}\orcidlink{0000-0001-6498-2945}\and
    F.~Bigiel \inst{\ref{ubonn}}\and
    A.~K.~Leroy \inst{\ref{ohio}, \ref{columbus}}\orcidlink{0000-0002-2545-1700} \and
    E.~Rosolowsky \inst{\ref{alberta}}\orcidlink{0000-0002-5204-2259} \and
    K.~Sandstrom \inst{\ref{UCSD}}\orcidlink{0000-0002-4378-8534}\and
    A.~Duarte-Cabral \inst{\ref{cardiff}}\orcidlink{0000-0002-5259-4774}\and
    H.~Faustino~Vieira \inst{\ref{stock}}\orcidlink{0000-0002-2199-0977}\and
    M.~I.~N.~Kobayashi \inst{\ref{cologne}}\orcidlink{0000-0003-3990-1204}\and
    H.~He \inst{\ref{ubonn}}\orcidlink{0000-0001-9020-1858} \and
    S.~E.~Meidt \inst{\ref{gent}}\orcidlink{0000-0002-6118-4048} \and
    A.~T.~Barnes \inst{\ref{eso}}\orcidlink{0000-0003-0410-4504} \and 
    R.~S.~Klessen \inst{\ref{ita},\ref{IWR},\ref{CfA},\ref{Radcliffe}}\orcidlink{0000-0002-0560-3172} \and
    S.~C.~O.~Glover \inst{\ref{ita}}\orcidlink{0000-0001-6708-1317} \and
    M.~D.~Thorp \inst{\ref{ubonn}}\orcidlink{0000-0003-0080-8547}\and
    H.-A.~Pan \inst{\ref{tku}}\orcidlink{0000-0002-1370-6964} \and
    R.~Chown \inst{\ref{ohio}}\orcidlink{0000-0001-8241-7704} \and
    R.~J.~Smith \inst{\ref{StAndrews}}\orcidlink{0000-0002-0820-1814} \and 
    D.~A.~Dale \inst{\ref{wyo}}\orcidlink{0000-0002-5782-9093} \and
    T.~G.~Williams \inst{\ref{Ox}, \ref{Manc}}\orcidlink{0000-0002-0012-2142} \and
    A.~Amiri \inst{\ref{UOA}}\orcidlink{0000-0002-8553-1964}\and
    S. Dlamini \inst{\ref{UCT}}\orcidlink{0000-0002-2885-6172} \and
    J.~Chastenet\inst{\ref{gent}}\orcidlink{0000-0002-5235-5589}\and
    S.~K.~Sarbadhicary \inst{\ref{jhu}}\orcidlink{0000-0002-4781-7291}\and
    A.~Hughes \inst{\ref{IRAP}}\orcidlink{0000-0002-9181-1161}\and
    J.~C.~Lee\inst{\ref{stsci}}\orcidlink{0000-0002-2278-9407}\and
    L.~Hands\inst{\ref{UCSD}}\orcidlink{0009-0005-0750-2956}\and
     the PHANGS collaboration}

    \institute{
        \label{ubonn}{Argelander-Institut f\"ur Astronomie, University of Bonn, Auf dem H\"ugel 71, 53121 Bonn, Germany} \and
        \label{ohio} Department of Astronomy, Ohio State University, 180 W. 18th Ave, Columbus, Ohio 43210 \and
        \label{columbus} Center for Cosmology and Astroparticle Physics, 191 West Woodruff Avenue, Columbus, OH 43210, USA \and
        \label{alberta}{Dept. of Physics, 4-183 CCIS, University of Alberta, Edmonton, AB, T6G 2E1, Canada} \and  
        \label{UCSD}{Department of Astronomy \& Astrophysics, University of California, San Diego, 9500 Gilman Dr., La Jolla, CA 92093, USA} \and
        \label{cardiff}{Cardiff Hub for Astrophysics Research and Technology (CHART), School of Physics \& Astronomy, Cardiff University, The Parade, CF24 3AA Cardiff, UK}\and
        \label{stock}{Department of Astronomy, Oskar Klein Center, Stockholm University, AlbaNova University Center, SE-106 91 Stockholm, Sweden}\and
        \label{cologne}{I. Physikalisches Institut, Universit\"at zu K\"oln, Z\"ulpicher Str 77, D-50937 K\"oln, Germany}\and
        \label{gent}{Sterrenkundig Observatorium, Universiteit Gent, Krijgslaan 281 S9, B-9000 Gent, Belgium} \and
        \label{eso} European Southern Observatory (ESO), Karl-Schwarzschild-Stra{\ss}e 2, 85748 Garching, Germany  \and
        \label{ita} Universit\"{a}t Heidelberg, Zentrum f\"{u}r Astronomie, Institut f\"{u}r Theoretische Astrophysik, Albert-Ueberle-Str.\ 2, 69120 Heidelberg, Germany \and
        \label{IWR} {Universit\"{a}t Heidelberg, Interdisziplin\"{a}res Zentrum f\"{u}r Wissenschaftliches Rechnen, Im Neuenheimer Feld 205, D-69120 Heidelberg, Germany} \and
        \label{CfA} {Harvard-Smithsonian Center for Astrophysics, 60 Garden Street, Cambridge, MA 02138, USA} \and
        \label{Radcliffe} {Elizabeth S.\ and Richard M.\ Cashin Fellow at the Radcliffe Institute for Advanced Studies at Harvard University, 10 Garden Street, Cambridge, MA 02138, USA} \and
        \label{tku}{Department of Physics, Tamkang University, No.151, Yingzhuan Road, Tamsui District, New Taipei City 251301, Taiwan} \and
        \label{StAndrews} School of Physics and Astronomy, University of, N Haugh, St Andrews KY16 9SS, United Kingdom \and
        \label{wyo} Department of Physics \& Astronomy, University of Wyoming, Laramie, WY 82071, USA \and
        \label{UOA}{Department of Physics, University of Arkansas, 226 Physics Building, 825 West Dickson Street, Fayetteville, AR 72701, USA} \and
        \label{UCT} Department of Astronomy, University of Cape Town, Rondebosch 7701, South Africa \and
        \label{Ox} Sub-department of Astrophysics, Department of Physics, University of Oxford, Keble Road, Oxford OX1 3RH, UK \and
        \label{Manc} UK ALMA Regional Centre Node, Jodrell Bank Centre for Astrophysics, Department of Physics and Astronomy, The University of Manchester, Oxford Road, Manchester M13 9PL, UK \and
        \label{jhu} Department of Physics and Astronomy, The Johns Hopkins University, Baltimore, MD 21218, USA \and
        \label{IRAP} IRAP/OMP/Universit\'e de Toulouse, 9 Av. du Colonel Roche, BP 44346, F-31028 Toulouse cedex 4, France \and
        \label{stsci} Space Telescope Science Institute, Baltimore, MD 21218, USA}

      \date{Received XXX; accepted XXX}
     \date{Received XXX; accepted XXX}
    
    \abstract
        {High-resolution JWST images of nearby spiral galaxies reveal polycyclic aromatic hydrocarbon (PAH) structures that potentially trace molecular clouds, even CO-dark regions. For this paper, we identified ISM cloud structures in PHANGS-JWST 7.7$\mu$m PAH emission maps for 66 galaxies, smoothed to a common physical resolution of 30 pc and at native resolution. We extracted 108,466 cloud structures in the 30 pc sample and 146,040 clouds in the native resolution sample. We then calculated their molecular properties following a linear conversion from PAH to CO. Given the tendency for clouds in galaxy centers to overlap in velocity space, we opted to flag these clouds and omit them from the analysis in this work. The remaining clouds correspond to giant molecular clouds, such as those detected in CO($2-1$) emission by ALMA, or lower surface density clouds that either fall below the ALMA detection limits of existing maps or genuinely have no molecular counterpart. We specifically used the homogenized sample for our analysis. Upon cross-matching the PAH clouds to the ALMA CO clouds at a homogenized resolution of 90 pc in 27 galaxies, we find that 41 $\%$ of the PAH clouds are associated with a CO counterpart. We also show that the converted molecular cloud properties of the PAH clouds do not differ much when compared in different galactic environments. However, outside the central environment, the highest molecular mass surface density clouds are preferentially found in spiral arms. We further apply a lognormal fit to the mass spectra to an unprecedented extragalactic completeness limit of $2 \times 10^{3} \, \msun $, and find that spiral arms contain the most massive clouds compared to other galactic environments. Our findings support the idea that spiral arm gravitational potentials foster the formation of high surface density clouds, and that lower surface density clouds form in the interarm regions. The cloud \sigmol\ values show a decline of a factor of $\sim 1.5-2$ toward the outer $2-3$ R$_{e}$. However, the trend largely varies in individual galaxies, with flat, decreasing, and even no trend as a function of R$_{\rm gal}$. Factors such as large-scale processes, galaxy types, and morphologies might influence the observed trends. We note that combining homogenized molecular properties of individual galaxies leads to the loss of information about the physical processes that are driving deviations in trends of those properties across different galactic environments. We published two catalogs at the CDS, one at the common resolution of 30 pc and another at the native resolution. We expect them to have broad utility for future studies of PAH clouds, molecular clouds, and star formation.}


    \titlerunning{Characterizing molecular clouds traced by PAH emission}

    \authorrunning{Z. Bazzi et al.}

    \maketitle
 
\section{Introduction}\label{S:introduction}

Giant molecular clouds (GMCs) and the stars they produce are fundamental components in the formation and evolution of galaxies. The physical characteristics and development of GMCs are closely intertwined with the larger-scale processes that shape galaxies. The conditions within the interstellar medium (ISM) significantly influence how GMCs emerge and evolve, with clear correlations observed between galactic properties and those of the clouds themselves, including gas pressure, surface density, and volume density \citep[e.g.,][]{Colombo_2014,Sun_2018,Chevance_2020}. Since star formation predominantly takes place within the cold, dense molecular phase of the ISM, characterizing GMCs is crucial for understanding the mechanisms that set those properties and ultimately for understanding the mechanisms that drive stellar birth, and by extension, galaxy evolution \citep{Bigiel_2008, Schruba_2011}. 

Over the past few decades, radio, millimeter, and infrared telescopes have advanced to achieve increasingly high resolution in the Milky Way and nearby galaxies, allowing us to resolve molecular clouds. With the successful launch of the James Webb Space Telescope (JWST), parsec-scale near- and mid-infrared imaging is now possible in nearby galaxies (e.g., distances < 20 Mpc). The Physics at High Angular Resolution in Nearby GalaxieS (PHANGS) survey \citep{Leroy_2021, Lee_2023, Williams_2024} has taken advantage of these advances in resolution to probe dust emission, specifically polycyclic aromatic hydrocarbons (PAHs), in galaxies at unprecedented sub-GMC scales of $5-30$ pc.

Several techniques are available to uncover the structures and properties of molecular clouds (e.g., gas mass, surface densities, velocity dispersion, radius). The CO emission, particularly from low-J transitions such as CO(1–0) and CO(2–1), is the most traditional tracer of molecular hydrogen, benefiting from extensive calibration of metallicity, gas surface density, and others \citep[e.g.,][]{Solomon_1987, Fukui_2001, Bolatto_2008, Bolatto_2013, Heyer_2009, Fukui_2010, Sun_2020a, schinnerer2024}. Cloud properties are often extracted from CO data using algorithms such as CPROPS \citep{Rosolowsky_2006,Rosolowsky_2021}, CLUMPFIND \citep{Williams_1994, Rosolowsky_2005b}, and Spectral Clustering for Molecular Emission Segmentation \citep[SCIMES;][]{Colombo_2015}, enabling the identification and characterization of GMCs across nearby galaxies, especially with high-resolution CO observations from the Atacama Large Millimeter Array (ALMA), specifically PHANGS-ALMA (e.g., \citealt{Leroy_2021,Rosolowsky_2021,Sun_2022}). Other methods include characterizing the intensity field of CO at fixed resolution \citep[e.g.,][]{Hughes_2013,Leroy_2016,Sun_2022}, which recovers similar information to object-finding decomposition approaches. Dust has also long been used to trace gas, starting from star count methods and extinction mapping \citep[e.g.,][]{Savage_1977,Bohlin_1978, Savage_1979}, offering an independent avenue to infer gas column densities. Recent advances using high-resolution optical imaging, such as Hubble Space Telescope (HST) observations, have enabled the derivation of sub-GMC high-resolution dust extinction maps across nearby galaxies, assuming a constant dust-to-gas ratio \citep[e.g.,][]{Vieira_2023,Vieira_2024,Vieira_2025}. 

Polycyclic aromatic hydrocarbons have emerged as a promising gas tracer thanks to the Spitzer Space telescope \citep{Houk_2004} and the wide sky coverage provided by WISE \citep{Wright_2010}. Studies using Spitzer have shown that, in nearby galaxies, PAH emission correlates with molecular gas on spatial scales ranging from several hundred parsecs to kiloparsecs \citep[e.g.,][]{Regan_2006, Cortzen_2018, Gao_2019, Chown_2021, Gao_2022}. Fortunately, the recent deployment of JWST opens a promising path forward. Its high-resolution and high-sensitivity imaging of mid-infrared PAH emission, which shows a strong correlation with CO emission (e.g., \citealt{Leroy_2023,Sandstrom_2023b,chown2025}), provides valuable insight into this topic. This offers the prospect to use the PAH emission to measure the structure of the cold ISM at high resolution and sensitivity \citep[e.g.,][]{Leroy_2023,sandstrom2023,Meidt_2023,Thilker_2023,Whitcomb_2023,chown2025}. Beyond the nearby Universe, JWST has also shown that the PAH--CO correspondence remains strong at intermediate- to high-redshift galaxies \citep[e.g.,][]{Shivaei_2024}.

Recent observations using JWST show that PAH emission can be decomposed into a first component tracing the molecular gas, where gas and dust column density variations dominate, and a second component tracing star formation, where interstellar radiation field intensity variations dominate \citep{Leroy_2023}. This result further offers new insights into the gas and dust structure at sub-GMC scales. \cite{chown2025} expanded this study to all PHANGS-JWST galaxies and further showed an excellent correspondence between the different JWST MIRI filters, specifically the F770W, and CO emission. They further suggested that PAH emission maps could be effectively converted to CO maps to obtain a more sensitive version than the already existing ALMA maps. However, JWST observations have further highlighted how PAH emission behaves differently in different environments. For instance, in active galactic nucleus (AGN) environments, PAHs can be partially destroyed or their emission suppressed due to strong radiation fields and shocks, whereas in star-forming galaxies without AGN activity, they remain a robust tracer of molecular material (e.g., \citealt{Bernete_2022,Bernete_2024}).

An important advantage of PAH emission is its ability to trace low-density regions where CO is faint or absent. Unlike CO, which requires sufficient shielding to avoid photodissociation \citep[e.g.,][]{Bolatto_2013, Saintonge_2022}, PAHs can emit in diffuse UV-irradiated environments. This allows PAH emission to probe the full extent of molecular cloud complexes, including CO-``dark'' molecular gas components \citep[e.g.,][]{Leroy_2023a, Sandstrom_2023b}. It is important to note, however, that the F770W filter also traces emission from hot dust continuum, stellar continuum, and weak ionic or molecular lines (e.g., \citealt{Draine_2007, whitcomb_dust}).

\cite{Sun_2022} (see also \citealt{Sun_2018,Sun_2020a,Sun_2020,Rosolowsky_2021}) analyzed the PHANGS-ALMA galaxy sample, examining molecular gas properties across different galactic environments at a fixed physical resolution of 60-150 pc. Their results suggest that kiloparsec-scale environmental conditions largely drive variations in cloud populations from galaxy to galaxy. They also find that cloud-scale surface densities, velocity dispersions, and turbulent pressures increase toward galactic centers, reaching exceptionally high values in the centers of barred galaxies, where the gas also appears to be less gravitationally bound, and are moderately elevated in spiral arms compared to interarm regions. However, the homogenized resolution for the full PHANGS-ALMA sample is 150 pc. Fortunately, with PHANGS-JWST, images resemble sharper, more sensitive versions of ALMA CO maps for the same galaxies. The resolution is also enhanced by a factor of five (homogenized resolution of 30 pc) for the F770W band. This means that fainter and smaller structures can now be investigated and might be associated with either molecular clouds or the atomic phase of the ISM \citep[see][]{sandstrom2023, Leroy_2023}.

Several other observations surveys investigated such properties of clouds in the Milky Way (e.g., \citealt{Roman_2010, Eden_2012, Cabral_2021}) and nearby galaxies (e.g., \citealt{Hirota_2011, Rebolledo_2012,Rebolledo_2015,Colombo_2014, Usero_2015, Rosolowsky_2021}). Some suggest that their star formation rates and/or efficiencies (e.g., \citealt{Rebolledo_2012, Rebolledo_2015}) and mass distributions (e.g., \citealt{Colombo_2014}) differ, for instance due to their crossing through spiral arms (e.g., \citealt{Cabral_2017}), and that the most massive clouds are mostly found in the spiral arms (e.g., \citealt{Rebolledo_2012, Vieira_2024}). This difference could be due to self gravity in the spiral arms, agglomeration of pre-existing molecular clouds (e.g., \citealt{Field_1965, Taff_1972, Scoville_1979, Casoli_1982, Dobbs_2008}), shock compression driven by spiral structures (e.g., \citealt{Meidt_2013}), or due to low shear effects (e.g., \citealt{Elmegreen_2011}) or other factors. Other authors have measured cloud mass spectra that are uniform and independent of the physical conditions in their surroundings (e.g., \citealt{Eden_2012, Meyer_2013}). 

Simulations have also investigated GMCs and their evolution across different environments (e.g., \citealt{Dobbs_2006, Nimori_2012, Fujimoto_2014, Dobbs_2015, Cabral_2016, Grand_2017, Hopkins_2018, Tress_2021, Smith_2020, Colman_2024}). \cite{Cabral_2016,Cabral_2017} found that most clouds exhibit properties largely independent of their location within the galaxy. However, some tails of the distributions depend on their location, indicating that more extreme clouds favor specific environments. In addition, \cite{Pettitt_2020} investigated different spiral arm models and saw differences in the interarm--arm mass spectra. Furthermore, analytical and numerical approaches suggest that the GMC lifecycle varies with the galactic environment (e.g., \citealt{Dobbs_2013, Fujimoto_2014, Dobbs_2015b, Jeffreson_2018, Meidt_2018}).  

For this paper, we identified PAH clouds in 66 homogenized 7.7$\mu$m emission PHANGS-JWST Cycle 1 \citep{Lee_2023, Williams_2024} and Cycle 2 \citep{chown2025} galaxy maps at a fixed resolution of 30~pc. We then used a linear conversion fit to convert the dust to CO emission maps using the prescription of \cite{chown2025}. This method provides insights into molecular clouds with higher resolution than current CO surveys and offers improved sensitivity for detecting smaller and fainter structures in the ISM. We also analyzed systematic environmental effects for a statistically significant sample of nearby galaxies, and present the molecular properties of PAH-to-CO converted structure down to an unprecedented extragalactic completeness limit of $\sim 2\times 10^{3} \, \msun$ which is 2.4 dex better than the previous ALMA-based CO approach \citep[e.g.,][]{Rosolowsky_2021}.  

The layout of the paper is as follows. In Sect.~\ref{S:data} we describe the data used in the study. In Sect.~\ref{S:SCIMES} we briefly explain the cloud extraction algorithm, SCIMES, and its input parameters. In  Sect.~\ref{S:MC_Props} we detail the different cloud properties and how we derived them. In Sect.~\ref{S:Results} we describe the distribution of identified structures and their properties as a function of different galactic environments, while highlighting the caveats of our approach. Finally, in Sect.~\ref{S:Summary} we summarize our findings and present our conclusions.

\section{Data and galaxy sample}\label{S:data}

We used a subset of 66 galaxies with high-resolution PHANGS-ALMA CO($2-1$) and JWST imaging tracing the PAH emission (see Table~\ref{T:sample}). The galaxies are star-forming and have specific star formation rates (SFR/$M_\star) \gtrsim 10^{-11}~\mathrm{yr^{-1}}$, stellar masses ($M_\star) \gtrsim 10^{9.5}~\mathrm{M_\odot}$ and CO luminosities (\Lco) $6.60 < \mathrm{log}_{10} (\Lco [\rm K\,km\,s^{-1}]) < 9.50$, have moderate inclination ($i \lesssim 70^\circ$), and are at distances ($D$) $\lesssim 20$~Mpc \citep{Leroy_2021}.

\subsection{JWST mid-IR data}\label{ss:JWST}

JWST MIRI filters provide high angular resolution and sensitivity imaging of dust emission maps, achieving sub-GMC scales in nearby galaxies. The full width at half maximum (FWHM) is 0.269\arcsec, 0.328\arcsec, 0.375\arcsec, and 0.674\arcsec\ for the F770W, F1000W, F1130W, and F2100W bands, respectively. Generally, those wavelengths capture stochastic emission from dust grains, including PAHs. Strong PAH features can be traced using the 7.7$\mu$m band (C--C stretching modes of PAHs) which are mainly due to ionized PAHs for a range of sizes, and 11.3$\mu$m band (C--H out-of-plane bending modes of PAHs) due to mostly larger and neutral PAHs. The 10$\mu$m band captures a mix of PAH and continuum emission, in addition to silicate features and prominent emission lines, while the 21$\mu$m band traces only continuum emission \citep{Draine_2007, Spoon_2007, Smith_2007, tielens2008}.

We used JWST MIRI and NIRCam imaging of all 19 galaxies from the PHANGS-JWST Cycle 1 Treasury \citep[GO~2107, PI: J. Lee;][]{Lee_2023} and 47 galaxies (available at the time of analysis, out of the full set of 51) from the PHANGS-JWST Cycle 2 Treasury (GO 3707, PI: A. Leroy; see \citealt{chown2025}). Observations, data reduction, and processing using the different JWST-MIRI bands are represented in \cite{Lee_2023}, \cite{Williams_2024}, and \cite{chown2025}. 

For the analysis presented in this paper, we used the F770W band to take advantage of the highest resolution MIRI band that captures PAH emission. The median physical resolution is $\sim$20 pc for the full sample with a $16{-}84\%$ range of $15{-}25$~pc. Given that the Rayleigh-Jeans tail of the stellar distribution contributes to the F770W band, it needs to be subtracted from the total surface brightness to obtain the emission from PAHs alone. In our analysis, we used stellar-continuum-subtracted F770W images obtained by subtracting the F200W (Cycle 1) or F300M (Cycle 2) times a scaling factor from the F770W filter following \cite{sutter2024}. The F770W stellar continuum correction (F770W$_\star$) $= 0.22\,\times\,$F300M for Cycle 2, or $0.12\,\times\,$F200W for Cycle 1. Additionally, we leverage empirical scaling relations that relate CO emission to the continuum-subtracted F770W maps. The dust continuum contribution might have a significant contribution to the F770W band. \cite{Whitcomb_2023} and \cite{Dale_2025} applied a method based on Spitzer Space Telescope mid-infrared spectra of nearby star-forming galaxies coupled with synthetic F770W, F1000W, and F1130W photometry (see \citealt{Whitcomb_2023} and Hands et al. in prep.). They find that the continuum-free PAH emission is $\sim 83~\%$. We further apply this method and find that across the Cycle 1 targets, the continuum-free PAH contribution is $\sim 81~\%$. However, there are regions within the galaxies, particularly around HII regions and toward the centers, where the PAH contribution decreases further, consistent with the known suppression of PAHs in these environments. In the central regions, the contribution can reach values of roughly $\sim20\,\%$ or lower on average (for more details, see Hands et al., in prep.).

To be able to make a comparison of the   66 galaxies in the sample at the same physical resolution, we smoothed our data to a common physical resolution of $\sim$ 30 pc. This corresponds to the MIRI F770W resolution for the farthest galaxy in our sample (NGC 3507). We used  JWST point spread functions (PSFs) generated by $\mathtt{webbpsf}$.\footnote{https://webbpsf.readthedocs.io/en/latest/} We then created a convolution kernel per galaxy per filter using $\mathtt{jwst\_kernels}$\footnote{https://github.com/francbelf/jwst${\_}$kernels} to achieve the required physical resolution. Finally, following \cite{Williams_2024}, we convolved our data and error maps using our corresponding convolution kernel. For our analysis, we used this common resolution sample. We also provide a catalog constructed at the native angular resolution of each map.

\subsection{Converting 7.7 micron to CO}\label{SS: 7.7 to CO}
The emission from PAHs shows a close link with CO emission on kiloparsec and parsec scales, revealing a strong correlation between both emissions over three orders of magnitude of intensity \citep{Regan_2004,Gao_2019,Chown_2021,Leroy_2021,Leroy_2023a,Leroy_2023, Whitcomb_2023,chown2025}. Following \cite{chown2025}, to first order, we expect
\begin{eqnarray}
\label{eq:dust}
 I_{\rm PAH} &\propto& \mathrm{DGR}\times \qpah  \times N_{\rm H_2}\times U \\
\nonumber &\propto&  \left(\mathrm{DGR}\times \qpah  \times X_\mathrm{CO}\times U \right)I_{\rm CO},
\end{eqnarray}
where $I_{\rm PAH}$ and $I_{\rm CO}$ are the observed intensities of PAH and CO emission in \mjysr and $\rm K \, km \, s^{-1}$, respectively. The dust-to-gas mass ratio is DGR, \qpah\ is the PAH-to-dust mass fraction, $U$ is the strength of the interstellar radiation field relative to that in the Solar neighborhood \citep[\qpah\ and $U$ are defined in][]{Draine_2007}, and $X_{\rm CO}$ is the CO-to-\htwo\ conversion factor. 

\cite{chown2025} also analyzed the resolved correlation between CO and the different PAH emission bands in 70 PHANGS galaxies, of which 66 are used here. They found the following relation

\begin{eqnarray}\label{E:copah_renorm}
\log_{10} I_\mathrm{CO(2-1)} &=& (0.88 \pm 0.06)\,(x - 1.44) + (1.36 \pm 0.06), \\
\nonumber x &\equiv& \log_{10}(\ipah) - \log_{10}(\norm), \\
\nonumber \ipah &=& I_{\rm F770W} - I_{\rm F770W_{\star}}, 
\end{eqnarray}

\noindent with scatter $\sigma=0.43$~dex. $I_{\rm F770W}$ is the nonstellar continuum subtracted intensity and $I_{\rm F770W_{\star}}$ is the stellar continuum correction which comes from NIRCAM, following the relations presented in Section~\ref{ss:JWST}, and \norm\ is the normalization of the best-fit CO($2-1$) versus F770W$_\mathrm{PAH}$ power law for each galaxy (see Equation 4 in \citealt{chown2025}). This normalization aims to remove the galaxy-to-galaxy scatter in the relationship. We relied on the \norm\ values provided by Table 3 in \cite{chown2025}. We note that this PAH-to-CO fit underestimates the CO emission in some galaxies (see \citealt{chown2025}), and the place where it most prominently breaks is in galaxy centers, which show an offset relation. Equation~\ref{E:copah_renorm} also does not correct for dust continuum emission, which might also have more prominent contributions toward central regions. However, since we directly use the galaxy intensity maps that the equation was calibrated for, we do not particularly care if the emission is due to PAHs or small dust grains. Thus, we do not do any dust continuum corrections and acknowledge that toward the central regions, we are tracing emission of PAHs with a significant contribution from the dust continuum.

The emission from PAHs could also emerge from dust mixed with atomic gas \citep{sandstrom2023}. Hence, the relation between CO and PAHs is only used in regions where the inclination-corrected $I_\mathrm{F770W}^{\rm PAH} \geq 1~\mathrm{MJy~sr^{-1}}$ and where the molecular mass surface density ($\Sigma_{\rm mol}$) $\gtrsim 4$~M$_\odot$~pc$^{-2}$ \citep[see][for further explanation]{Leroy_2023a, chown2025}. 

Dust also exists in regions where CO is dark, typically the outer layers of molecular clouds where CO is photodissociated by ultraviolet radiation into ionized carbon (C\,\textsc{ii}) \citep[e.g.,][]{Wolfire_2010,Glover_2016}. These regions still contain H$_2$, which survives due to effective self-shielding and the presence of dust \citep{Van_1988}. Dust is also present in low-metallicity environments, where the reduced dust-to-gas ratio and lower carbon abundance limit the formation and survival of CO, causing it to trace only a small fraction of the total H$_2$ mass \citep{Leroy_2011, Bolatto_2008, Smith_2014}. Since the conversion from PAH to CO is applied to regions regardless of whether CO exists, the converted intensity might also trace CO-dark regions.  

\subsection{Environmental masks}

A key part of this work involves studying the properties of the (giant) molecular clouds with respect to the galactic environment. To categorize the galactic environment of each GMC, we employed the PHANGS environmental masks developed by \cite{Querejeta_2021}. These masks were created using the 3.6 $\mu$m Spitzer Survey of Stellar Structures in Galaxies (S$^{4}$G; \citealt{Sheth_2010}) along with other Near Infrared (NIR) observations. This approach produced detailed morphological masks of subgalactic environments for galaxies within the PHANGS survey. Notably, these masks are purely morphological and do not include kinematic information, which might lead to alternative definitions of the environments. 

For our study, we employ those simple masks to categorize the galactic environments into the following regions: center, which denotes the small bulge or nucleus; bar, encompassing the bar feature along with its ends (and any overlap with spiral arms); spiral arm, extending from the interbar region to the full extent of the spiral structure; interarm, covering the space between the bar and the spiral arms as well as in-between spiral arms, and the outer disk in galaxies lacking distinct spiral features or masks; and disk, which includes the region outside the bar.

\section{SCIMES cloud extraction}\label{S:SCIMES}

To identify GMCs, we adopted a machine learning algorithm called SCIMES\footnote{https://github.com/Astroua/SCIMES} \cite[][hereafter \citetalias{Colombo_2015}, see also \citealt{Colombo_2019}, and Appendix~\ref{a:scimes_update} for information about the updated version of the algorithm]{Colombo_2015}. This method considers the dendrogram of the emission in the framework of graph theory and utilizes spectral clustering to find regions with similar emission properties. Various other segmentation methods exist, from simple brightness thresholding \citep{Sanders_1985, Solomon_1987, Dame_2001} to more sophisticated approaches that identify characteristic geometries \citep[GAUSSCLUMPS,][]{Stutzki_1990}, or associate neighboring voxels by their values \citep[Clumpfind and CPROPS,][]{Williams_1994, Rosolowsky_2006, Rosolowsky_2021}

As described in \citetalias{Colombo_2015}, SCIMES classifies molecular clouds by first identifying dendrogram structures and then constructing a similarity (or affinity) matrix based on selected properties, which in this study are \Mmol\ and radius. Next, SCIMES computes the spectral embedding, applies the $k$-means algorithm, and determines the optimal clustering configuration for each galaxy. The parameters used to build the dendrograms using the \ipah\ maps of 66 galaxies and run SCIMES on this galaxy sample are described below. 

\subsection{Dendrogram structures}\label{SS:Dendro}

Dendrograms represent hierarchical structures within intensity maps, where emission regions are nested at different intensity levels. In this context, they provide a tree-like representation of cloud substructures based on spatial (position-position) information in two-dimensional data or spatial and spectral (position-position-velocity) information in three-dimensional data cubes. They are tree-like structures composed of leaves, branches, and trunks. Following the definition of \cite{Houlan_1992}, the leaves are the local maxima in the data; they are on top of the dendrogram and have no substructure. On the other hand, the branches can contain multiple substructures and split into other branches and leaves. A third structure, the trunk, is the largest structure with no parent structure, and represents the base of the dendrogram where all the branches and leaves eventually merge (i.e., the lowest contour level). 

The local maxima in this publication refer to the position-position (PP) maxima in \ipah\ maps at a homogenized resolution of 30$\,$pc. The structures that are due to noise are suppressed by ensuring that only emission above a given threshold ($\mathtt{ min\_value}$, typically taken to be a multiple of the noise rms) is considered in constructing the dendrogram, and that local maxima are eliminated if they cover an area lower than a certain number of pixels ($\mathtt{min\_npix}$, usually limited by the spatial resolution), or if its local maximum value is lower than a certain flux difference ($\mathtt{min\_delta}$, also refers to the step size for the intensity levels, usually set as a multiple of the noise) above the level at which that maximum merges with another local maximum. SCIMES uses the dendrogram implementations from \cite{Rosolowsky_2008}. The dendrogram and catalog of the structures within SCIMES are constructed using the $\mathtt{Python}$ package $\mathtt{Astrodendro}$.\footnote{https://dendrograms.readthedocs.io/en/stable/} It requires four parameters as input: $\mathtt{data}$, which is the data cube or in our case the \ipah\ map; $\mathtt{min\_value}$, in our case this is set to be the three times the worst sensitivity level of the data, $\sigma_{\rm rms}$, to make sure that our structures are significant; $\mathtt{min\_delta}$, also set to be three times the sensitivity level; $\mathtt{min\_npix}$, set to be the number of pixels per beam ($\Omega_{\rm beam}/\Omega_{\rm pix}$, where $\Omega_{\rm beam}$ and $\Omega_{\rm pix}$ are the solid angles of the resolution element and the pixel, respectively). We use a common $\sigma_{\rm rms}$ input for all 66 galaxies, which refers to the maximum $\sigma_{\rm rms}$ value of our sample (3$\sigma_{\rm rms} \sim 0.19 ~\mathrm{MJy~sr^{-1}}$  $\sim 0.21~ \rm K \,km \,s^{-1}$ as per Eq.~\ref{E:copah_renorm}).

\subsection{SCIMES}\label{SS:SC}

The SCIMES algorithm deploys spectral embedding and clustering techniques to enhance the identification of molecular clouds within a dendrogram. This approach leverages the properties of the graph Laplacian to map data into a space where clustering properties are more pronounced, followed by clustering in this transformed space using the $k$-means algorithm.

We used the $\mathtt{SpectralCloudstering}$ class in $\mathtt{SCIMES}$, which deals with embedding, clustering, and choosing the best clustering configuration. This class takes different input parameters, some crucial ones are $\mathtt{dendrogram}$, which is the dendrogram structure of the data generated by $\mathtt{Astrodendro}$; $\mathtt{catalog}$, the catalog that contains property (e.g., flux, radius) information of each structure created by the dendrogram; $\mathtt{header}$, corresponding to the header of the data FITS file; $\mathtt{criteria}$, that specifies which affinity matrix criteria to be used and can use multiple criteria (e.g., flux, radius, volume); $\mathtt{user\_scalpars}$, which is an optional scaling parameter that can be used to Gaussian smooth the affinity matrix. It should be noted that each affinity criterion has an associated scaling parameter that can be set. We also set $\mathtt{save\_all = True}$, which retains discarded structures, including both isolated leaves and intra-clustered leaves that are typically removed as noise. This ensures that small PAH cloud structures, which may still hold physical significance, are preserved in the final catalog. Additionally, this setting retains unassigned branches within clusters, allowing a more complete representation of the cloud hierarchy.

For our study, we used the molecular mass (see Sect.~\ref{ss:quantities}) and radius of the structures as the clustering affinity criteria, and the $\mathtt{pp\_catalog}$ function from $\mathtt{Astrodendro}$ to create the catalog. We also manually set the scaling parameters to 100 pc for the radius (roughly the sizes of large GMCs; e.g., \citealt{Rosolowsky_2021, Demachi_2024}), and consistent with \cite{Vieira_2025}. We also set the molecular mass scaling parameter to $5\times10^{6}\, \rm \msun$ (within the upper limit of a GMC; e.g., \citealt{Demachi_2024}). Manually setting the scaling parameters is crucial in spectral clustering, as this scaling parameter essentially determines the weighting of radius and mass when computing similarities between clouds, and removes structures that show affinity connections on scales larger than typical GMC scales (see Appendix~\ref{ss:scaling_appendix} for details on our choice of scaling parameters).

\section{Molecular cloud properties}\label{S:MC_Props}

In this section, we present the different methods used to calculate the sizes and fluxes of the clouds identified by SCIMES. We directly infer the radius based on the exact footprint area of the cloud (e.g., \citealt{Williams_1994, Heyer_2001}), and following \cite{Rosolowsky_2006} (hereafter \citetalias{Rosolowsky_2006}), we also use moment measurements to compare our findings to results from CO-based GMC catalogs for the same PHANGS galaxies (Hughes et al. in prep.).

\subsection{Directly measured properties} \label{SS:method}

One direct way of finding the radius of a structure is by directly inferring it from the area (e.g., \citealt{Williams_1994, Heyer_2001}). Consider a cloud with $N$ pixels; then the area of the cloud is simply

\begin{equation} \label{E:Area}
    A_{\rm cloud} = \sum^{N}_{i}A_{\rm pix}\;,
\end{equation}

\noindent where $A_{\rm pix}$ represents the pixel area. Subsequently, the deconvolved equivalent radius of the cloud can be found as

\begin{equation}\label{E:Reff}
    \mathrm{R_{eq}} = \sqrt{\frac{A_{\rm cloud} - A_{\rm beam}}{\pi}},
\end{equation}

\noindent where $A_{\rm beam}[{\rm pc}^{2}] = 1.18^{2} \sigma_{\rm b,maj}[{\rm deg}]\sigma_{\rm b,min}[{\rm deg}] (\frac{\pi \times D [{\rm pc}]}{180})^{2}$ is the beam area of the observation, and $\sigma_{\rm b, maj}$ and $\sigma_{\rm b, min}$ are the beam major and minor axes expressed in degrees (see Appendix~\ref{ss:radii} for more information). 

Following \citetalias{Rosolowsky_2006}, the radius of the cloud can also be assessed by intensity-weighted moment-based measurements 

\begin{eqnarray}
\label{E:rad_mom}
\mathrm{R} &=& \eta \, \sigma_{\rm r}, \\
\sigma_{\rm r} &=& \sqrt{\sigma_{\rm maj} 
\sigma_{\rm min}}, \nonumber
\end{eqnarray}

\noindent where $\eta$ depends on the light distribution within the cloud (see \citetalias{Rosolowsky_2006}). In this paper, we use a value of $\eta = \sqrt{2 \ln 2} = 1.18$ corresponding to the half-width at half-maximum of a Gaussian distribution (e.g., \citealt{Rosolowsky_2021}). The size of the cloud is $\sigma_{\rm r}$, and $\sigma_{\rm maj}$ and $\sigma_{\rm min}$ are the second spatial moments (see \citetalias{Rosolowsky_2006} for further information).

We converted the \ipah\ to $I_\mathrm{CO(2-1)}$ (see Sect.~\ref{SS: 7.7 to CO}) and obtained the converted luminosity in CO of a cloud, which can then be defined as

\begin{equation}\label{E:lum}
    L_{\rm CO} = A_{\rm pix}\sum_{i}^{N} F_{i} \, .
\end{equation}

\noindent $F_{i}$ represents here the flux in units of $\rm K\,km\,s^{-1}$ of an element in the cloud obtained from Eq.~\ref{E:copah_renorm}, and it is summed over all cloud pixels. $A_{\rm pix}$ is the projected physical area of the pixel in units of $\rm pc^{2}$.

\subsection{Derived physical properties} \label{ss:quantities}

In this section, we present the physical properties that we derived from either the moments method or the direct estimation of the radius from the beam-deconvolved area of the cloud. The radii of the clouds are converted from arcsec to parsec using $D$ measurements from Table~3 in \cite{Leroy_2021}.

After converting \ipah\ to $I_{\rm CO(2-1)}$ following Eq.~\ref{E:copah_renorm}, the molecular mass of CO can be derived from the luminosity (Eq. \ref{E:lum}) as

\begin{equation}\label{E:Mass}
   \mathrm{ M_{\rm mol}}~[\mathrm{M_{\odot}}] = \alpha_{\rm CO}\,[\mathrm{M_{\odot}(K\,km\,s^{-1}\,pc^{2})^{-1}}] \times L_{\rm CO}\, [\mathrm{K\,km\,s^{-1}\,pc^{2}}],
\end{equation}

\noindent where \aco\ (and previously defined $X_{\rm CO}$) refer to the CO-to-H$_{2}$ conversion factor (see \citealt{Bolatto_2013} for detailed definition). For each identified cloud, we take a median \aco\ value within its boundary, and then multiply it by the $L_{\rm CO}$ of the cloud to obtain \Mmol .

In the case of PHANGS-ALMA, the CO transition observed is CO($2-1$), thus, we refer to the conversion factor as $\alpha_{\rm CO(2-1)}$. We also rely on an updated version of the PHANGS-ALMA $\alpha_{\rm CO(2-1)}$ estimates presented by Sun et al. (in prep.) (see also \citealt{Sun_2022,Sun_2023}), based on the recommended \aco\ from \cite{schinnerer2024}, which incorporates corrections for excitation, CO-dark gas, and emissivity variations. Following \cite{schinnerer2024}, we define $\alpha_{\rm CO(2-1)}$ as:

\begin{eqnarray}\label{E:alpha_sl}
    \alpha_{\rm CO(2-1)} &\approx& 4.35 \times f (Z) \times g (\Sigma_{\star}) \times R_{21}(\Sigma_{\rm SFR})^{-1},\\
     R_{21}(\Sigma_{\rm SFR}) &=&  0.65(\sigsfr/0.018)^{0.125}, \nonumber
\end{eqnarray}

\noindent where $f (Z) = (Z/Z_{\rm solar})^{-1.5}$ is the CO-dark factor that depends on the metallicity ($Z$) for 0.2 < $Z/Z_{\rm solar}$ < 2 (see \citealt{schinnerer2024} for further information), where $Z_{\rm solar}$ is the solar metallicity ($\rm 12+log(O/H) =8.69$ as per \citealt{Asplund_2009}). This prescription complements observations of dust and C\,\textsc{ii}, where a higher \aco\ is needed in regions of low mass and low metallicity (e.g., \citealt{Leroy_2011,Jameson_2016}). It is also an accompaniment to simulations which reveal a strong dependence of \aco\ on metallicity, with significantly suppressed CO emission at low metallicity and low extinction (e.g., \citealt{Glover_2011,Hu_2022}). However, $f (Z)$ does not take into consideration additional factors such as the dust-to-metals ratio, interstellar radiation field, cosmic ray ionization rate, and the structure of the clouds themselves, which all play an important role, and further add to the uncertainty of the \Mmol\ estimation. The starburst emissivity factor is $g(\Sigma_{\star}) = \mathrm{max}(\Sigma_{\star}/100, 1)^{-0.25}$, where $\Sigma_{\star}$ is the stellar mass surface density in units of $\mathrm{M_{\odot}\, pc^{-2}}$. Additionally, $R_{21}(\Sigma_{\rm SFR})$ is the line ratio between CO($2-1$) and CO($1-0$), and $\Sigma_{\rm SFR}$ is the star formation rate surface density (see \citealt{Leroy_2022} and \citealt{schinnerer2024} for more information). The metallicity is approximated as a function of galactocentric radius based on the global mass-metallicity relation of \cite{Sanchez_2019}, adopting the PP04 O3N2 calibration \citep{Pettini_2004} and extrapolating the predictions to the whole PHANGS-ALMA footprint using a metallicity gradient as per \cite{Sanchez_2014} (see Sun et al. in prep. for more information).

We then calculate the molecular mass surface density as follows

\begin{eqnarray}
\label{E:sigma_1}
    \Sigma_{\rm mol}~[\rm M_{\odot}pc^{-2}] &=& \frac{\rm M_{mol}}{\rm \pi R_{eq}^{2}}, \\
\label{E:sigma_2}
    \Sigma_{\rm mol,R}~[\rm M_{\odot}pc^{-2}] &=& \frac{\rm M_{mol}}{2\pi \rm R^{2}}. 
\end{eqnarray}

\noindent The model in Eq.~\ref{E:sigma_1} follows that \sigmol\ is directly inferred from the area of the cloud. Meanwhile, the model in Eq.~\ref{E:sigma_2} follows the two-dimensional Gaussian cloud model in which half the mass is contained inside the FWHM. We used the first model throughout the paper and the second model only to compare PAH and CO clouds in Sect.~\ref{SS:compare}.

\subsection{Error estimation}\label{ss:error}

We applied morphological alterations to the shapes of individual clouds to estimate the errors in their properties, given that many of their properties depend on the exact cloud footprint, and thus on the number of pixels assigned in the cloud segmentation process. The SCIMES-defined structures exhibit only slight variations depending on the input parameters of the dendrograms, except for $\mathtt{min\_value}$, where more variations are observed \citep{Colombo_2015}. Varying the scaling parameters by a small fraction ($\sim 20\,\%$) also leads to slight variations in the properties of the clouds. Therefore, to quantify the potential uncertainties in the cloud properties due to the choice of assignment mask, we used the $\mathtt{binary\_erosion}$ and $\mathtt{binary\_dilation}$ functions of the $\mathtt{scipy.ndimage}$$\footnote{https://docs.scipy.org/doc/scipy/reference/ndimage.html}$ $\mathtt{Python}$ package. We applied a dilation and erosion with one-third the number of pixels per beam ($\sim$ 10 pc) and calculated the cloud properties of both as upper and lower limits, respectively.

Following \citetalias{Rosolowsky_2006}, we also estimate the errors for the properties of the clouds by bootstrapping. This involves generating several trial clouds from the original data by randomly sampling the data points within the cloud, with some points being repeated. A cloud in this case is considered to be a collection of data $\{x_{i}, y_{i}, T_{i}\}$, for $i = 1, ...\,, N,$ where $N$ is the number of points in the cloud. We measured the properties of each trial cloud and estimated the uncertainty as the $84{\rm th} - 50{\rm th}$ and $50{\rm th} - 16{\rm th}$ percentiles of the distributions. This bootstrapping considers the errors from $D$, \ipah, and the fit error (including the scatter) in Eq.~\ref{E:copah_renorm}. 

To assess the bias in the cloud \Mmol\ according to \aco , we use five different \aco\ prescriptions. The first prescription is represented in Eq.~\ref{E:alpha_sl}, and it is our preferred prescription used in the analysis. The second one is a constant Galactic $\alpha_{\rm CO(2-1)} = \frac{4.35}{0.65} = 6.69\, \rm \msunperpcsq (K\,km\,s^{-1})^{-1}$, where $R_{21}= 0.65$ is based on \cite{Leroy_2013} and \cite{denbrok_2020}, measured at kpc scales, and $\alpha_{\rm CO(1-0)} = 4.35~\rm \msunperpcsq (K\,km\,s^{-1})^{-1}$ is the standard Galactic value at solar metallicity (i.e., \citealt{Bolatto_2013}). The third description is according to a varying metallicity and gas surface density \aco\ based on Eq. 31 in \cite{Bolatto_2013}. The fourth one also varies according to Eq. 2 in \cite{Teng_2024}, which relies on the intensity-weighted mean molecular gas velocity dispersion measured at $150$ pc scale. The last prescription depends only on the metallicity \citep[see][]{Sun_2020a}. The exact creation of each \aco\ map is further described in Sun et al. (in prep.). 

We calculate luminosity-weighted averages of both the cloud properties and their uncertainties within the FOV of all galaxies. This method is motivated by \cite{Leroy_2016} (see also \citealt{Sun_2022}), where they calculated the intensity-weighted average for clouds within an aperture encompassing several GMCs. Our bootstrapping technique yields a luminosity-weighted uncertainty average of the cloud mass measurement of $\sim 20\,\%$ and that of the radius measurement is $\sim 7\,\%$. However, erosion and dilation yield a luminosity-weighted uncertainty average of $\sim 54\,\%$ and $\sim 58\,\%$ for the mass and radius measurements, respectively. To assess the \aco\ bias of our prescription (see Eq.~\ref{E:alpha_sl}), we compare the luminosity-weighted \sigmol\ average value using our adopted \aco\ with the other prescriptions. In spiral arms, interarms, and disks, the \sigmol\ variation due to adopting another \aco\ prescription is on average  $\sim 23\, \%$. Meanwhile, in bars, the variation is $\sim 49\, \%$, and in centers  $\sim 125\, \% $. This highlights the uncertainty of the measurements toward the central regions of the galaxy. For the final error on the cloud properties, we apply Gaussian error propagation on the bootstrapping and morphological alteration methods, and provide the \Mmol\ calculated using the different \aco\ prescriptions in our cloud catalog.

\begin{figure*}[h]
    \centering
    \includegraphics[width=0.85\paperwidth]{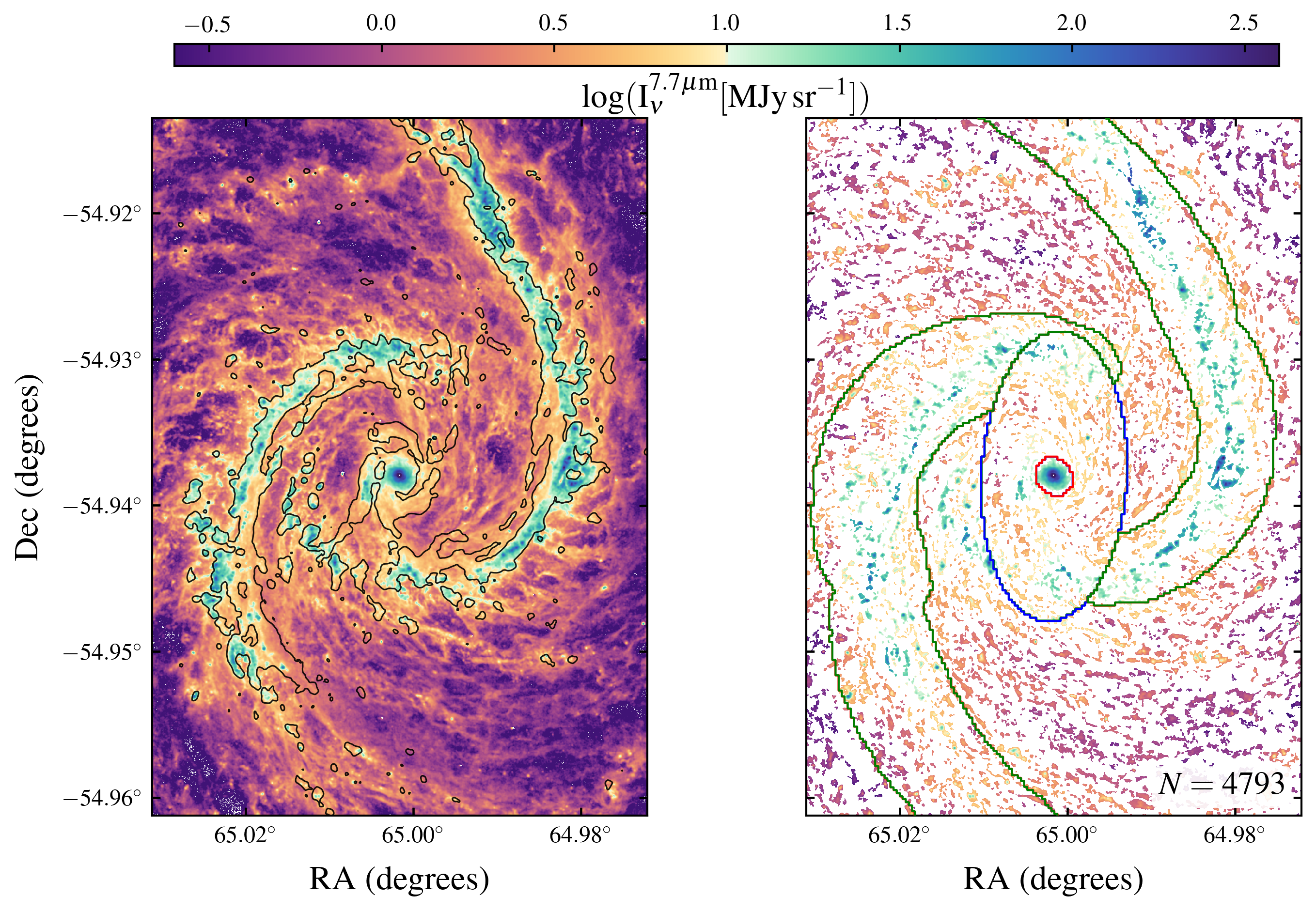}
    \caption{A zoomed-in view of NGC1566, one of the 66 galaxies. \textit{Left}: Continuum-subtracted intensity image of the galaxy in the F770W MIRI band. The 2 $\sigma$ CO intensity contours from PHANGS-ALMA are represented in black. \textit{Right}: PAH cloud structures identified by SCIMES  color-coded by their F770W intensities. The green, blue, and red contours indicate the spiral arm, bar, and central region masks, respectively. The interarm region in this case consists of the remaining clouds that are not enclosed by the contours. The number of PAH clouds identified in this galaxy is represented in the bottom right. The color bar at the top of the image shows the 7.7$\mu$m intensity range of the identified clouds.}
    \label{fig:NGC1566}
\end{figure*}

\subsection{Cloud population}\label{SS:CG}

Figure \ref{fig:NGC1566} shows an example of the PAH clouds extracted with SCIMES for one of our galaxies (NGC 1566). In this figure, we show this cloud segmentation at 30 pc resolution to show the performance of SCIMES in recognizing structures. We further provide both a common-resolution (30 pc) and sensitivity (0.19$~\mathrm{MJy~sr^{-1}}$) cloud catalog, and a native resolution and sensitivity catalog.

We selected 77,884 clouds for analysis that meet our selection criteria. Initially, a total of 108,466 clouds were identified across the 66 galaxies, and all of them are included in the final cloud catalog (see information about native resolution in Appendix~\ref{s:catalog}). However, since we do not have velocity information, clouds with different velocities might overlap if they are along the same line of sight. For that, we used assignment cubes of GMCs identified in CO($2-1$) from the PHANGS-ALMA survey using CPROPS (\citealt{Rosolowsky_2021}; Hughes et al. in prep.) to check, flag, and exclude overlapping 7.7$\mu$m-identified clouds in velocity space from our analysis. Also, we flagged and excluded clouds on the edge of the maps ($f\_edge = 0$ to exclude in the final catalog). To ensure that most of the chosen PAH clouds are tracing the molecular phase, we only included in our analysis clouds that have an average $\ipah\ > 1 \,$ \mjysr\ (twice the threshold suggested by \citealt{chown2025} since the $0.5$ \mjysr\ threshold still includes a significant amount of $\sigmol < 4 \,\msunperpcsq$ clouds) or $ \sigmol > 4\, \msunperpcsq$. We set a flag, $f\_mol = 1$, to include the latter from the final catalog.

We projected the native-resolution CPROPS GMC assignment cubes onto the same grid space as the SCIMES assignment maps to exclude overlapping clouds in velocity space. Then, after applying a 2D projection of the clouds, we checked how many CO overlapping cloud pixels exist in a specific SCIMES-identified cloud. Finally, we flagged clouds that have more than $30\%$ contribution from multiple CO clouds. In the final catalog, we include $f\_overlap$ as a binary flag, where a value of one corresponds to overlapping structure, and zero to non-overlapping ones. We also include $overlap\_ratio$ to check the ratio of overlap (e.g., a value of 0.3 corresponds to $30\, \%$ overlap). Once we match those clouds with our clouds, we find that $\sim12\,\%$ of the full cloud sample comprises overlapping clouds. This poses a challenge in central regions as multiple velocity elements and a high-velocity dispersion exist in those regions \citep{Rosolowsky_2021}; we report that $\sim 65\, \%$ of clouds in the central regions, $\sim 24\,\%$ bar clouds, $\sim 13\,\%$ spiral arm clouds, $\sim 5\,\%$ interarm clouds, and $\sim 12\,\%$ disk clouds contain overlapping counterparts that are flagged out in the analysis.

The summary of the flagging is represented in Table~\ref{T:flag}. Also, galaxy centers have well-connected leaf structures and high branch weights in the dendrograms (i.e., extremely bright). As mentioned before, those structures are large and massive, and the central regions mainly comprise overlapping clouds in velocity. Therefore, we flag out all central clouds (1303 clouds) as explained in this section, but show them in Fig.~\ref{fig:CO_compare_all} to emphasize the bias of including them.

In Table~\ref{T:sample}, we show that the full cloud sample covers a median of $40^{+4}_{-5}\,\%$,\footnote{The upper limit is the difference between the 84th and 50th percentiles, and the lower limit is between the 50th and 16th percentiles of the fraction of flux within the clouds in the 66 galaxies.} and the filtered subsample covers a median of $26^{+12}_{-11}\,\%$ of the emission from the \ipah\ maps. This highlights that most of the flagged clouds are high \sigmol\ clouds ( $> 100 \, \msunperpcsq$) and poses a bias in our analysis toward lower \sigmol\ clouds.

\setlength{\tabcolsep}{2pt} 
\renewcommand{\arraystretch}{1.7}
\begin{table}
\centering
\caption{Number of clouds excluded from our analysis using each flagging method.}
\begin{tabular}{|c|c|c|c|c|c|c|}
\hline
\multicolumn{1}{|c}{$N\_edge$} & 
\multicolumn{1}{|c}{$N\_mol$} & 
\multicolumn{1}{|c}{$N\_overlap$ } &
\multicolumn{1}{|c}{$N\_beam$ } &
\multicolumn{1}{|c}{$N\_center$ } &
\multicolumn{1}{|c}{$N_{\rm flag}$ } &
\multicolumn{1}{|c|}{$N_{\rm tot}$ } \\
 (1) & (2) & (3) & (4) & (5) & (6) & (7) \\
\hline
3633 & 10 237 & 12 844 & 6641 & 1303 & 30 278 & 77 884\\
\hline
\end{tabular}
\label{T:flag}
\tablefoot{(1) Number of clouds overlapping the edge of our field of view. (2) Number of clouds with mean \ipah\ < 1 \mjysr\ and \sigmol\ < 4 \msunperpcsq . (3) Number of overlapping clouds in velocity space. (4) Number of clouds with sizes comparable to the beam size. (5) Number of clouds in galaxy centers. (6) Total number of flagged clouds. (7) Final number of clouds used in the analysis.}
\end{table}
\setlength{\tabcolsep}{6pt}

\section{Results and discussion}\label{S:Results}

In this section, we investigate how well PAH-identified clouds using the F770W JWST band (see Appendix~\ref{ss:F1130W} for a comparison between cloud properties extracted using the F770W and F1130W bands) could resemble CO-identified GMCs. We further rely on the common-resolution data to compare the molecular cloud properties in different galactic environments according to the \cite{Querejeta_2021} environmental masks. This was previously done on the PHANGS-ALMA sample (e.g., \citealt{Rosolowsky_2021,Sun_2022}). We then present the cloud mass-radius scaling relation and mass spectrum per environment. Finally, we discuss how the cloud properties vary with respect to galactocentric radius and highlight the caveats.

\subsection{\ipah\ and CO  cloud property comparison}\label{SS:compare}

We compare the properties of the PAH clouds identified by SCIMES at 30~pc resolution to cross-matched CO clouds identified by CPROPS at 90~pc resolution in 27 PHANGS galaxies. The galaxies in CO have a common sensitivity of $0.15 \, \rm K$  
(\citealt{Rosolowsky_2021}, Hughes et al. in prep.). We note that CO clouds were identified using position-position-velocity (PPV) data. We find that 41$~\%$ of the PAH clouds in the 27 galaxies could be associated with CO counterparts in the same FOV as JWST. We note that the completeness limit of PHANGS-ALMA is $4.7\times10^{5}~\msun$, which is $\sim 2.4$ dex higher than our lowest \Mmol\ clouds.

For comparison, we use \Mmol\ measurements from the GMC catalog provided by Hughes et al. (in prep.) and based on the \cite{schinnerer2024} \aco\ prescription (see Equation~\ref{E:alpha_sl}). We also use second-moment measurements for the radii (as described in Sect.~\ref{SS:method}) for both PAH and CO clouds to maintain consistency. 

\begin{figure}
    \centering
    \includegraphics[width=0.43
    \paperwidth]{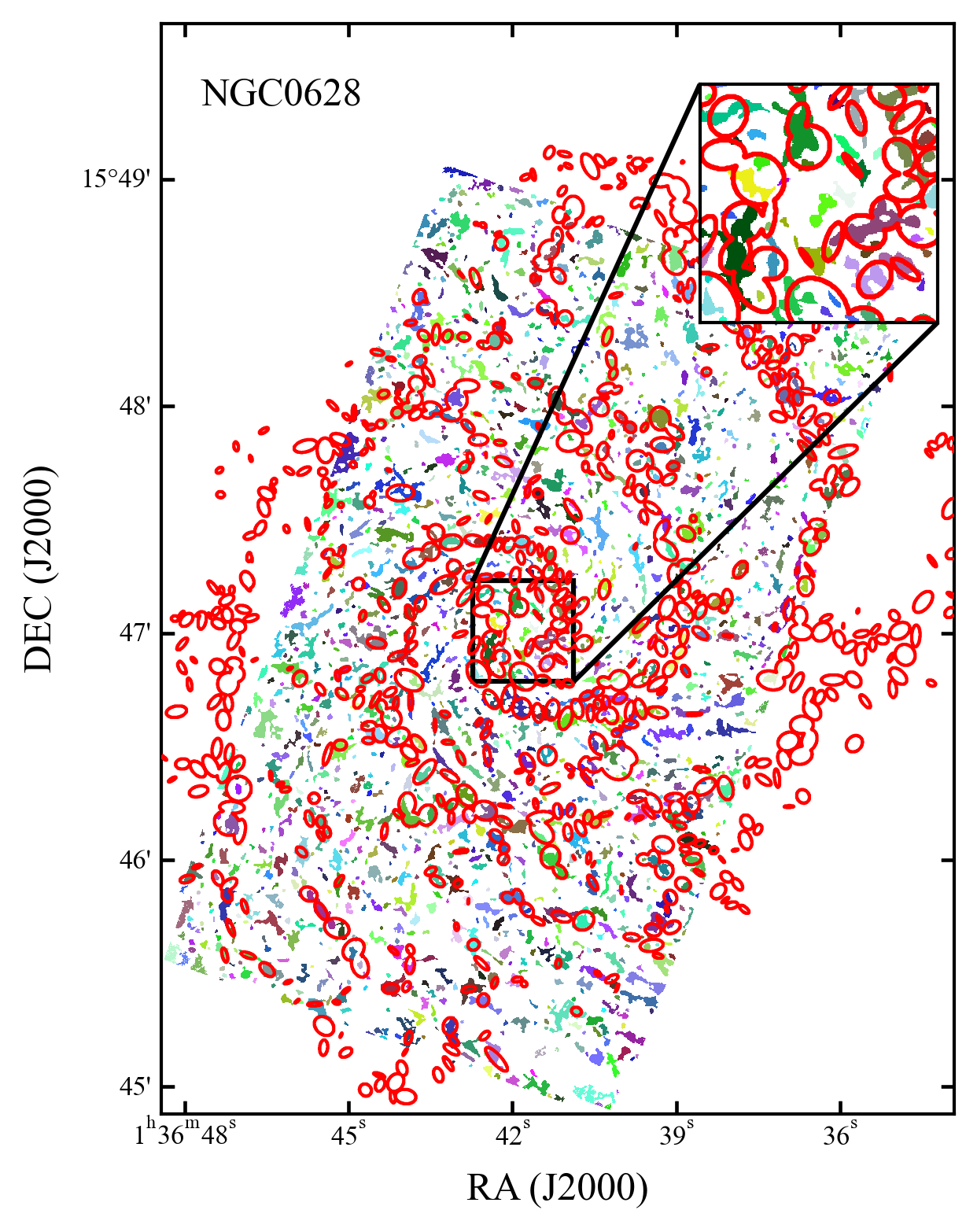}
    \caption{GMCs in NGC0628. CO-identified GMCs using CPROPS are shown as red ellipses (Hughes et al. in prep.), and the 7.7$\mu$m identified PAH clouds using SCIMES are shown in the background in different colors. A zoomed-in view of the central region is shown in the inset (upper right corner); it focuses on the structures identified by both PAH and CO. This image highlights the resolution advantage of PAH clouds and the better sensitivity compared to CO, which allows the detection of fainter and smaller clouds where CO is not detected.}
    \label{fig:clust_compare}
\end{figure}

\begin{figure*}
\begin{center}
\includegraphics[width=0.33\textwidth]{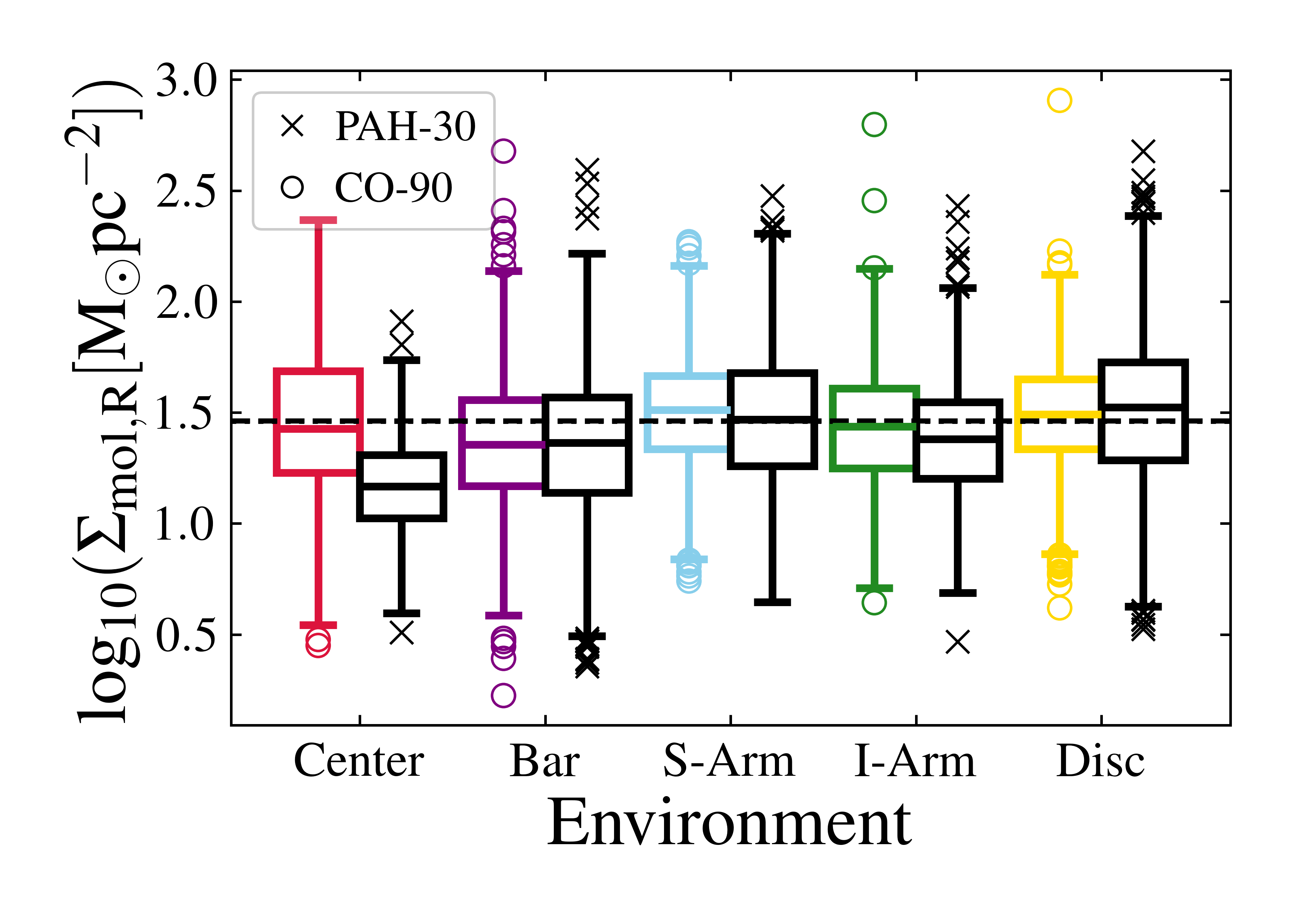}
\includegraphics[width=0.33\textwidth]{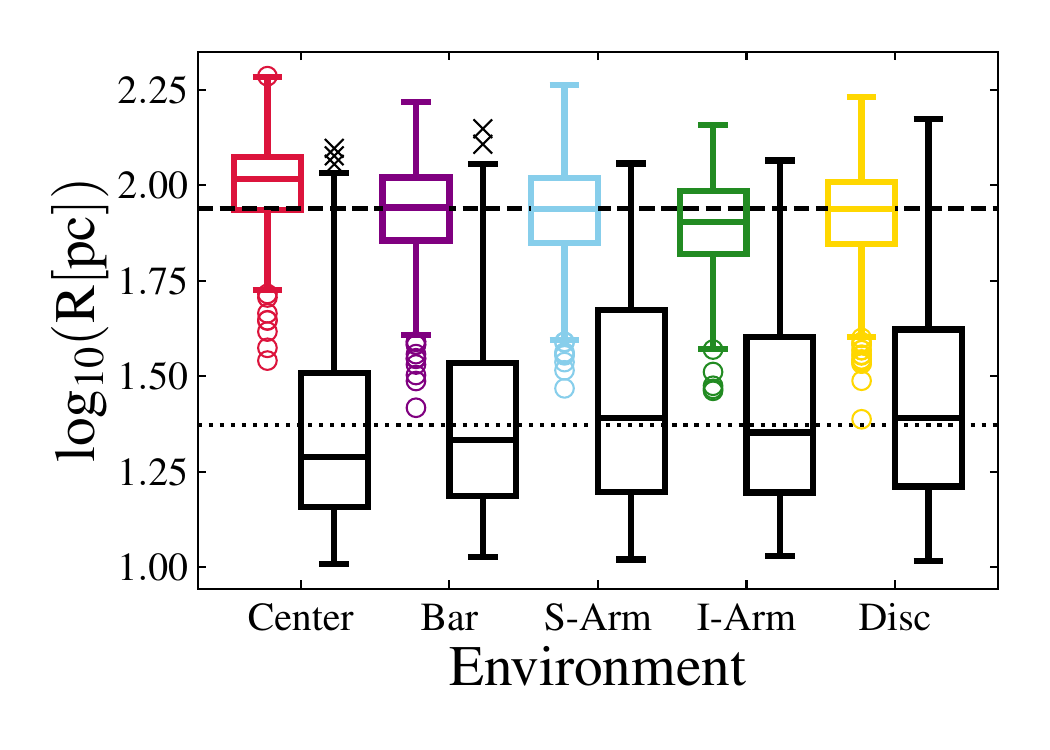}
\includegraphics[width=0.33\textwidth]{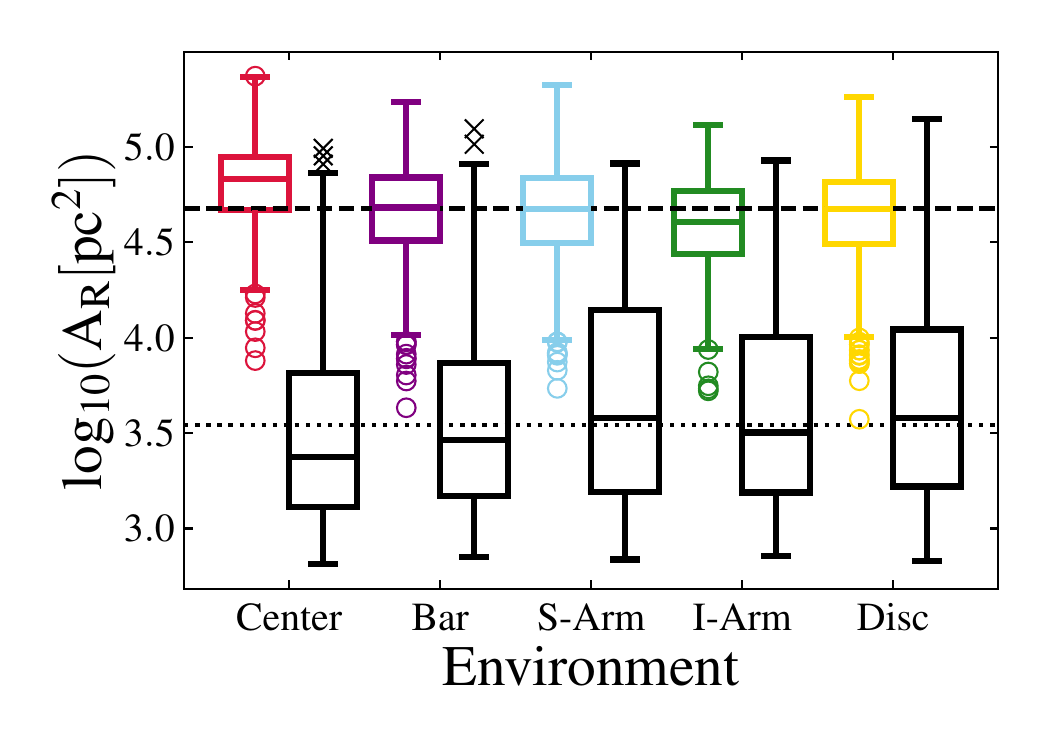}
\caption{
Box plots with quantiles and outliers comparing the $\Sigma_{\rm mol,R}$ (\textit{left}), cloud radius (\textit{middle}), and cloud area ($\mathrm{A_{R}}$; \textit{right}) distributions of cross-matched PAHs at 30 pc physical resolution and CO clouds at 90 pc in a subsample of 27 galaxies. The colored boxes represent the PAH cloud property distributions without overlapping clouds. The black boxes represent the property distributions of the cross-matched CO clouds. The dashed and dotted horizontal lines represent the median property of the full sample of CO clouds and PAH clouds, respectively.}
\label{fig:CO_compare_all}
\end{center}
\end{figure*}

The median \Mmol\ is $8.3 (\pm 0.2)\times 10^{4}\,\msun$ in the PAH-cloud sample; this value is one dex lower than the completeness limit of PHANGS CO-identified GMCs \citep{Rosolowsky_2021}. Also, the median cloud radius is 34.7 pc. This highlights the better sensitivity and physical resolution of JWST that allows the detection of fainter and smaller clouds than CO-identified GMCs as seen in Fig.~\ref{fig:clust_compare} and \ref{fig:CO_compare_all}. However, this does not test how well PAH clouds recover CO-traced clouds. Instead, we use $\Sigma_{\rm mol,R}$ (see Eq.~\ref{E:sigma_2}) to compare the two cloud samples, reducing the effect of the different resolutions between the studies. We therefore compare the $\Sigma_{\rm mol,R}$ distributions of the matched CO and PAH clouds represented in Fig.~\ref{fig:CO_compare_all}. The median $\Sigma_{\rm mol,R}$ of the PAH clouds is 28.7 $\msunperpcsq$, which is the same as that of the CO cloud sample. Also, no differences are observed in the $\Sigma_{\rm mol,R}$ distributions of both matched PAH and CO clouds in the different environments, except in the central regions. There, we notice a decrease of $\sim$ 0.3 dex in PAH-cloud $\Sigma_{\rm mol,R}$, which introduces a caveat in our cloud identification in the central regions. This is due to the removal of overlapping clouds in velocity space from our analysis, and because the PAH-to-CO relationship most prominently breaks in galaxy centers (see \citealt{chown2025}).

\begin{figure}
\begin{center}
\includegraphics[width=0.48\textwidth]{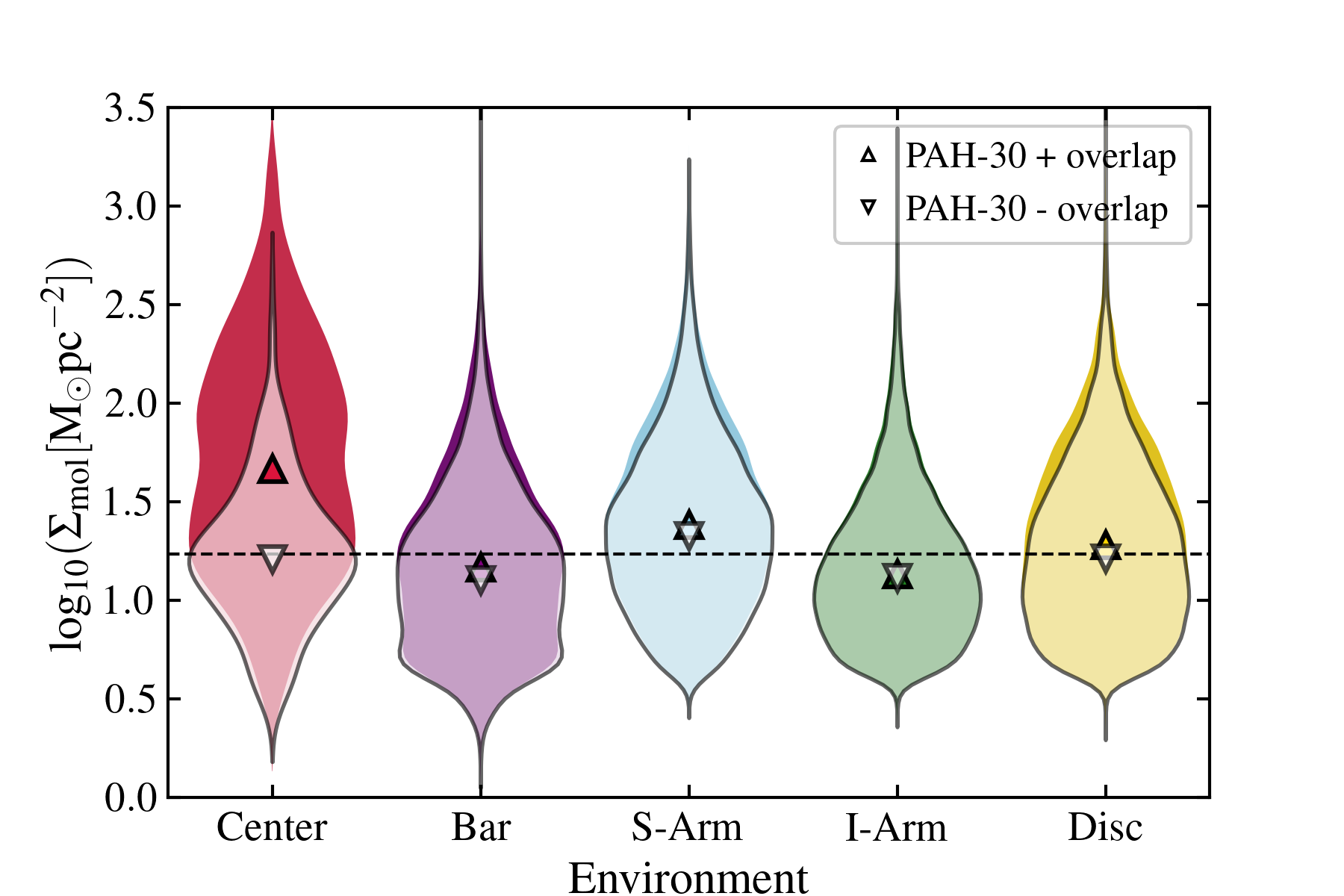}
\caption{Violin plots showing the distribution and medians of \sigmol\ in each galactic environment for the full sample of PAH clouds (colored) and the 77,844 clouds without overlap (transparent) in the 66 galaxies. The dashed line represents the median \sigmol\ for the full PAH cloud sample.}
\label{fig:CO_compare_pah}
\end{center}
\end{figure}

\begin{figure*}
    \centering
    \includegraphics[width=0.327
    \paperwidth]{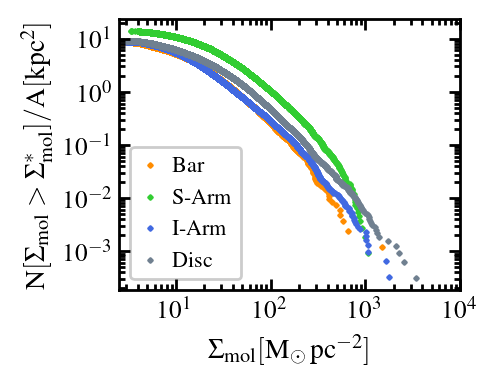}
    \includegraphics[width=0.25
    \paperwidth]{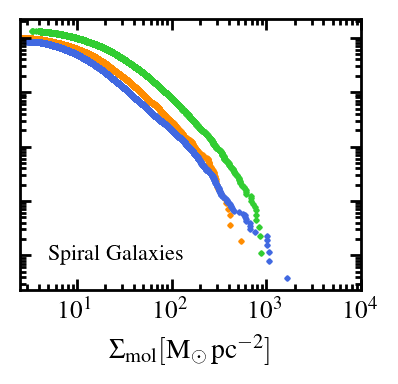}
    \includegraphics[width=0.25
    \paperwidth]{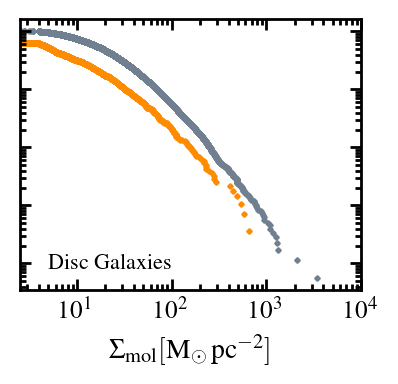}
    \caption{\textit{Left:} Cumulative distributions of the molecular mass surface densities from the full cloud sample. The different colors represent the different environments. The y-axis is the fraction of clouds with a surface density greater than a given value. All distributions are normalized by the total area of their specific environment, $\mathrm{A}$. \textit{Middle:} Same as the left plot, but only considering barred spiral galaxies and excluding disks. \textit{Right:} Same as the left plot, but only considering barred disk galaxies and excluding spirals. We removed the central region from the PDFs due to overlapping cloud bias. }
    \label{fig:PDF_SD}
\end{figure*}

\begin{figure}
    \centering
    \includegraphics[width=0.40
    \paperwidth]{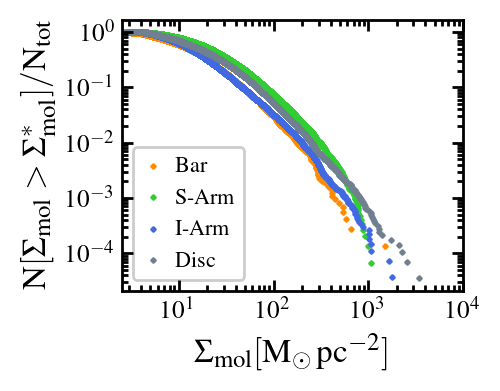}
    \caption{Cumulative distributions of the molecular mass surface densities from the full cloud sample. The different colors represent the different environments. The y-axis is the fraction of clouds with a surface density greater than a given value. All distributions are normalized by the total number of clouds in their corresponding environment, $\mathrm{N_{tot}}$. We removed the central region from the PDFs due to overlapping cloud bias. }
    \label{fig:PDF_SD_clouds}
\end{figure}

\renewcommand{\arraystretch}{1.4}
\begin{table*}[h]
\caption{Summary of PAH cloud properties in various galactic environments.}
\centering
\begin{tabular}{|c|c|c|c|c|c|c|}
\hline
Property & Statistic & Global & Bar & Spiral Arm & Interarm & Disk \\
\hline
$ N_{tot}$ & - & 77\,884 & 7298 & 14\,902 & 27\,120 & 28\,564 \\
\hline
$ N_{tot}/A~[\rm kpc^{-2}]$ &  - &  8.3 &  8.6 &  13.4  & 7.4 & 7.6 \\
\hline
 & Median & $6.6^{+24.6}_{-4.6}$ & $4.6^{+16.2}_{-3.2}$ & $7.1^{+30.7}_{-4.8}$ & $5.9^{+21.2}_{-6.7}$ & $7.7^{+27.0}_{-5.6}$ \\
$\Mmol \, [10^{4}\,\msun]$ & Mean & 20.8 & 14.2 & 25.9 & 17.6 & 22.8 \\
 & Weighted& 137.6 & 75.7 & 196.9 & 111.7 & 131.8 \\
\hline
 & Median & $37.4^{+46.3}_{-20.0}$ & $34.9^{+41.8}_{-18.4}$ & $34.7^{+44.2}_{-18.6}$ & $38.2^{+47.0}_{-20.2}$ & $38.9^{+47.5}_{-21.1}$ \\
$\Reff \, [\mathrm{pc}]$ & Mean & 48.3 & 45.1 & 45.5 & 49.4 & 49.69\\
 & Weighted& 92.2 & 85.2 & 92.7& 95.2 &91.2 \\
\hline
 & Median & $15.3^{+28.7}_{-8.7}$ & $12.7^{+21.5}_{ -7.5}$ & $21.3^{+38.7}_{-12.8}$ & $13.0^{+20.8}_{-6.7}$ & $16.3^{+32.0}_{-9.6}$ \\
$\sigmol \, [\msunperpcsq]$ & Mean & 30.0 & 23.3 & 39.3 & 24.2 & 32.4 \\
 & Weighted & 48.4 & 33.0 & 62.0 & 40.3 & 49.1 \\
\hline
\end{tabular}
\tablefoot{Each property is shown with its median, mean, and luminosity-weighted mean values. Medians with the 84th - 50th percentile and 50th - 16th percentile displayed in superscript and subscript, respectively. Here, $N_{tot}/A$ refers to the total number of clouds within an environment divided by the total galaxy-by-galaxy summed area of a specific environment.}
\label{T:Prop_Summary}
\end{table*}

\subsection{Masses, radii, and surface densities}\label{SS:Distributions}

The properties of the clouds, such as \Reff, \Mmol\, and \sigmol\ analyzed in this paper are presented in Table~\ref{T:Prop_Summary}. We highlight that various methods exist for calculating the properties, as outlined in Sect.~\ref{S:MC_Props}, and these methods can produce differing results, which may impact comparisons. Therefore, caution should be taken when calculating and comparing properties with other cloud catalogs. A summary of the cloud properties listed in our catalogs can be found in Table~\ref{T:property_description}.

The lowest and highest cloud \sigmol\ medians are 5.8 and 49 \msunperpcsq\ corresponding to NGC 4941 and NGC 4781, respectively. This shows that the sensitivity of the \ipah\ images allows us to locate faint structures (< 10 \msunperpcsq; see also \citealt{sandstrom2023, Leroy_2023, chown2025}) that could be associated with atomic or molecular clouds and are not detected by ALMA observations (e.g., \citealt{Rosolowsky_2021}). We speculate that those structures could be associated with faint clouds that CO does not detect. Notably, 51$~\%$ of those clouds are located in the disk, and the rest are equally spread in the other galactic environments.

To investigate whether clouds in spiral galaxies have distinctive properties, we split our sample into ``spiral'' and ``disk'' galaxies (i.e., with and without strong spiral arms present, respectively) according to the environmental classification of \cite{Querejeta_2021}. We associate 30,362 clouds with 40 disk galaxies and 47,522 clouds with 26 spiral galaxies. As seen in Table~\ref{T:Prop_Summary}, clouds in the spiral arm have the highest \sigmol\ compared to other environments, followed by disk clouds, and the least dense clouds are in the interarm and bar regions (see also Fig.~\ref{fig:CO_compare_pah}). While the PAH cloud $\sigmol$ values in bars appear similar to those of CO clouds (e.g., Fig.~\ref{fig:CO_compare_all}), CO emission might be systematically underestimated in bar ends across the full PHANGS–JWST sample.

When we look at the \sigmol\ probability distribution function (PDF) of the interarm clouds in Fig.~\ref{fig:PDF_SD} and \ref{fig:PDF_SD_clouds}, we see fewer clouds with densities > 10 \msunperpcsq\ compared to spiral-arm clouds. This indicates that spiral arms favor denser clouds, which agrees with the picture proposed by \cite{Koda_2009}, where the potential well of the spiral arm assists in the formation of massive, dense structures. The contrast in surface densities between spiral arms and interarm is seen in the PHANGS-ALMA galaxies \citep[e.g.,][]{Sun_2020, Sun_2022, Meidt_2021, Querejeta_2024}. This picture is also backed by other observations \citep{Ragan_2014} and simulations \citep{Cabral_2016, Cabral_2017, Tress_2021}, where they report an abundance of long filamentary objects in the interarms, and massive clouds in the spiral arms \citep{Dobbs_2011, Colombo_2014}. 

Figures~\ref{fig:PDF_SD} and \ref{fig:PDF_SD_clouds} reveal differences in the molecular gas distribution across galactic environments. In spiral galaxies, the number density of PAH clouds with $\sigmol < 10~\msunperpcsq$ is similar in spiral arm and interarm regions, and 1.2 times lower than in spiral arm compared to bar regions. In disk galaxies, cloud number densities in the disk are also 1.4 times lower than those in the bar for the same \sigmol\ threshold. For clouds with $\sigmol > 10~\msunperpcsq$, spiral arms show 1.6 times higher number densities than bar regions, and 2.3 times higher than interarm regions. In this higher \sigmol\ regime, the disk exhibits 2.1 times higher cloud number densities than bar regions. Overall, spiral arm clouds exhibit the highest number density across all \sigmol\ values when compared to clouds in other environments (0.2 dex higher; see Table~\ref{T:Prop_Summary}), and the other environments have similar cloud number densities. Additionally, disk galaxies are, on average, $\sim 0.5$ dex less massive than spiral galaxies. Since bars tend to have a more pronounced impact on the ISM in more massive systems (e.g., \citealt{Verwilghen_2025}), this may explain the observed similarity in the cloud $\sigmol$ distribution between bars and disks (e.g., Fig.~\ref{fig:PDF_SD}) in the disk galaxies. 
On the other hand, Fig.~\ref{fig:PDF_SD_clouds} illustrates that the shape of the \sigmol\ distribution is generally similar across all environments, with the exception that spiral arms host a slightly greater number of high-\sigmol\ clouds compared to disk clouds for $\sigmol\ < 10^{3} ~\msunperpcsq$, and relatively more than bar and interarm regions.

In the PHANGS-ALMA sample, \cite{Querejeta_2021} report that, on kpc-scales, and using the \cite{Sun_2020a} \aco\ prescription, \sigmol\ values of interarm regions are comparable to those in disk regions, with interarm properties resembling those of disks in nonspiral galaxies \citep[see also,][]{Meidt_2021, Querejeta_2024}. Expanding on this, we find that on scales of tens of parsecs, molecular clouds in the disk regions show distributions and median values of \Reff, \Mmol, and \sigmol\ that resemble a combination of those found in both interarm and spiral arm regions (e.g., Fig.~\ref{fig:CO_compare_all} and \ref{fig:CO_compare_pah}). Also, using CO maps for the same galaxies presented here, previous studies \citep[e.g.,][]{Sun_2018,Sun_2020,Sun_2022, Leroy_2021, Querejeta_2021, Leroy_2025} report higher \sigmol\ toward the central regions of galaxies, with a more pronounced increase in barred galaxies. They attribute this to bar-driven gas inflows. Here, we see a decline of \sigmol\ in bars compared to disks. We note that toward central regions (i.e., in bars and centers), the CO emission is underestimated because the CO-to-PAH relationship is $\sim0.2$ dex higher there than in galactic disks \citep[e.g.,][]{chown2025}. Also, the stellar continuum is too bright, and subtracting it becomes more difficult \citep[e.g.,][]{sutter2024,Baron_2024}. Finally, the \aco\ conversion factor choice does affect the measurements, especially toward central regions. The usage of an \aco\ that depends on $\Sigma_{\star}$, \sigsfr\, and the metallicity does lower the \sigmol\ in bars and centers more than using one that does not account for all. Adopting a different \aco\ measurement does not affect our analysis or conclusions when comparing \sigmol\ of clouds in spiral arms, interarms, and disks. However, when adopting another prescription, \sigmol\ values in bars become comparable to or higher than those of galactic disks. It is worth noting that in low-metallicity regions (12+log(O/H) $<$ 8.2), the PAH abundance, traced by the F770W/F2100W ratio drops sharply; meanwhile, at higher metallicities, the ratio reaches a plateau \citep{Egorov_2025}. This highlights that PAHs could be more efficiently destroyed in the low-metallicity HII regions due to the higher UV hardness.

Finally, we divided the galaxies into active and non-active following \cite{Cetty_2010}, where active galaxies are defined as quasars (starlike nuclei with broad emission lines and absolute magnitude $M_B < -22.25$), BL Lac objects, and Seyfert galaxies (types 1–2, including LINERs), while normal galaxies are considered non-active. In our sample, 15 galaxies are classified as active. The trends in \sigmol, \Mmol, and \Reff\ across galactic environments are consistent between active and non-active galaxies, differing by only $\sim$0.1 dex. Therefore, the impact of the AGN in our sample might be small.

\begin{figure*}[h]
    \centering
    \includegraphics[width=0.93
    \paperwidth]{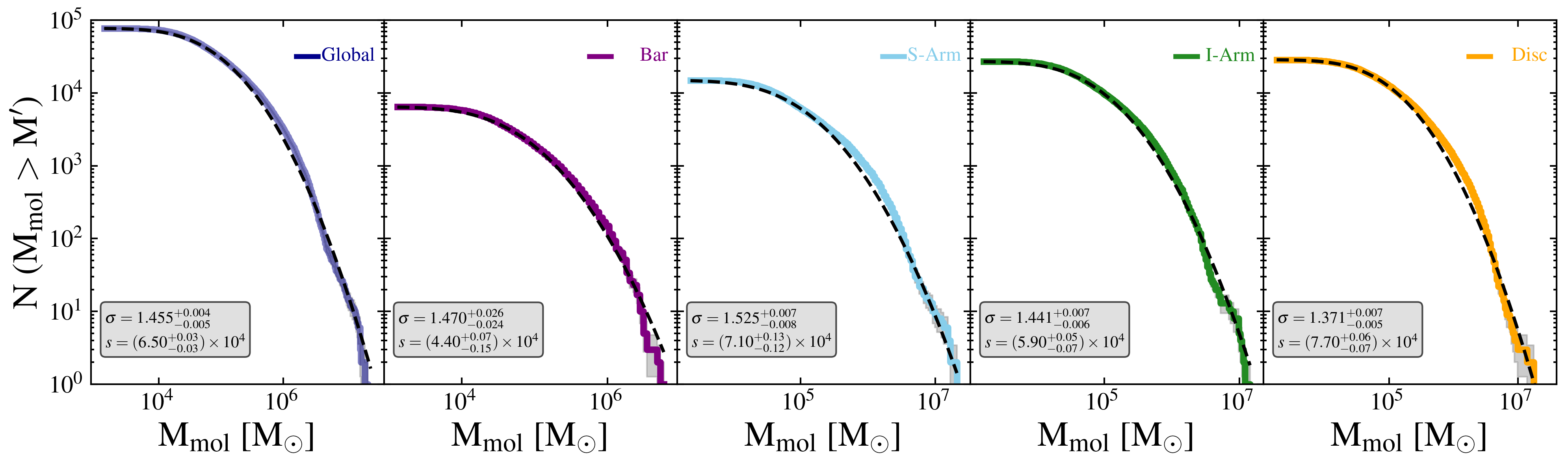}
    \caption{Mass spectra for the PAH clouds in different environments, as labeled in each plot. The dashed black curves show the survival function fits. The fit parameters are displayed in the bottom left corner of each figure, where $\sigma$ is the shape parameter and $s$ is the scale parameter. The gray region represents the Poisson errors on the counts $(\sqrt{N})$.}
    \label{fig:CDF}
\end{figure*}

\subsection{Cloud mass spectrum}

In this section, we focus on the cloud mass spectrum, providing a more comprehensive view of the mass distribution of clouds by quantifying the fraction of clouds above or below a given threshold. This approach is motivated by previous works, such as \cite{Rosolowsky_2005} \citep[see also][]{Blitz_2007,Fukui_2010,Hughes_2013,Colombo_2014,Mok_2020}. For example, \cite{Colombo_2014} used mass spectra to highlight environmental differences in M51, showing steeper distributions in interarm regions than spiral arms. Similarly, cumulative and differential techniques have been used to explore the effects of feedback and dynamical processes on the molecular cloud population \citep{Mazumdar_2021} and in simulations \citep[e.g.,][]{Colman_2024}. Following these methods, we employ the cloud mass spectrum to investigate the cloud population and formation. The fit itself is an important result to be matched by theories of cloud formation and evolution. By doing so, we aim to uncover how environmental factors influence the entire molecular cloud population.

Previous work has often used a single truncated or normal power law to identify the shape, steepness, or shallowness of the cloud mass spectrum \citep{Rosolowsky_2005, Colombo_2014, Mok_2020}. However, when analyzing a large sample of clouds, we find that they fail to catch the full distribution of clouds, especially the tail of the distribution, since it departs from a power law. \cite{Pathak_2024} show that two components could represent the PDF of mid-infrared intensities in individual PHANGS-JWST Cycle 1 galaxies. A diffuse lognormal part that peaks at low intensities and strongly correlates with SFR and gas surface density, and a power law tail at high intensities that traces HII regions. The lognormal component dominates the 7.7$\mu$m emission. Therefore, we test whether a survival function of a lognormal distribution can be used to define the mass spectra of clouds inferred from the same maps.

\begin{figure}[h]
    \centering
    \includegraphics[width=0.38
    \paperwidth]{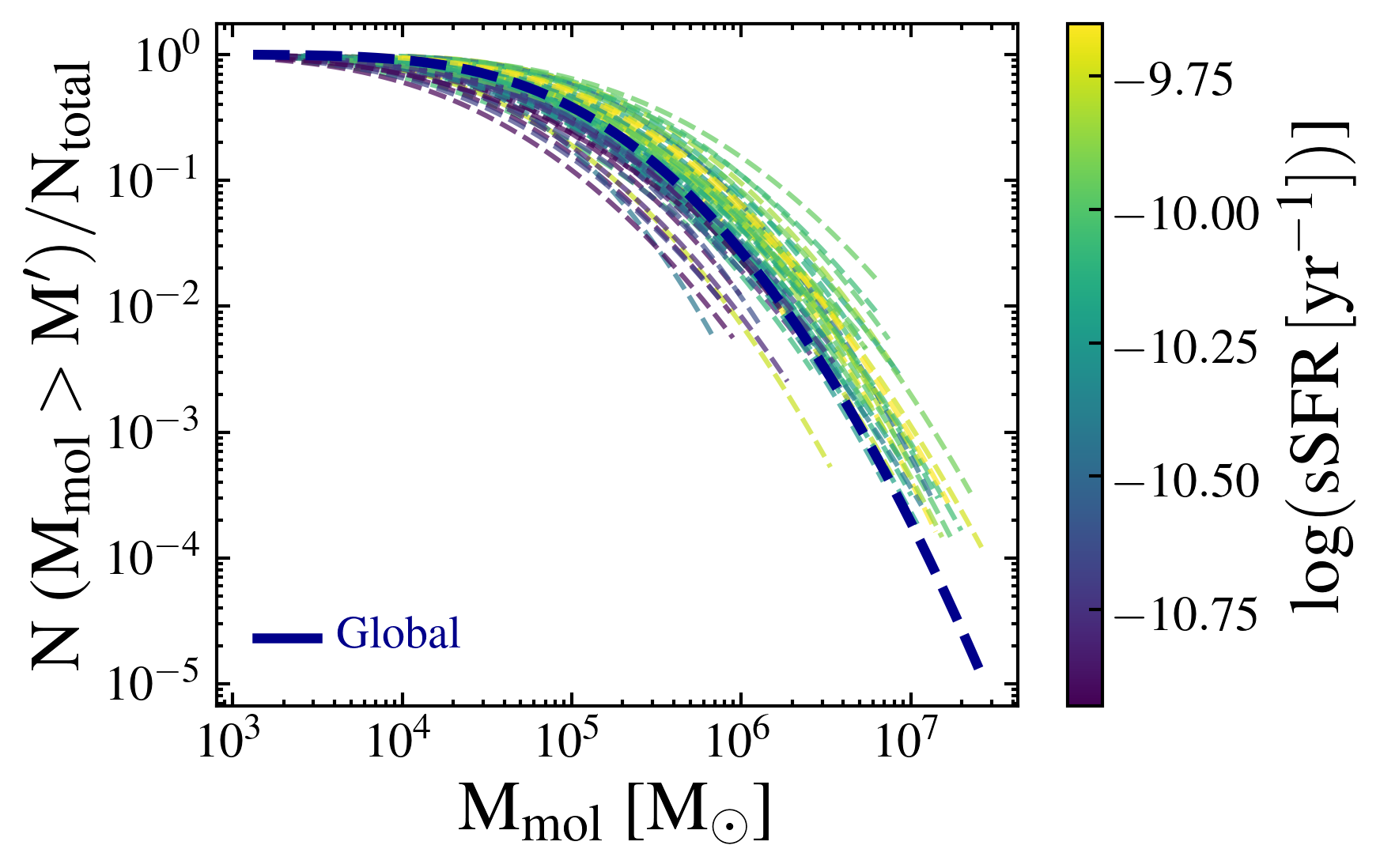}
    \caption{Normalized survival function fits for PAH clouds per galaxy. The dark blue dashed line represents the global fit to the entire sample. The background thin dashed lines are color-coded by specific star formation rate (sSFR) and show the global fits to each galaxy. }
    \label{fig:CDF_Gals}
\end{figure}

The lognormal distribution can be represented as

\begin{equation}
    f(M ;\sigma, s) = \frac{1}{M \sigma \sqrt{2\pi}} 
    \exp \left( -\frac{(\ln M - \ln s)^2}{2\sigma^2} \right), \quad M > 0, \quad \sigma > 0
,\end{equation}

\noindent where $M$ (or \Mmol) is the mass of the cloud. Two important parameters are the shape parameter, which refers to the standard deviation ($\sigma$), and the scale parameter ($s$), which refers to the $e^{\mu}$, where $\mu$ is the mean of the lognormal distribution. 

The cumulative distribution function is

\begin{equation}
    F(M; \sigma, s) = \Phi \left( \frac{\ln M - \ln s}{\sigma} \right),
\end{equation}
\noindent where \( \Phi(M) \) is the standard normal CDF:

\begin{equation}
    \Phi(M) = \frac{1}{2} \left[ 1 + \operatorname{erf} \left( \frac{M}{\sqrt{2}} \right) \right].
\end{equation}

\noindent Here erf is the standard error function defined as $\operatorname{erf}(M) = \frac{2}{\sqrt{\pi}} \int_0^M e^{-t^2} dt$, and $t$ is the mass element.

Therefore, the complementary cumulative distribution function (CCDF) or survival function is

\begin{equation}
    S(M; \sigma, s) = 1 - F(x) = 1 - \Phi \left( \frac{\ln M - \ln s}{\sigma} \right),
\end{equation}

\noindent where $S(M; \sigma, s)$ is the normalized form of the survival function. It should be multiplied by the total number of clouds to replicate the CCDFs shown in Fig.~\ref{fig:CDF}.

We rely on the CCDF description of SCIPY $\mathtt{lognorm.sf}\footnote{https://docs.scipy.org/doc/scipy/reference/generated/$\newline$scipy.stats.lognorm.html}$ package. For that, we optimize the lognormal parameters by minimizing the negative log-likelihood for the lognormal distribution using the $\mathtt{ minimize}\footnote{https://docs.scipy.org/doc/scipy/reference/generated/$\newline$scipy.optimize.minimize.html}$ function in SCIPY. We apply a bootstrap of 100 iterations on the minimization to find the fit error. We fit the survival function for the full sample of clouds and per galactic environment for a completeness limit of $\sim 2\times 10^{3}~\msun$ as seen in Fig.~\ref{fig:CDF}. This means we only consider \Mmol\ values corresponding to more than 8 times the 3~$\sigma_{\rm rms}$ level.

We present the fit parameters in Table~\ref{T:Plaw}. Overall, the spiral arm and disk environments have the highest $s$ values. The shape parameter indicates that a larger $\sigma$ corresponds to a shallower distribution slope, implying the presence of more massive structures. This parameter is the highest in the spiral arm region, which, alongside the highest $s$ value, suggests that higher-mass clouds are more prevalent in spiral arms. This trend is also evident in Fig.~\ref{fig:CDF_Norm}, where the fits of individual environments closely follow each other, but spiral arm clouds have a shallower slope, appearing more prominent at higher masses.

Furthermore, Fig.~\ref{fig:CDF_Gals} shows significant scatter ($\sim$ 1 dex) toward the high masses in the mass spectra. To investigate this, we compare the distributions of the individual galaxies using a Kolmogorov–Smirnov (KS) two-sample test. Of the 66 galaxies, we form 2145 pairs and check if the $p$-value decreases or increases when comparing the full \Mmol\ distribution of the pairs versus the sample excluding the high-mass clouds (> $10^{6}\, \msun$). In $78~\%$ of the cases, we see an increase in $p$-value when excluding the high-mass clouds. This indicates that the high-mass clouds are driving differences in the distributions. Figure~\ref{fig:CDF_Norm} further shows that this deviation is most prominent in the bar and disk regions. This suggests that molecular cloud formation and evolution differ more significantly between different galactic bars or disks than between different spiral arms or interarm regions.

The Spearman correlation coefficient in Table~\ref{T:correlations} shows that the $s$ of the lognormal fits reflects the median of the cloud \Mmol\ and strongly correlates with the \sigmol\ median. There is also a positive correlation with the sSFR (see also Fig.~\ref{fig:CDF_Gals} and \ref{fig:CDF_Norm}), number density of clouds, inclination, and HI mass of the galaxy. This means that galaxies with a higher value of $s$ tend to have more clouds within their area and more ``active'' star formation. Also, this reflects the nature of the PAHs tracing heating by star formation \citep[e.g.,][]{Peeters_2004, Calzetti_2007, Belfiore_2023, Leroy_2023}. Therefore, $s$ is a metric that mainly relates the cloud properties to their star formation capability. Additionally, the total mass within the clouds per galaxy positively correlates with the HI mass, SFR, sSFR, and the total number of clouds within the galaxy (see Table~\ref{T:correlations}), indicating that star formation is more prominent in galaxies having a higher number and more massive clouds. Also, the correlation between both $s$ and the total mass of clouds with the mass of HI hints that the atomic gas acts as a reservoir for molecular clouds, and the more atomic gas present, the more molecular clouds are forming.

To investigate how both galactic environment and host galaxy influence the variation in molecular cloud mass distributions, we compare the lognormal fit parameters obtained globally per environment (i.e., Fig.~\ref{fig:CDF}) with those derived on a galaxy-by-galaxy basis. The results are presented in Table~\ref{T:Plaw}. Across environments, the global fit parameters, particularly $\sigma$, vary in a relatively narrow range (from $\sim$1.37 in the disk to $\sim$1.53 in the spiral arms). Also, the distribution of $\sigma$ values obtained from the galaxy-by-galaxy fits within each environment exhibits a similar spread. It is worth noting that the small range that $\sigma_{\rm gal}$ varies within galaxies implies that galaxies generally exhibit similar cloud \Mmol\ PDF width, which explains why $\sigma$ shows little to no correlations with the global galactic properties. Additionally, the scale parameter $s$, where the distribution of values from the galaxy-by-galaxy fits (with 84th to 16th percentile ranges on the order of $2$ to $7 \times 10^4~\msun$) is wider than the overall shift in $s$ across environments (ranging from $\sim 4$ to $8 \times 10^4~\msun$). These results indicate that the differences in cloud mass distributions are not fully captured by environment-based classification alone. Instead, variation between host galaxies, even within the same environment, contributes significantly to the overall distribution.

\renewcommand{\arraystretch}{1.5}
\setlength{\tabcolsep}{6pt} 
\begin{table}
\caption{Survival function parameters ($\sigma$, $s$) for the Global sample of clouds, per galactic environment, and galaxy-by-galaxy ($\sigma_{\rm gal}$, $s_{\rm gal}$).}
\centering
\begin{tabular}{ccccc}
\hline
\multicolumn{1}{c}{Env.} & 
\multicolumn{1}{c}{$\sigma$} & 
\multicolumn{1}{c}{$s$} & 
\multicolumn{1}{c}{$\sigma_{\rm gal}$} & 
\multicolumn{1}{c}{$s_{\rm gal}$} \\
   &   &  $10^{4} \, \msun$ &  & $10^{4} \, \msun$\\
\hline
Global & $1.455^{+0.004}_{-0.005}$ & $6.50^{+0.03}_{-0.03}$ & $1.553^{+0.144}_{-0.117}$& $5.78^{+5.05}_{-2.41}$   \\
Bar & $1.470^{+0.026}_{-0.024}$ & $4.40^{+0.07}_{-0.15}$ & $1.594^{+0.204}_{-0.196}$ & $4.49^{+2.73}_{-2.06}$ \\
Spiral Arm & $1.525^{+0.007}_{-0.008}$ & $7.10^{+0.13}_{-0.12}$ & $1.628^{+0.120}_{-0.117}$ & $7.11^{+4.32}_{-2.49}$  \\
Interarm & $1.441^{+0.007}_{-0.006}$ & $5.90^{+0.05}_{-0.07}$ & $1.573^{+0.137}_{-0.124}$ & $5.32^{+3.41}_{-1.79}$  \\
Disk & $1.371^{+0.007}_{-0.005}$ & $7.70^{+0.06}_{-0.07}$ & $1.552^{+0.173}_{-0.176}$ & $5.93^{+7.19}_{-2.70}$  \\
\hline
\end{tabular}
\tablefoot{The errors listed in the first two columns represent bootstrapped 84th - 50th and 50th - 16th percentiles as superscript and subscript, respectively. Meanwhile, the values in the last two columns represent the medians of the galaxy-by-galaxy fits, and the errors display the 84th - 50th and 50th - 16th percentiles of the distribution of galaxy-by-galaxy fits.}
\label{T:Plaw}
\end{table}
\setlength{\tabcolsep}{6pt}

\renewcommand{\arraystretch}{1.3}
\setlength{\tabcolsep}{3pt}
\begin{table*}
\caption{Spearman correlation coefficients ($r$) and p-values ($p$) between the scale parameter of the lognormal distribution $s_{\rm gal}$, the shape parameter $\sigma_{\rm gal}$, the total mass within clouds ($\sum \Mmol$), and various galactic and/or cloud parameters across the 66 galaxies.}
\centering
\begin{tabular}{lcc|cc|cc}
\hline
\multicolumn{1}{c}{Parameter} & 
\multicolumn{2}{c|}{$s_{\rm gal}$} & 
\multicolumn{2}{c|}{$\sigma_{\rm gal}$} &
\multicolumn{2}{c}{$\sum \Mmol$} \\
& $r$ & $p$ & $r$ & $p$ & $r$ & $p$ \\
\hline
$N_{clouds}$ & 0.20 & 0.10 & -0.17 & 0.18 & 0.88 & $7.24 \times 10^{-22}$ \\
$N_{clouds}/A ~ [\rm kpc^{-2}]$ & 0.38 & $1.5\times 10^{-3}$ & 0.08 & 0.52 & 0.59 & $1.56\times10^{-7}$ \\
$\sigmol~ [\msunperpcsq]$     & 0.78 & $2.06\times10^{-14}$ & -0.08 & 0.53             & 0.52 & $9.40\times10^{-6}$ \\
$\Mmol~   [\msun]$               & 1.00 & $1.32\times10^{-94}$ & -0.27 & 0.03             & 0.58 & $3.36\times10^{-7}$ \\
$\Reff~[\rm pc]$              & 0.31 & 0.01                  & -0.16 & 0.19             & 0.06 & 0.63 \\
$\log \rm SFR~ [\msunperyr]$  & 0.24 & 0.05                  & -0.31 & 0.01             & 0.69 & $1.68\times10^{-10}$ \\
$\log \rm M_{\star}~[\msun]$  & -0.28 & 0.02                 & -0.28 & 0.02             & 0.35 & $3.68\times10^{-3}$ \\
$\log \rm sSFR~[yr^{-1}]$     & 0.65 & $4.95\times10^{-9}$   & -0.10 & 0.44             & 0.42 & $4.37\times10^{-4}$ \\
$\log \rm M_{H_{I}}~[\msun]$  & 0.37 & $2.11\times10^{-3}$                  & -0.21 & 0.09             & 0.56 & $1.15\times10^{-6}$ \\
R$_{e}~ [\rm kpc]$            & 0.04 & 0.74                 & -0.25 & 0.05             & 0.33 & $7.44\times10^{-3}$ \\
$i ~[\rm deg]$                & 0.30 & 0.01 & -0.28 & 0.02             & -0.31 & 0.01 \\
\hline
\end{tabular}
\tablefoot{Correlations are computed between the scale $s_{\rm gal}$ and shape $\sigma_{\rm gal}$ of the cloud mass spectra (from lognormal fits), total molecular cloud mass $\sum \Mmol$, and various galactic properties across 66 galaxies. $N_{clouds}$ is the total number of clouds within a galaxy, and $N_{clouds}/A_{env}$ is the number density of clouds within a galaxy. \sigmol, \Mmol, and \Reff\ are median cloud values per galaxy. Global galaxy properties (SFR, M$_{\star}$, atomic hydrogen mass $\rm M_{H_{I}}$, sSFR, R$_{e}$, $i$) are taken from \cite{Leroy_2021}.}
\label{T:correlations}
\end{table*}
\setlength{\tabcolsep}{6pt}

\renewcommand{\arraystretch}{1.5}

\begin{figure*}
    \centering
    \includegraphics[width=0.32\textwidth]{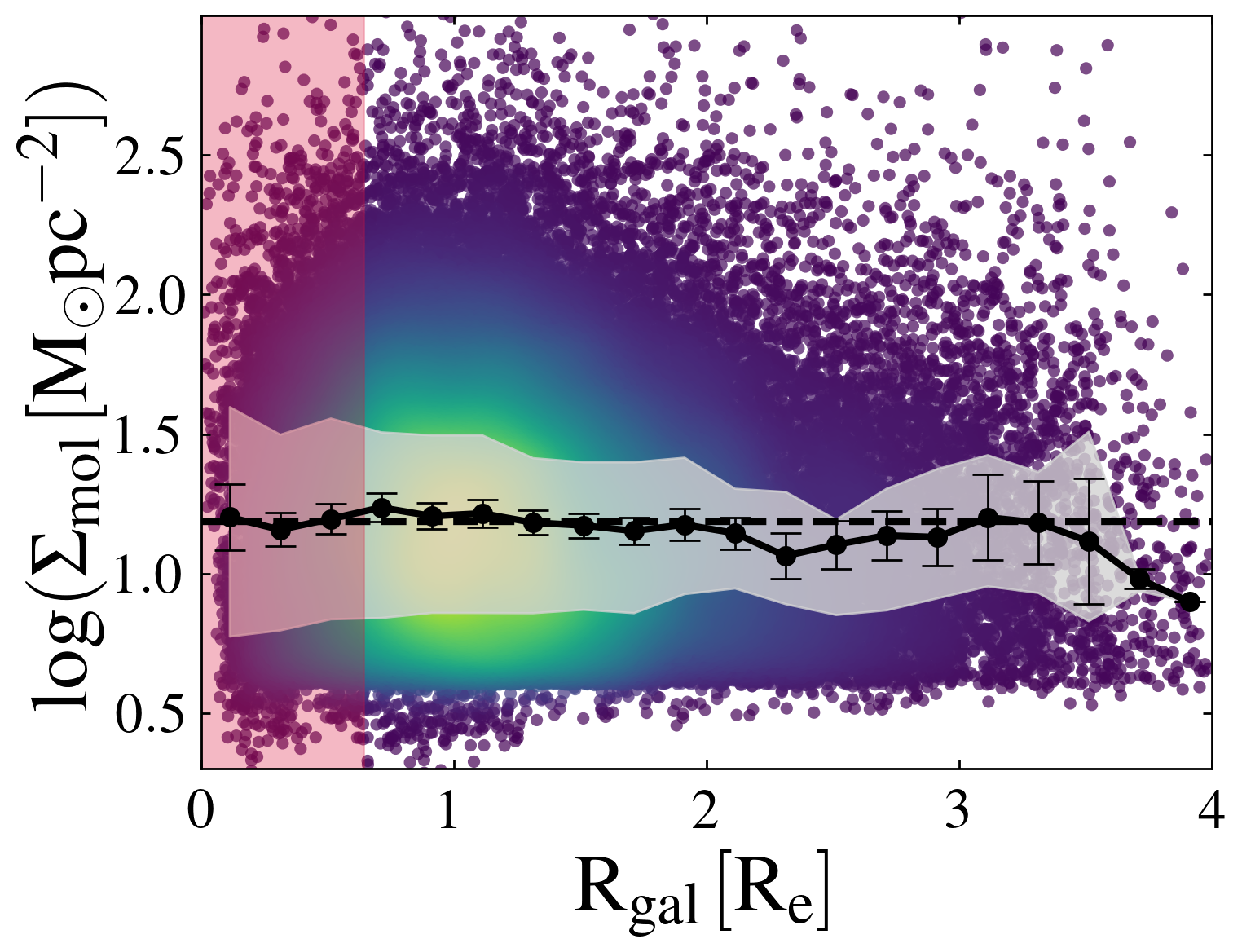}
    \includegraphics[width=0.33\textwidth]{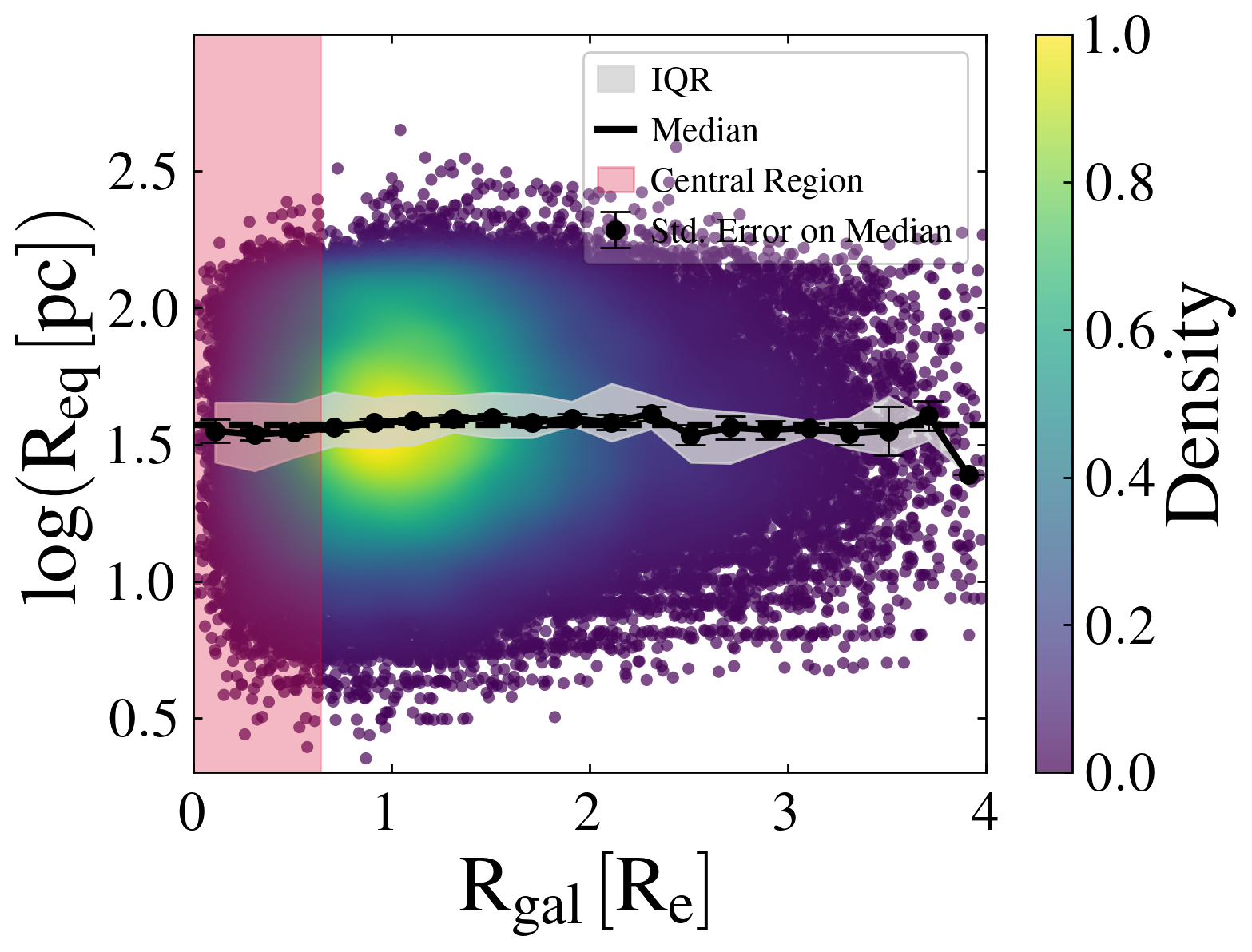}
    \includegraphics[width=0.31\textwidth]{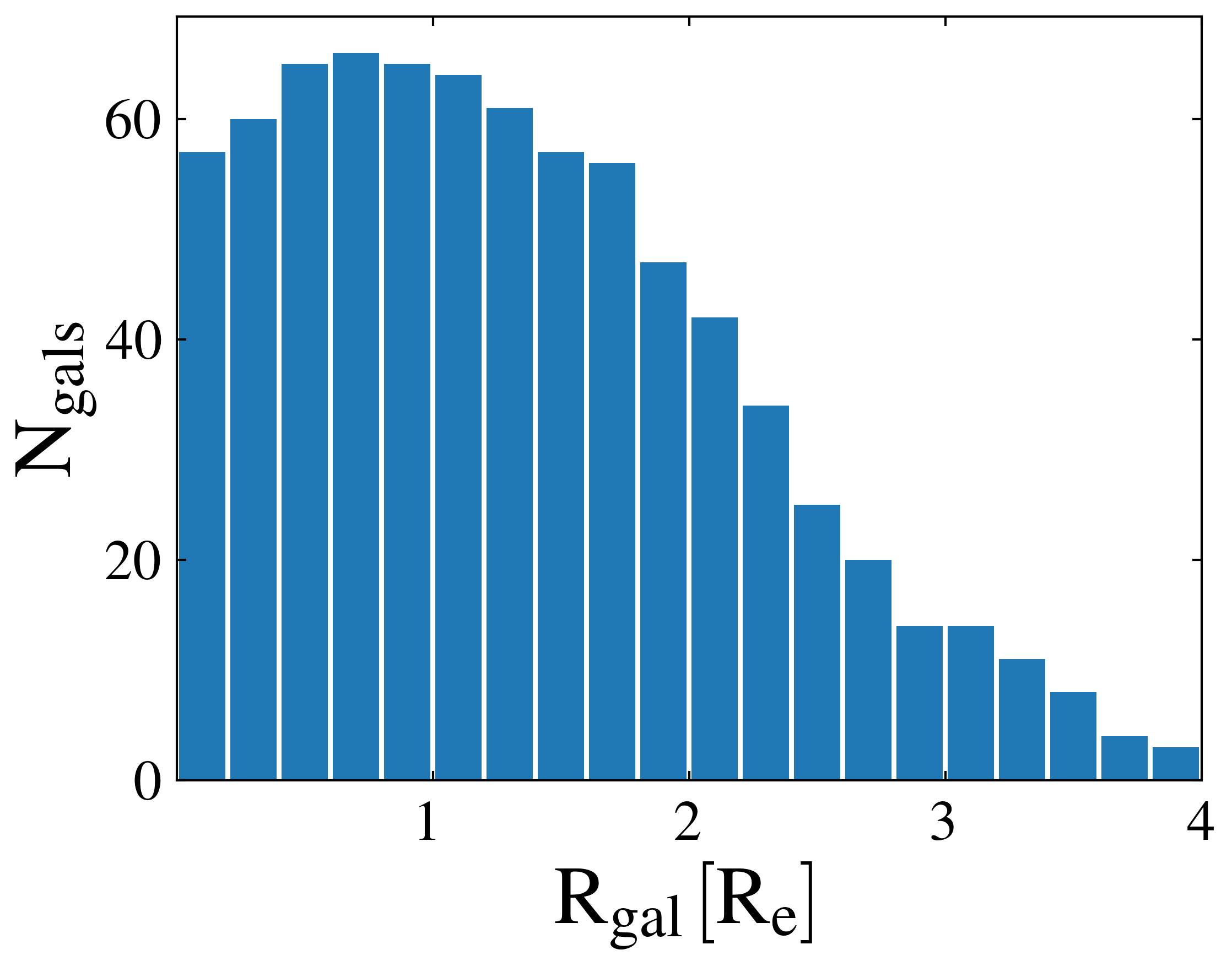}
    \caption{Properties of the PAH clouds vs $\mathrm{R_{gal}}$ for all the clouds in the 66 galaxies: \sigmol\ of the cloud (\textit{left}), \Reff\ (\textit{middle}), and the total number of galaxies contributing to a specific bin (\textit{right}). The running galaxy median (filled black circles) is plotted for a bin width of 0.2 R$_{e}$. The gray-shaded region represents the interquartile range of the medians per galaxy. The error bars on the median are the standard errors (1.253$\sigma$/$\sqrt{N}$), where N is the number of galaxies contributing to a specific bin. The background data points represent a scatterplot of the full sample of GMCs colored by the normalized density of clouds. The red-shaded region depicts the maximum extent of the central clouds. The horizontal dashed black line in each plot corresponds to the median of the plotted property.}
    \label{fig:radial}
\end{figure*}

\subsection{Cloud property distributions as a function of galactocentric radius}\label{SS:rad_prof}

Examining the distribution of all PAH-identified clouds as a function of galactocentric radius ($\rm R_{gal}$) provides insight into whether local cloud properties reflect the broader structure of the gas reservoir from which they form. While large-scale processes such as gravitational torques, spiral density waves, and hydrodynamic shocks (e.g., \citealt{Lin_1964, Roberts_1979, Sormani_2019, Yu_2022}) act on longer timescales than the lifetime of individual clouds, they significantly influence the spatial arrangement and surface density of the molecular gas. Over time, these mechanisms facilitate angular momentum loss and drive gas radially inward, giving rise to the well-known exponential decline in molecular gas surface density with radius. If cloud-scale properties such as \sigmol\ and \Reff\ are coupled to the large-scale galactic processes, we may expect to detect systematic radial variations as a result. In this section, we investigate how \sigmol\ and \Reff\ vary with $\rm R_{gal}$ across all 66 galaxies in our sample. PAH emission enables us to trace a broader range of cloud masses, including small clouds often missed in CO studies, allowing a more complete census of the \sigmol\ and \Reff\ variation as a function of $\rm R_{gal}$.

In Fig.~\ref{fig:radial}, we present the radial profiles of \sigmol\ and \Reff\ (see Eq.~\ref{E:sigma_1} and \ref{E:Reff}, respectively). We further fit a Gaussian for the distribution at each radial bin and find that the Gaussian $\sigma$ values are consistent between the bins. This generally indicates that the distributions span similar values in all bins. The scatter around the median is approximately $0.5 - 1$ dex, while the individual galaxy scatter is lower, around $0.2 - 0.3$ dex. The inner $\sim 0.5$ R$_{e}$ (stellar effective radius; obtained from \citealt{Leroy_2021}) is an ambiguous region due to removing high-mass clouds associated with overlapping structures. Beyond $\sim 0.5$ R$_{e}$, the radial profile has a near-constant median for \Reff. Meanwhile, for \sigmol\, there is a decline of a factor of $1.5 - 2$ toward higher R$_{e}$, and a bump at 3 R$_{e}$ due to spiral arm clouds in a few galaxies. We note that $80\%$ of the cloud contribution after 2 R$_{e}$ comes from spiral galaxies, and the number of galaxies contributing per bin starts dropping after $\sim 1.5$ R$_{e}$ to reach less than 20 galaxies after 3 R$_{e}$. 

In the PHANGS–ALMA sample (e.g., \citealt{Sun_2020, Leroy_2025}), cloud-scale \sigmol\ shows little variation with $\rm R_{gal}$ beyond the central regions, in agreement with our results. Also, at fixed $\rm R_{gal}$/R$_e$, PAH clouds located in spiral arms exhibit \sigmol\ values approximately $1.5 - 2.5$ times higher than those in interarm regions (see Fig.~\ref{fig:radial_envs}), consistent with the spiral arm–interarm contrast observed in PHANGS–ALMA.

We find that the galaxy-by-galaxy \sigmol\ in Fig.~\ref{fig:appendix} and \ref{fig:radial_all} show considerable variation with $\rm R_{gal}$, with flat, declining, and ambiguous profiles observed across different galaxies. The \sigmol\ behavior in some CO-based cloud analysis (e.g., \citealt{Rebolledo_2015,Faesi_2018}) show no clear trend with $\rm R_{gal}$, highlighting that, in some galaxies, the galactic environment might play a bigger role in determining the GMC properties than a radial-based approach. An example of this in our sample would be NGC 0628, NGC 1365, NGC 2090, and NGC 2997 (see Fig.~\ref{fig:appendix}), where the spiral arm-interarm contrast exists, but there is a flat \sigmol\ trend with $\rm R_{gal}$. However, in other galaxies, such as NGC 1385, NGC 1546, NGC 1559, and NGC 3059, that lack spiral arm features, we see a $\sim 0.5 - 1$ dex decrease of \sigmol\ toward the outer regions of the galaxies. Together, these examples highlight that radial trends in cloud \sigmol\ are not universal, but instead vary strongly with different galaxies.

\subsection{Extreme clouds} \label{ss:Extreme}

In this section, we focus on clouds at the extremes of both \sigmol\ and \Mmol\ within our sample. An overabundance of high \sigmol\ and high \Mmol\ clouds in specific large-scale galactic environments suggests that localized physical processes in these regions preferentially drive the formation of distinct, dense, and massive cloud populations, potentially enhancing star formation activity. Conversely, the prevalence of low \sigmol\ and low-mass clouds in certain environments may indicate the presence of mechanisms that inhibit efficient gas compression and cloud growth, such as strong shear, elevated turbulence, or low external pressure. These processes act to suppress the formation of gravitationally bound and massive structures, ultimately limiting star formation efficiency.

We define low-mass \sigmol\ clouds as clouds with $\sigmol\ \leq 10\msunperpcsq$, representing $\sim 32\%$ of our sample size, and extremely low \sigmol\ clouds are the 1000 least dense clouds. The highest \sigmol\ clouds are clouds with $\sigmol\ \geq 100 \msunperpcsq$, representing $\sim 5\%$ of our sample size, and the extremely highest \sigmol\ clouds are the 1000 highest dense clouds.

We rely on fractional differences between the full sample and low- or high-density clouds per galactic environment to assess where those clouds prevail more. The fractional difference is then defined as

\begin{equation}
    \Delta f = \bigg(\frac{N^{\rm env}_{\rm sub}}{N_{\rm sub}} -  \frac{N^{\rm env}_{\rm sample}}{N_{\rm sample}}\bigg)\times 100 ,
\end{equation}

\noindent where $N^{\rm env}_{\rm sub}$ is the number of extreme clouds in a specific environment from the extreme subsample ($N_{\rm sub}$), and $N^{\rm env}_{\rm sample}$ is the number of clouds in a specific environment in the full sample ($N_{\rm sample}$).

A positive $\Delta f$ value would indicate, probabilistically, higher prevalence in a specific environment. The values of $\Delta f_{\rm low}$, $\Delta f_{\rm high}$, $\Delta f^{e}_{\rm low}$, and $\Delta f^{e}_{\rm high}$  are provided in Table~\ref{T:fvals}. Here, $\Delta f_{\rm low}$, $\Delta f_{\rm high}$ are the fractional differences between the full sample and the low or high  \sigmol\ regimes, respectively. The notation ``$e$'' is for the extreme samples. 

The $\Delta f$ values presented in Table~\ref{T:fvals} indicate that the low \sigmol\ clouds are most frequent in bar and interarm regions and are the least frequent in spiral arm regions. Also, the highest \sigmol\ clouds are most and least prevalent in spiral arm and interarm regions, respectively. We note that extremely low \sigmol\ in bars could be due to the under-approximation of the CO emission in bar ends and the \aco\ prescription used here, and due to the existence of low \sigmol\ clouds in bar lanes. Upon using the other \aco\ prescriptions, we notice that our results are consistent in all the environments except the bar region, where interarm clouds take over as the lowest density structures. 
 
\cite{Sun_2022} further demonstrate that in the PHANGS-ALMA sample, CO cloud properties correlate strongly with environmental conditions, particularly \sigsfr\ and \sigmol. Together with our results, these studies support the picture where spiral arms are key sites for the formation of dense, high \sigmol\ and high-mass clouds, while the interarm regions are mainly populated by diffuse and lower-mass clouds. Again, we emphasize that central regions were excluded from our main analysis due to the removal of overlapping structures. However, when we include the central clouds, they emerge as the primary hosts of the extremely highest density clouds, consistent with both Galactic and extragalactic observations \citep[e.g.,][]{Longmore_2012, Mills_2017, Sun_2018, Sun_2020}, followed by the spiral arms.

Many galactic factors play a role in shaping the clouds across the mass distribution and spectra. The high-mass clouds are intrinsically rarer than lower-mass clouds ($\sim$4\,\% of our sample size). \cite{Kobayashi_2017} (see also \citealt{Tasker_2009, Kobayashi_2018}) show that in simulated clouds, cloud-cloud collisions mostly affect the tail of the cloud mass spectra. Those collisions lead to the formation of more massive GMCs ($\Mmol\ \gtrsim\ 10^{6}~ \msun$). \cite{Kruijssen_2014} suggested that the maximum GMC mass may correspond to the maximum mass that could collapse against centrifugal forces (i.e., Toomre mass; \citealt{Toomre_1964}). Models that predict the maximum GMC mass (e.g., \citealt{Campos_2017}) explain that those masses change from shear-limited to feedback-limited as galaxies become less gas-rich and evolve toward low shear. The $s$ parameter of the lognormal has a strong positive correlation with the high-mass cloud fraction, implying that more massive clouds exist at higher $s$ values. At lower masses, cloud self-growth by accumulating surrounding HI gas and destruction by massive star radiative feedback (e.g., due to photo-ionization, photo-dissociation) shape the cloud mass spectra. Upon binning the \Mmol\ distribution, we examine the correlation between the total mass in each bin and various global galaxy properties across our sample (see Fig.~\ref{fig:sfr_corr}). We find that only clouds with \Mmol\ between $10^{4}$ and $10^{6}~\msun$, which make up $90~\%$ of the sample, show a significant positive correlation with both the HI mass and the star formation rate (SFR) of their host galaxies (see Fig.~\ref{fig:sfr_corr}). In contrast, clouds with \Mmol\ below $10^{4}~\msun$ show no apparent correlation with global SFR or HI mass. This suggests that massive star formation is not prominent in these lower-mass clouds.

\renewcommand{\arraystretch}{1.5}
\setlength{\tabcolsep}{2pt} 
\begin{table}
\caption{Pearson $\chi^{2}$ and fractional difference statistical tests for the extreme cloud subsamples.}
\centering
\begin{tabular}{ccc|cc}
\hline
\multicolumn{1}{c}{Env.} & 
\multicolumn{1}{c}{$\Delta f_{high}$} &
\multicolumn{1}{c|}{$\Delta f^{e}_{high}$} &
\multicolumn{1}{c}{$\Delta f_{low}$} &
\multicolumn{1}{c}{$\Delta f^{e}_{low}$} \\
  (1) & (2) [$\%$] & (3) [$\%$] & (4) [$\%$] & (5) [$\%$]  \\
\hline
Bar  & $-3.84$ & $-4.66$ &   $2.87$ & $36.34$   \\
Spiral Arm & $11.14$ & $12.55$ & $-6.54$ & $-14.05$  \\
Interarm & $-11.49$  & $-12.21$ & $5.34$ & $-9.01$ \\
Disk & $4.18$ & $4.32$ & $-1.66$ &  $-13.28$\\

\hline
\end{tabular}
\tablefoot{(1) The galactic environment. (2) The fractional difference between the environmental counts in the high \sigmol\ regime ($\sigmol\ \geq 100 ~\msunperpcsq$) and the full sample of clouds ($\Delta f_{high}$). (3) The fractional difference between the environmental counts in the 1000 highest \sigmol\ subsample and the full sample of clouds ($\Delta f^{e}_{high}$). (4) The fractional difference between the environmental counts in the low \sigmol\ regime ($\sigmol\ \leq 10 ~\msunperpcsq$) and the full sample of clouds ($\Delta f_{low}$). (5) The fractional difference between the environmental counts in the 1000 lowest \sigmol\ subsample and the full sample of clouds ($\Delta f^{e}_{low}$).}
\label{T:fvals}
\end{table}
\setlength{\tabcolsep}{6pt} 

\section{Summary and conclusion}\label{S:Summary}

In this paper, we utilized SCIMES (\citetalias{Colombo_2015}), an unsupervised clustering algorithm, to identify cloud structures in 66 nearby PHANGS-JWST galaxies \citep{Lee_2023, Williams_2024}. Using stellar continuum-corrected \ipah\ maps, we identified 108,466 and 146,040 clouds in the common-resolution and native resolution samples, respectively. This represents the largest extragalactic cloud catalog to date. We used the common resolution sample for our analysis. We also flagged and excluded from our analysis clouds dominated by \ipah\ < 1 \mjysr, those located at the edges of the maps, and those overlapping in velocity space when cross-matched with CO-identified clouds using CPROPS (Hughes et al., in prep.). After these exclusions, the final sample consists of 77,884 clouds. Those strict measures were taken to avoid any biases. This results in a significant loss of high \sigmol\ clouds, mostly toward galactic centers, which we therefore excluded from our analysis. 

Upon comparing the \ipah\ identified clouds to CO-identified clouds, we notice an agreement in the \sigmol\ between both methods in the different environments. We refer to the identified clouds as GMCs; however, regions with \sigmol\ $\leq 10 \msunperpcsq$ may correspond to either diffuse atomic gas or faint molecular clouds that remain undetectable in CO observations. Upon examining these regions, we find that such clouds are predominantly located in the interarm and disk regions, reinforcing our previous assertion. 

When investigating the \sigmol\ CDFs and distributions across different environments, we find that the spiral arms contain the highest number density of clouds (including high \sigmol\ clouds), and the interarm clouds show a sharper decline in \sigmol\ values after 10 \msunperpcsq . This observation is further backed by our fractional difference test, which confirms that spiral arms preferentially host the highest \sigmol\ clouds. Our findings generally align with hydrodynamical simulations (e.g., \citealt{Cabral_2016}) where the highest \sigmol\ GMC complexes are in the spiral arms. This also agrees with the idea that the gravitational potential of spiral arms aids in the formation of high \sigmol\ clouds, which subsequently fragment into less dense structures as they drift into interarm regions (e.g., \citealt{Dobbs_2006, Koda_2009}).

We also fit a lognormal to the mass spectra with a completeness limit extending down to $2\times 10^{3} \, \msun$, well below previously obtained depths with CO observations, and find that it represents a good fit. Spiral arm clouds tend to have more massive clouds than the other environments according to the $s$ and $\sigma$ of the lognormal distribution fits. Positive correlations exist between both the $s_{\rm gal}$ and $\sigma_{\rm gal}$ of clouds, sSFR, and the median \sigmol\ of the galaxies, reflecting that galaxies that host more massive clouds have more star formation with respect to their stellar mass.

The cloud properties show minimal variation as a function of $\rm R_{gal}$. The \Reff\ of the clouds is consistent toward the outskirts of the galaxies. Meanwhile, the cloud \sigmol\ values decrease by a factor of $\sim 1.5 - 2$ toward the outer parts of the galaxies. Also, at a fixed $\rm R_{gal}$/R$_e$, spiral arms have \sigmol\ values approximately $1.5 - 2.5$ times higher than those in interarm regions. It is worth noting that \sigmol\ trend varies largely between different galaxies, with flat, decreasing, and even no trend as a function of $\rm R_{gal}$. Factors such as large-scale processes, galaxy types, and morphologies might influence the observed trends. 

The cloud mass spectra, radial profiles, and properties vary from galaxy to galaxy depending on their physical conditions and local environments. In contrast, combining all clouds across galaxies averages out this local information, emphasizing only global environmental differences between galaxies.

We list a few key points to summarize our findings:

\begin{enumerate}
    \item A total of 108,466 PAH clouds were identified across 66 galaxies using SCIMES. Of these, 77,844 clouds met or selection criteria and were included in the analysis, while others were flagged in the catalog (e.g., for velocity overlap, edge effects, low molecular gas content) and excluded from the analysis. This led to a bias toward lower \sigmol\ clouds, evidenced by a reduced PAH cloud flux recovery (a median of 26\% in the flagged sample vs 40\% in the full sample).
    
    \item The PAH clouds reveal fainter, smaller structures, especially in the interarm and bar regions, compared to CO-based ALMA clouds ($\sim 2$ dex better completeness limit). Those clouds may correspond to either faint molecular clouds or diffuse atomic gas clouds, or be sensitivity-limited in CO observations.

    \item Both PAH and CO identified clouds show consistent $\rm \Sigma_{mol, R}$ distributions across most environments. However, toward central regions, PAH cloud $\rm \Sigma_{mol, R}$ is 0.3 dex lower. This could be due to the PAH-to-CO fit, \aco\ prescription, and overlapping clouds in velocity pose a challenge in our analysis, making it challenging to derive any conclusions there.
    
    \item The cloud \sigmol\ varies with galactic environment, with spiral arms hosting the highest \sigmol\ and \Mmol\ clouds, interarms and bars the least, and disks showing intermediate properties of spiral arms and interarms. This supports the view that spiral arm potentials favor the formation of massive, dense clouds consistent with both observations and simulations.

    \item The mass spectra are better described by a lognormal distribution than a single power law, especially when considering a large cloud sample, with lognormal $\mu$ and $\sigma$ indicating that spiral arms host more massive clouds and a shallower mass spectrum compared to other environments.

    \item Variations in the cloud mass spectrum are more strongly influenced by differences between galaxies than by intra-galactic environments. This indicates that global galaxy properties such as gas content, star formation activity, and dynamics are the primary factors shaping the distribution of cloud masses.

    \item The cloud \sigmol\ values shows a decline of a factor of $\sim 1.5 - 2$ toward the outskirts of the galaxies ($2-3$ R$_{e}$), and at fixed $\rm R_{gal}$/R$_e$, PAH clouds located in spiral arms exhibit \sigmol\ values approximately $1.5 - 2.5$ times higher than those in interarm regions.

    \item The clouds with $\Mmol\ > 10^{4}~ \msun $ show a positive correlation with SFR, indicating that these clouds are key contributors to star formation. Meanwhile, clouds with \Mmol\ < $10^{4}~\msun$ do not show a correlation, indicating limited involvement in large-scale star-forming activity, or that their star formation is below detection limits.

    \item We publish our PAH cloud catalog at a homogenized resolution of 30 pc and native resolution. The catalog includes measurements of \Mmol , \Reff , \sigmol , and other parameters presented in Table~\ref {T:property_description}.

\end{enumerate}
We emphasize the importance of handling PP intensity images with care when identifying structure. We showed how including overlapping clouds could bias the results, and excluding them could push the analysis toward lower \sigmol\, and \Mmol\ clouds. Our findings provide valuable insights and calibrations for molecular cloud simulations, especially the mass spectra depicted here. Future work could include calibrating the CO to PAH relationship while subtracting the emission from small dust grains. Those grains could play a significant role, especially toward the central regions of the galaxies.

\section*{Data availability}

The full figure of Appendix D (Fig.~\ref{fig:appendix}) is available on Zenodo at \href{https://zenodo.org/records/15428261}{https://zenodo.org/records/15428261}. Table~\ref{T:property_description} is available at the CDS via anonymous ftp to \href{ftp://130.79.128.5}{cdsarc.u-strasbg.fr (130.79.128.5)} or via \href{http://cdsweb.u-strasbg.fr/cgi-bin/qcat/J/A+A/}{http://cdsweb.u-strasbg.fr/cgi-bin/qcat/J/A+A/}

\begin{acknowledgements}
This work has been carried out as part of the PHANGS collaboration. This work is based on observations made with the NASA/ESA/CSA JWST. The data were obtained from the Mikulski Archive for Space Telescopes at the Space Telescope Science Institute, which is operated by the Association of Universities for Research in Astronomy, Inc., under NASA contract NAS 5-03127 for JWST. These observations are associated with programs 2107 and 3707.
ZB, DC, and FB gratefully acknowledge the Collaborative Research Center 1601 (SFB 1601 sub-project B3) funded by the Deutsche Forschungsgemeinschaft (DFG, German Research Foundation) – 500700252.  DC acknowledges support by the \emph{Deut\-sche For\-schungs\-ge\-mein\-schaft, DFG\/} project number SFB956-A3. A.K.L., S.S., and R.C. gratefully acknowledge support from NSF AST AWD 2205628, JWST-GO-02107.009-A, and JWST-GO-03707.001-A. A.K.L. also gratefully acknowledges support by a Humboldt Research Award. H.F.V. acknowledges support from the Swedish National Space Agency (SNSA) through the grant 2023-00260, and support from RS 31004975. A.D.C. acknowledges the support from the Royal Society University Research Fellowship URF/R1/191609. E.R. acknowledges support from the Canadian Space Agency, funding reference 23JWGO2A07. R.S.K. and S.C.O.G. acknowledge financial support from ERC via Synergy Grant “ECOGAL” (project ID 855130), from the German Excellence Strategy via the “STRUCTURES” Cluster of Excellence (EXC 2181—390900948), and from the German Ministry for Economic Affairs and Climate Action in project “MAINN” (funding ID 50OO2206). R.S.K. also thanks the 2024/25 Class of Radcliffe Fellows for their company and for highly stimulating discussions. K.S. acknowledges funding support from grants JWST-GO-02107.006-A and JWST-GO-03707.005-A. Finally, we thank the anonymous referee for the helpful suggestions, which improved the quality of this paper.
This research made use of Astropy (\href{http://www.astropy.org}{http://www.astropy.org}) a community-developed core Python package for Astronomy \citep{astropy2013, astropy2018}; matplotlib \citep{matplotlib2007}; numpy and scipy \citep{scipy2020}.

\end{acknowledgements}

\footnotesize{
\bibliographystyle{aa}
\bibliography{Ref}

@ARTICLE{Solomon_1987,
       author = {{Solomon}, P.~M. and {Rivolo}, A.~R. and {Barrett}, J. and {Yahil}, A.},
        title = "{Mass, Luminosity, and Line Width Relations of Galactic Molecular Clouds}",
      journal = {\apj},
         year = 1987,
        month = aug,
       volume = {319},
        pages = {730},
          doi = {10.1086/165493},
       adsurl = {https://ui.adsabs.harvard.edu/abs/1987ApJ...319..730S}
}

@ARTICLE{Cetty_2010,
       author = {{V{\'e}ron-Cetty}, M. -P. and {V{\'e}ron}, P.},
        title = "{A catalogue of quasars and active nuclei: 13th edition}",
      journal = {\aap},
         year = 2010,
        month = jul,
       volume = {518},
          eid = {A10},
        pages = {A10},
          doi = {10.1051/0004-6361/201014188},
       adsurl = {https://ui.adsabs.harvard.edu/abs/2010A&A...518A..10V}
}

@ARTICLE{whitcomb_dust,
       author = {{Whitcomb}, Cory M. and {Sandstrom}, Karin and {Smith}, John-David T.},
        title = "{JWST-MIRI Synthetic Photometry Composition using 460 Spitzer-IRS Spectra of Nearby Galaxies}",
      journal = {Research Notes of the American Astronomical Society},
         year = 2023,
        month = mar,
       volume = {7},
       number = {3},
          eid = {38},
        pages = {38},
          doi = {10.3847/2515-5172/acc073},
       adsurl = {https://ui.adsabs.harvard.edu/abs/2023RNAAS...7...38W}
}

@ARTICLE{Egorov_2025,
       author = {{Egorov}, Oleg V. and {Leroy}, Adam K. and {Sandstrom}, Karin and {Kreckel}, Kathryn and {Baron}, Dalya and {Belfiore}, Francesco and {Chown}, Ryan and {Sutter}, Jessica and {Boquien}, M{\'e}d{\'e}ric and {Saguer}, Mar Canal i and {Congiu}, Enrico and {Dale}, Daniel A. and {Egorova}, Evgeniya and {Huber}, Michael and {Li}, Jing and {Williams}, Thomas G. and {Chastenet}, J{\'e}r{\'e}my and {Chiang}, I-Da and {Gerasimov}, Ivan and {Hassani}, Hamid and {Kim}, Hwihyun and {Koziol}, Hannah and {Lee}, Janice C. and {McClain}, Rebecca L. and {M{\'e}ndez Delgado}, Jos{\'e} Eduardo and {Pan}, Hsi-An and {Pathak}, Debosmita and {Rosolowsky}, Erik and {Sarbadhicary}, Sumit K. and {Schinnerer}, Eva and {Thilker}, David and {Ubeda}, Leonardo and {Weinbeck}, Tony},
        title = "{Polycyclic aromatic hydrocarbons destruction in star-forming regions across 42 nearby galaxies}",
      journal = {arXiv e-prints},
         year = 2025,
        month = sep,
          eid = {arXiv:2509.13845},
        pages = {arXiv:2509.13845},
          doi = {10.48550/arXiv.2509.13845},
archivePrefix = {arXiv},
       eprint = {2509.13845},
 primaryClass = {astro-ph.GA},
       adsurl = {https://ui.adsabs.harvard.edu/abs/2025arXiv250913845E}
}

@ARTICLE{Verwilghen_2025,
       author = {{Verwilghen}, Pierrick and {Emsellem}, Eric and {Renaud}, Florent and {Agertz}, Oscar and {Valentini}, Milena and {Fraser-McKelvie}, Amelia and {Meidt}, Sharon and {Neumann}, Justus and {Schinnerer}, Eva and {Klessen}, Ralf S. and {Glover}, Simon C.~O. and {Barnes}, Ashley. T. and {Dale}, Daniel A. and {Gleis}, Damian R. and {Smith}, Rowan J. and {Stuber}, Sophia K. and {Williams}, Thomas G.},
        title = "{Simulating nearby disc galaxies on the main star formation sequence: II. The gas structure transition in low and high stellar mass discs}",
      journal = {\aap},
         year = 2025,
        month = aug,
       volume = {700},
          eid = {A3},
        pages = {A3},
          doi = {10.1051/0004-6361/202554056},
archivePrefix = {arXiv},
       eprint = {2506.12923},
 primaryClass = {astro-ph.GA},
       adsurl = {https://ui.adsabs.harvard.edu/abs/2025A&A...700A...3V}
}

@ARTICLE{Bernete_2024,
       author = {{Garc{\'\i}a-Bernete}, I. and {Rigopoulou}, D. and {Donnan}, F.~R. and {Alonso-Herrero}, A. and {Pereira-Santaella}, M. and {Shimizu}, T. and {Davies}, R. and {Roche}, P.~F. and {Garc{\'\i}a-Burillo}, S. and {Labiano}, A. and {Hermosa Mu{\~n}oz}, L. and {Zhang}, L. and {Audibert}, A. and {Bellocchi}, E. and {Bunker}, A. and {Combes}, F. and {Delaney}, D. and {Esparza-Arredondo}, D. and {Gandhi}, P. and {Gonz{\'a}lez-Mart{\'\i}n}, O. and {H{\"o}nig}, S.~F. and {Imanishi}, M. and {Hicks}, E.~K.~S. and {Fuller}, L. and {Leist}, M. and {Levenson}, N.~A. and {Lopez-Rodriguez}, E. and {Packham}, C. and {Ramos Almeida}, C. and {Ricci}, C. and {Stalevski}, M. and {Villar Mart{\'\i}n}, M. and {Ward}, M.~J.},
        title = "{The Galaxy Activity, Torus, and Outflow Survey (GATOS): V. Unveiling PAH survival and resilience in the circumnuclear regions of AGNs with JWST}",
      journal = {\aap},
         year = 2024,
        month = nov,
       volume = {691},
          eid = {A162},
        pages = {A162},
          doi = {10.1051/0004-6361/202450086},
archivePrefix = {arXiv},
       eprint = {2409.05686},
 primaryClass = {astro-ph.GA},
       adsurl = {https://ui.adsabs.harvard.edu/abs/2024A&A...691A.162G}
}

@ARTICLE{Bernete_2022,
       author = {{Garc{\'\i}a-Bernete}, I. and {Rigopoulou}, D. and {Alonso-Herrero}, A. and {Donnan}, F.~R. and {Roche}, P.~F. and {Pereira-Santaella}, M. and {Labiano}, A. and {Peralta de Arriba}, L. and {Izumi}, T. and {Ramos Almeida}, C. and {Shimizu}, T. and {H{\"o}nig}, S. and {Garc{\'\i}a-Burillo}, S. and {Rosario}, D.~J. and {Ward}, M.~J. and {Bellocchi}, E. and {Hicks}, E.~K.~S. and {Fuller}, L. and {Packham}, C.},
        title = "{A high angular resolution view of the PAH emission in Seyfert galaxies using JWST/MRS data}",
      journal = {\aap},
         year = 2022,
        month = oct,
       volume = {666},
          eid = {L5},
        pages = {L5},
          doi = {10.1051/0004-6361/202244806},
archivePrefix = {arXiv},
       eprint = {2208.11620},
 primaryClass = {astro-ph.GA},
       adsurl = {https://ui.adsabs.harvard.edu/abs/2022A&A...666L...5G}
}

@ARTICLE{Shivaei_2024,
       author = {{Shivaei}, Irene and {Boogaard}, Leindert A.},
        title = "{The tight correlation between PAH and CO emission from z {\ensuremath{\sim}} 0 to 4}",
      journal = {\aap},
         year = 2024,
        month = nov,
       volume = {691},
          eid = {L2},
        pages = {L2},
          doi = {10.1051/0004-6361/202451826},
archivePrefix = {arXiv},
       eprint = {2409.05710},
 primaryClass = {astro-ph.GA},
       adsurl = {https://ui.adsabs.harvard.edu/abs/2024A&A...691L...2S}
}

@article{Sun_2023,
doi = {10.3847/2041-8213/acbd9c},
url = {https://dx.doi.org/10.3847/2041-8213/acbd9c},
year = {2023},
month = {mar},
publisher = {The American Astronomical Society},
volume = {945},
number = {2},
pages = {L19},
author = {Sun, Jiayi and Leroy, Adam K. and Ostriker, Eve C. and Meidt, Sharon and Rosolowsky, Erik and Schinnerer, Eva and Wilson, Christine D. and Utomo, Dyas and Belfiore, Francesco and Blanc, Guillermo A. and Emsellem, Eric and Faesi, Christopher and Groves, Brent and Hughes, Annie and Koch, Eric W. and Kreckel, Kathryn and Liu, Daizhong and Pan, Hsi-An and Pety, Jérôme and Querejeta, Miguel and Razza, Alessandro and Saito, Toshiki and Sardone, Amy and Usero, Antonio and Williams, Thomas G. and Bigiel, Frank and Bolatto, Alberto D. and Chevance, Mélanie and Dale, Daniel A. and Gensior, Jindra and Glover, Simon C. O. and Grasha, Kathryn and Henshaw, Jonathan D. and Jiménez-Donaire, María J. and Klessen, Ralf S. and Kruijssen, J. M. Diederik and Murphy, Eric J. and Neumann, Lukas and Teng, Yu-Hsuan and Thilker, David A.},
title = {Star Formation Laws and Efficiencies across 80 Nearby Galaxies},
journal = {\apjl}
}

@article{Tasker_2009,
doi = {10.1088/0004-637X/700/1/358},
url = {https://dx.doi.org/10.1088/0004-637X/700/1/358},
year = {2009},
month = {jul},
publisher = {The American Astronomical Society},
volume = {700},
number = {1},
pages = {358},
author = {Tasker, Elizabeth J. and Tan, Jonathan C.},
title = {STAR FORMATION IN DISK GALAXIES. I. FORMATION AND EVOLUTION OF GIANT MOLECULAR CLOUDS VIA GRAVITATIONAL INSTABILITY AND CLOUD COLLISIONS},
journal = {\apj}
}

@ARTICLE{Kobayashi_2017,
       author = {{Kobayashi}, Masato I.~N. and {Inutsuka}, Shu-ichiro and {Kobayashi}, Hiroshi and {Hasegawa}, Kenji},
        title = "{Evolutionary Description of Giant Molecular Cloud Mass Functions on Galactic Disks}",
      journal = {\apj},
         year = 2017,
        month = feb,
       volume = {836},
       number = {2},
          eid = {175},
        pages = {175},
          doi = {10.3847/1538-4357/836/2/175},
archivePrefix = {arXiv},
       eprint = {1701.03781},
 primaryClass = {astro-ph.GA},
       adsurl = {https://ui.adsabs.harvard.edu/abs/2017ApJ...836..175K}
}

@ARTICLE{Rosolowsky_2008,
       author = {{Rosolowsky}, E.~W. and {Pineda}, J.~E. and {Kauffmann}, J. and {Goodman}, A.~A.},
        title = "{Structural Analysis of Molecular Clouds: Dendrograms}",
      journal = {\apj},
         year = 2008,
        month = jun,
       volume = {679},
       number = {2},
        pages = {1338-1351},
          doi = {10.1086/587685},
archivePrefix = {arXiv},
       eprint = {0802.2944},
 primaryClass = {astro-ph},
       adsurl = {https://ui.adsabs.harvard.edu/abs/2008ApJ...679.1338R}
}

@article{schinnerer2024,
   author = "Schinnerer, E. and Leroy, A.K.",
   title = "Molecular Gas and the Star-Formation Process on Cloud Scales in Nearby Galaxies", 
   journal= "\araa",
   year = "2024",
   volume = "62",
   number = "Volume 62, 2024",
   pages = "369-436",
   doi = "https://doi.org/10.1146/annurev-astro-071221-052651",
   url = "https://www.annualreviews.org/content/journals/10.1146/annurev-astro-071221-052651",
   publisher = "Annual Reviews",
   issn = "1545-4282",
   type = "Journal Article",
  }

@article{Kruijssen_2014,
doi = {10.1088/0264-9381/31/24/244006},
url = {https://dx.doi.org/10.1088/0264-9381/31/24/244006},
year = {2014},
month = {dec},
publisher = {IOP Publishing},
volume = {31},
number = {24},
pages = {244006},
author = {Kruijssen, J. M. Diederik},
title = {Globular cluster formation in the context of galaxy formation and evolution},
journal = {\cqg}
}

@article{Chevance_2020,
   title={The Molecular Cloud Lifecycle},
   volume={216},
   ISSN={1572-9672},
   url={http://dx.doi.org/10.1007/s11214-020-00674-x},
   DOI={10.1007/s11214-020-00674-x},
   number={4},
   journal={\ssr},
   publisher={Springer Science and Business Media LLC},
   author={Chevance, Mélanie and Kruijssen, J. M. Diederik and Vazquez-Semadeni, Enrique and Nakamura, Fumitaka and Klessen, Ralf and Ballesteros-Paredes, Javier and Inutsuka, Shu-ichiro and Adamo, Angela and Hennebelle, Patrick},
   year={2020},
   month=apr }

@ARTICLE{Smith_2020,
       author = {{Smith}, Rowan J. and {Tre{\ss}}, Robin G. and {Sormani}, Mattia C. and {Glover}, Simon C.~O. and {Klessen}, Ralf S. and {Clark}, Paul C. and {Izquierdo}, Andr{\'e}s F. and {Duarte-Cabral}, Ana and {Zucker}, Catherine},
        title = "{The Cloud Factory I: Generating resolved filamentary molecular clouds from galactic-scale forces}",
      journal = {\mnras},
         year = 2020,
        month = feb,
       volume = {492},
       number = {2},
        pages = {1594-1613},
          doi = {10.1093/mnras/stz3328},
archivePrefix = {arXiv},
       eprint = {1911.05753},
 primaryClass = {astro-ph.GA},
       adsurl = {https://ui.adsabs.harvard.edu/abs/2020MNRAS.492.1594S}
}

@ARTICLE{Pettini_2004,
       author = {{Pettini}, Max and {Pagel}, Bernard E.~J.},
        title = "{[OIII]/[NII] as an abundance indicator at high redshift}",
      journal = {\mnras},
         year = 2004,
        month = mar,
       volume = {348},
       number = {3},
        pages = {L59-L63},
          doi = {10.1111/j.1365-2966.2004.07591.x},
archivePrefix = {arXiv},
       eprint = {astro-ph/0401128},
 primaryClass = {astro-ph},
       adsurl = {https://ui.adsabs.harvard.edu/abs/2004MNRAS.348L..59P}
}

@ARTICLE{Colman_2024,
       author = {{Colman}, Tine and {Brucy}, No{\'e} and {Girichidis}, Philipp and {Glover}, Simon C.~O. and {Benedettini}, Milena and {Soler}, Juan D. and {Tress}, Robin G. and {Traficante}, Alessio and {Hennebelle}, Patrick and {Klessen}, Ralf S. and {Molinari}, Sergio and {Miville-Desch{\^e}nes}, Marc-Antoine},
        title = "{Cloud properties across spatial scales in simulations of the interstellar medium}",
      journal = {\aap},
         year = 2024,
        month = jun,
       volume = {686},
          eid = {A155},
        pages = {A155},
          doi = {10.1051/0004-6361/202348983},
archivePrefix = {arXiv},
       eprint = {2403.00512},
 primaryClass = {astro-ph.GA},
       adsurl = {https://ui.adsabs.harvard.edu/abs/2024A&A...686A.155C}
}

@article{Campos_2017,
    author = {Reina-Campos, Marta and Kruijssen, J. M. Diederik},
    title = {A unified model for the maximum mass scales of molecular clouds, stellar clusters and high-redshift clumps},
    journal = {\mnras},
    volume = {469},
    number = {2},
    pages = {1282-1298},
    year = {2017},
    month = {03},
    issn = {0035-8711},
    doi = {10.1093/mnras/stx790},
    url = {https://doi.org/10.1093/mnras/stx790},
    eprint = {https://academic.oup.com/mnras/article-pdf/469/2/1282/17272390/stx790.pdf},
}

@article{Cabral_2016,
    author = {Duarte-Cabral, A. and Dobbs, C. L.},
    title = {What can simulated molecular clouds tell us about real molecular clouds?},
    journal = {\mnras},
    volume = {458},
    number = {4},
    pages = {3667-3683},
    year = {2016},
    month = {03},
    issn = {0035-8711},
    doi = {10.1093/mnras/stw469},
    url = {https://doi.org/10.1093/mnras/stw469},
    eprint = {https://academic.oup.com/mnras/article-pdf/458/4/3667/13454299/stw469.pdf},
}

@article{Field_1965,
  title={A Statistical Model of the Formation of Stars and Interstellar Clouds.},
  author={Field, George B and Saslaw, William C},
  journal={\apj},
  volume={142},
  pages={568},
  year={1965}
}

@article{Taff_1972,
    author = {Taff, L. G. and Savedoff, Malcolm P.},
    title = {The Mass Spectrum of Interstellar Clouds and the Assumption of Total Coalescence},
    journal = {\mnras},
    volume = {160},
    number = {1},
    pages = {89-97},
    year = {1972},
    month = {10},
    issn = {0035-8711},
    doi = {10.1093/mnras/160.1.89},
    url = {https://doi.org/10.1093/mnras/160.1.89},
    eprint = {https://academic.oup.com/mnras/article-pdf/160/1/89/8079447/mnras160-0089.pdf},
}

@ARTICLE{Scoville_1979,
       author = {{Scoville}, N.~Z. and {Hersh}, K.},
        title = "{Collisional growth of giant molecular clouds.}",
      journal = {\apj},
         year = 1979,
        month = apr,
       volume = {229},
        pages = {578-582},
          doi = {10.1086/156991},
       adsurl = {https://ui.adsabs.harvard.edu/abs/1979ApJ...229..578S}
}

@article{Dobbs_2006,
    author = {Dobbs, C. L. and Bonnell, I. A. and Pringle, J. E.},
    title = {The formation of molecular clouds in spiral galaxies},
    journal = {\mnras},
    volume = {371},
    number = {4},
    pages = {1663-1674},
    year = {2006},
    month = {08},
    issn = {0035-8711},
    doi = {10.1111/j.1365-2966.2006.10794.x},
    url = {https://doi.org/10.1111/j.1365-2966.2006.10794.x},
    eprint = {https://academic.oup.com/mnras/article-pdf/371/4/1663/3614912/mnras0371-1663.pdf},
}

@article{Roman_2010,
  title={Physical properties and galactic distribution of molecular clouds identified in the galactic ring survey},
  author={Roman-Duval, Julia and Jackson, James M and Heyer, Mark and Rathborne, Jill and Simon, Robert},
  journal={\apj},
  volume={723},
  number={1},
  pages={492},
  year={2010},
  publisher={IOP Publishing}
}

@article{Eden_2012,
  title={Star formation towards the Scutum tangent region and the effects of Galactic environment},
  author={Eden, DJ and Moore, TJT and Plume, R and Morgan, LK},
  journal={\mnras},
  volume={422},
  number={4},
  pages={3178--3188},
  year={2012},
  publisher={The Royal Astronomical Society}
}

@article{Hirota_2011,
  title={Giant Molecular Clouds in the Spiral Arm of IC 342},
  author={Hirota, Akihiko and Kuno, Nario and Sato, Naoko and Nakanishi, Hiroyuki and Tosaki, Tomoka and Sorai, Kazuo},
  journal={\apj},
  volume={737},
  number={1},
  pages={40},
  year={2011},
  publisher={IOP Publishing}
}

@article{Rebolledo_2012,
doi = {10.1088/0004-637X/757/2/155},
url = {https://dx.doi.org/10.1088/0004-637X/757/2/155},
year = {2012},
month = {sep},
publisher = {The American Astronomical Society},
volume = {757},
number = {2},
pages = {155},
author = {Rebolledo, David and Wong, Tony and Leroy, Adam and Koda, Jin and Meyer, Jennifer Donovan},
title = {GIANT MOLECULAR CLOUDS AND STAR FORMATION IN THE NON-GRAND DESIGN SPIRAL GALAXY NGC 6946},
journal = {\apj}
}

@article{Rebolledo_2015,
doi = {10.1088/0004-637X/808/1/99},
url = {https://dx.doi.org/10.1088/0004-637X/808/1/99},
year = {2015},
month = {jul},
publisher = {The American Astronomical Society},
volume = {808},
number = {1},
pages = {99},
author = {Rebolledo, David and Wong, Tony and Xue, Rui and Leroy, Adam and Koda, Jin and Meyer, Jennifer Donovan},
title = {SCALING RELATIONS OF THE PROPERTIES FOR CO RESOLVED STRUCTURES IN NEARBY SPIRAL GALAXIES},
journal = {\apj}
}

@article{Usero_2015,
  title={Variations in the star formation efficiency of the dense molecular gas across the disks of star-forming galaxies},
  author={Usero, Antonio and Leroy, Adam K and Walter, Fabian and Schruba, Andreas and Garc{\'\i}a-Burillo, Santiago and Sandstrom, Karin and Bigiel, Frank and Brinks, Elias and Kramer, Carsten and Rosolowsky, Erik and others},
  journal={\aj},
  volume={150},
  number={4},
  pages={115},
  year={2015},
  publisher={IOP Publishing}
}

@article{Meyer_2013,
  title={Resolved giant molecular clouds in nearby spiral galaxies: insights from the CANON CO (1--0) survey},
  author={Meyer, Jennifer Donovan and Koda, Jin and Momose, Rieko and Mooney, Thomas and Egusa, Fumi and Carty, Misty and Kennicutt, Robert and Kuno, Nario and Rebolledo, David and Sawada, Tsuyoshi and others},
  journal={\apj},
  volume={772},
  number={2},
  pages={107},
  year={2013},
  publisher={IOP Publishing}
}

@article{Nimori_2012,
    author = {Nimori, M. and Habe, A. and Sorai, K. and Watanabe, Y. and Hirota, A. and Namekata, D.},
    title = {Dense cloud formation and star formation in a barred galaxy},
    journal = {\mnras},
    volume = {429},
    number = {3},
    pages = {2175-2182},
    year = {2012},
    month = {12},
    issn = {0035-8711},
    doi = {10.1093/mnras/sts487},
    url = {https://doi.org/10.1093/mnras/sts487},
    eprint = {https://academic.oup.com/mnras/article-pdf/429/3/2175/3223605/sts487.pdf},
}

@article{Fujimoto_2014,
    author = {Fujimoto, Yusuke and Tasker, Elizabeth J. and Wakayama, Mariko and Habe, Asao},
    title = {Do giant molecular clouds care about the galactic structure?},
    journal = {\mnras},
    volume = {439},
    number = {1},
    pages = {936-953},
    year = {2014},
    month = {02},
    issn = {0035-8711},
    doi = {10.1093/mnras/stu014},
    url = {https://doi.org/10.1093/mnras/stu014},
    eprint = {https://academic.oup.com/mnras/article-pdf/439/1/936/5617459/stu014.pdf},
}

@ARTICLE{Kobayashi_2018,
       author = {{Kobayashi}, Masato I.~N. and {Kobayashi}, Hiroshi and {Inutsuka}, Shu-ichiro and {Fukui}, Yasuo},
        title = "{Star formation induced by cloud-cloud collisions and galactic giant molecular cloud evolution}",
      journal = {\pasj},
         year = 2018,
        month = may,
       volume = {70},
          eid = {S59},
        pages = {S59},
          doi = {10.1093/pasj/psy018},
archivePrefix = {arXiv},
       eprint = {1708.07952},
 primaryClass = {astro-ph.GA},
       adsurl = {https://ui.adsabs.harvard.edu/abs/2018PASJ...70S..59K}
}

@article{Dobbs_2015,
    author = {Dobbs, Clare L.},
    title = {The interstellar medium and star formation on kpc size scales},
    journal = {\mnras},
    volume = {447},
    number = {4},
    pages = {3390-3401},
    year = {2015},
    month = {01},
    issn = {0035-8711},
    doi = {10.1093/mnras/stu2585},
    url = {https://doi.org/10.1093/mnras/stu2585},
    eprint = {https://academic.oup.com/mnras/article-pdf/447/4/3390/5685597/stu2585.pdf},
}

@ARTICLE{Dale_2025,
       author = {{Dale}, Daniel A. and {Graham}, Gabrielle B. and {Barnes}, Ashley T. and {Baron}, Dalya and {Bigiel}, Frank and {Boquien}, M{\'e}d{\'e}ric and {Chandar}, Rupali and {Chastenet}, J{\'e}r{\'e}my and {Chown}, Ryan and {Egorov}, Oleg V. and {Glover}, Simon C.~O. and {Hands}, Lindsey and {Henny}, Kiana F. and {Indebetouw}, Remy and {Klessen}, Ralf S. and {Larson}, Kirsten L. and {Lee}, Janice C. and {Leroy}, Adam K. and {Maschmann}, Daniel and {Pathak}, Debosmita and {Rodr{\'\i}guez}, M. Jimena and {Rosolowsky}, Erik and {Sandstrom}, Karin and {Schinnerer}, Eva and {Sutter}, Jessica and {Thilker}, David A. and {Weinbeck}, Tony D. and {Whitmore}, Bradley C. and {Williams}, Thomas G. and {Wofford}, Aida},
        title = "{PAH Feature Ratios around Stellar Clusters and Associations in 19 Nearby Galaxies}",
      journal = {\aj},
         year = 2025,
        month = mar,
       volume = {169},
       number = {3},
          eid = {133},
        pages = {133},
          doi = {10.3847/1538-3881/ada89f},
archivePrefix = {arXiv},
       eprint = {2501.10539},
 primaryClass = {astro-ph.GA},
       adsurl = {https://ui.adsabs.harvard.edu/abs/2025AJ....169..133D}
}

@article{Baron_2024,
doi = {10.3847/1538-4357/ad39e5},
url = {https://dx.doi.org/10.3847/1538-4357/ad39e5},
year = {2024},
month = {jun},
publisher = {The American Astronomical Society},
volume = {968},
number = {1},
pages = {24},
author = {Baron, Dalya and Sandstrom, Karin M. and Rosolowsky, Erik and Egorov, Oleg V. and Klessen, Ralf S. and Leroy, Adam K. and Boquien, Médéric and Schinnerer, Eva and Belfiore, Francesco and Groves, Brent and Chastenet, Jérémy and Dale, Daniel A. and Blanc, Guillermo A. and Méndez-Delgado, José E. and Koch, Eric W. and Grasha, Kathryn and Chevance, Mélanie and Thilker, David A. and Colombo, Dario and Williams, Thomas G. and Pathak, Debosmita and Sutter, Jessica and Brown, Toby and Wu, John F. and Peek, Josh E. G. and Emsellem, Eric and Larson, Kirsten L. and Neumann, Justus},
title = {PHANGS-ML: Dissecting Multiphase Gas and Dust in Nearby Galaxies Using Machine Learning},
journal = {\apj}
}

@article{Pettitt_2020,
    author = {Pettitt, Alex R and Dobbs, Clare L and Baba, Junichi and Colombo, Dario and Duarte-Cabral, Ana and Egusa, Fumi and Habe, Asao},
    title = {How do different spiral arm models impact the ISM and GMC population?},
    journal = {\mnras},
    volume = {498},
    number = {1},
    pages = {1159-1174},
    year = {2020},
    month = {08},
    issn = {0035-8711},
    doi = {10.1093/mnras/staa2242},
    url = {https://doi.org/10.1093/mnras/staa2242},
    eprint = {https://academic.oup.com/mnras/article-pdf/498/1/1159/33731268/staa2242.pdf},
}

@article{Whitcomb_2023,
doi = {10.3847/1538-4357/acc316},
url = {https://dx.doi.org/10.3847/1538-4357/acc316},
year = {2023},
month = {may},
publisher = {The American Astronomical Society},
volume = {948},
number = {2},
pages = {88},
author = {Whitcomb, C. M. and Sandstrom, K. and Leroy, A. and Smith, J.-D. T.},
title = {Star Formation and Molecular Gas Diagnostics with Mid- and Far-infrared Emission},
journal = {\apj}
}

@ARTICLE{Rosolowsky_2005b,
       author = {{Rosolowsky}, E. and {Blitz}, L.},
        title = "{Giant Molecular Clouds in M64}",
      journal = {\apj},
         year = 2005,
        month = apr,
       volume = {623},
       number = {2},
        pages = {826-845},
          doi = {10.1086/428897},
archivePrefix = {arXiv},
       eprint = {astro-ph/0501387},
 primaryClass = {astro-ph},
       adsurl = {https://ui.adsabs.harvard.edu/abs/2005ApJ...623..826R}
}

@article{Peeters_2004,
doi = {10.1086/423237},
url = {https://dx.doi.org/10.1086/423237},
year = {2004},
month = {oct},
publisher = {},
volume = {613},
number = {2},
pages = {986},
author = {Peeters, E. and Spoon, H. W. W. and Tielens, A. G. G. M.},
title = {Polycyclic Aromatic Hydrocarbons as a Tracer of Star Formation?},
journal = {\apj}
}

@article{Belfiore_2023,
   title={Calibrating mid-infrared emission as a tracer of obscured star formation on H II-region scales in the era of JWST},
   volume={678},
   ISSN={1432-0746},
   url={http://dx.doi.org/10.1051/0004-6361/202347175},
   DOI={10.1051/0004-6361/202347175},
   journal={\aap},
   publisher={EDP Sciences},
   author={Belfiore, Francesco and Leroy, Adam K. and Williams, Thomas G. and Barnes, Ashley T. and Bigiel, Frank and Boquien, Médéric and Cao, Yixian and Chastenet, Jérémy and Congiu, Enrico and Dale, Daniel A. and Egorov, Oleg V. and Eibensteiner, Cosima and Emsellem, Eric and Glover, Simon C. O. and Groves, Brent and Hassani, Hamid and Klessen, Ralf S. and Kreckel, Kathryn and Neumann, Lukas and Neumann, Justus and Querejeta, Miguel and Rosolowsky, Erik and Sanchez-Blazquez, Patricia and Sandstrom, Karin and Schinnerer, Eva and Sun, Jiayi and Sutter, Jessica and Watkins, Elizabeth J.},
   year={2023},
   month=oct, pages={A129} }

@article{Calzetti_2007,
doi = {10.1086/520082},
url = {https://dx.doi.org/10.1086/520082},
year = {2007},
month = {sep},
publisher = {},
volume = {666},
number = {2},
pages = {870},
author = {Calzetti, D. and Kennicutt, R. C. and Engelbracht, C. W. and Leitherer, C. and Draine, B. T. and Kewley, L. and Moustakas, J. and Sosey, M. and Dale, D. A. and Gordon, K. D. and Helou, G. X. and Hollenbach, D. J. and Armus, L. and Bendo, G. and Bot, C. and Buckalew, B. and Jarrett, T. and Li, A. and Meyer, M. and Murphy, E. J. and Prescott, M. and Regan, M. W. and Rieke, G. H. and Roussel, H. and Sheth, K. and Smith, J. D. T. and Thornley, M. D. and Walter, F.},
title = {The Calibration of Mid-Infrared Star Formation Rate Indicators*},
journal = {\apj}
}

@article{Grand_2017,
    author = {Grand, Robert J. J. and Gómez, Facundo A. and Marinacci, Federico and Pakmor, Rüdiger and Springel, Volker and Campbell, David J. R. and Frenk, Carlos S. and Jenkins, Adrian and White, Simon D. M.},
    title = {The Auriga Project: the properties and formation mechanisms of disc galaxies across cosmic time},
    journal = {\mnras},
    volume = {467},
    number = {1},
    pages = {179-207},
    year = {2017},
    month = {01},
    issn = {0035-8711},
    doi = {10.1093/mnras/stx071},
    url = {https://doi.org/10.1093/mnras/stx071},
    eprint = {https://academic.oup.com/mnras/article-pdf/467/1/179/10327352/stx071.pdf},
}

@article{Dobbs_2015b,
    author = {Dobbs, C. L. and Pringle, J. E. and Duarte-Cabral, A.},
    title = {The frequency and nature of ‘cloud–cloud collisions’ in galaxies},
    journal = {\mnras},
    volume = {446},
    number = {4},
    pages = {3608-3620},
    year = {2014},
    month = {12},
    issn = {0035-8711},
    doi = {10.1093/mnras/stu2319},
    url = {https://doi.org/10.1093/mnras/stu2319},
    eprint = {https://academic.oup.com/mnras/article-pdf/446/4/3608/9381757/stu2319.pdf},
}

@article{Jeffreson_2018,
    author = {Jeffreson, Sarah M R and Kruijssen, J M Diederik},
    title = {A general theory for the lifetimes of giant molecular clouds under the influence of galactic dynamics},
    journal = {\mnras},
    volume = {476},
    number = {3},
    pages = {3688-3715},
    year = {2018},
    month = {03},
    issn = {0035-8711},
    doi = {10.1093/mnras/sty594},
    url = {https://doi.org/10.1093/mnras/sty594},
    eprint = {https://academic.oup.com/mnras/article-pdf/476/3/3688/24508053/sty594.pdf},
}

@article{Meidt_2018,
doi = {10.3847/1538-4357/aaa290},
url = {https://dx.doi.org/10.3847/1538-4357/aaa290},
year = {2018},
month = {feb},
publisher = {The American Astronomical Society},
volume = {854},
number = {2},
pages = {100},
author = {Meidt, Sharon E. and Leroy, Adam K. and Rosolowsky, Erik and Kruijssen, J. M. Diederik and Schinnerer, Eva and Schruba, Andreas and Pety, Jerome and Blanc, Guillermo and Bigiel, Frank and Chevance, Melanie and Hughes, Annie and Querejeta, Miguel and Usero, Antonio},
title = {A Model for the Onset of Self-gravitation and Star Formation in Molecular Gas Governed by Galactic Forces. I. Cloud-scale Gas Motions},
journal = {\apj}
}

@article{Hopkins_2018,
    author = {Hopkins, Philip F and Wetzel, Andrew and Kereš, Dušan and Faucher-Giguère, Claude-André and Quataert, Eliot and Boylan-Kolchin, Michael and Murray, Norman and Hayward, Christopher C and Garrison-Kimmel, Shea and Hummels, Cameron and Feldmann, Robert and Torrey, Paul and Ma, Xiangcheng and Anglés-Alcázar, Daniel and Su, Kung-Yi and Orr, Matthew and Schmitz, Denise and Escala, Ivanna and Sanderson, Robyn and Grudić, Michael Y and Hafen, Zachary and Kim, Ji-Hoon and Fitts, Alex and Bullock, James S and Wheeler, Coral and Chan, T K and Elbert, Oliver D and Narayanan, Desika},
    title = {FIRE-2 simulations: physics versus numerics in galaxy formation},
    journal = {\mnras},
    volume = {480},
    number = {1},
    pages = {800-863},
    year = {2018},
    month = {06},
    issn = {0035-8711},
    doi = {10.1093/mnras/sty1690},
    url = {https://doi.org/10.1093/mnras/sty1690},
    eprint = {https://academic.oup.com/mnras/article-pdf/480/1/800/25368704/sty1690.pdf},
}

@article{Cabral_2021,
    author = {Duarte-Cabral, A and Colombo, D and Urquhart, J S and Ginsburg, A and Russeil, D and Schuller, F and Anderson, L D and Barnes, P J and Beltrán, M T and Beuther, H and Bontemps, S and Bronfman, L and Csengeri, T and Dobbs, C L and Eden, D and Giannetti, A and Kauffmann, J and Mattern, M and Medina, S-N X and Menten, K M and Lee, M-Y and Pettitt, A R and Riener, M and Rigby, A J and Traficante, A and Veena, V S and Wienen, M and Wyrowski, F and Agurto, C and Azagra, F and Cesaroni, R and Finger, R and Gonzalez, E and Henning, T and Hernandez, A K and Kainulainen, J and Leurini, S and Lopez, S and Mac-Auliffe, F and Mazumdar, P and Molinari, S and Motte, F and Muller, E and Nguyen-Luong, Q and Parra, R and Perez-Beaupuits, J-P and Montenegro-Montes, F M and Moore, T J T and Ragan, S E and Sánchez-Monge, A and Sanna, A and Schilke, P and Schisano, E and Schneider, N and Suri, S and Testi, L and Torstensson, K and Venegas, P and Wang, K and Zavagno, A},
    title = {The SEDIGISM survey: molecular clouds in the inner Galaxy},
    journal = {\mnras},
    volume = {500},
    number = {3},
    pages = {3027-3049},
    year = {2020},
    month = {09},
    issn = {0035-8711},
    doi = {10.1093/mnras/staa2480},
    url = {https://doi.org/10.1093/mnras/staa2480},
    eprint = {https://academic.oup.com/mnras/article-pdf/500/3/3027/43389727/staa2480.pdf},
}

@article{Meidt_2013,
doi = {10.1088/0004-637X/779/1/45},
url = {https://dx.doi.org/10.1088/0004-637X/779/1/45},
year = {2013},
month = {nov},
publisher = {The American Astronomical Society},
volume = {779},
number = {1},
pages = {45},
author = {Meidt, Sharon E. and Schinnerer, Eva and García-Burillo, Santiago and Hughes, Annie and Colombo, Dario and Pety, Jérôme and Dobbs, Clare L. and Schuster, Karl F. and Kramer, Carsten and Leroy, Adam K. and Dumas, Galle and Thompson, Todd A.},
title = {GAS KINEMATICS ON GIANT MOLECULAR CLOUD SCALES IN M51 WITH PAWS: CLOUD STABILIZATION THROUGH DYNAMICAL PRESSURE},
journal = {\apj}
}

@article{Vieira_2025,
    author = {Faustino Vieira, Helena and Duarte-Cabral, Ana and Smith, Matthew W L and Colombo, Dario and Davis, Timothy A and Bazzi, Zein},
    title = {Cloud populations versus galactic environment in NGC 4689, NGC 628, NGC 1566 and NGC 4321},
    journal = {\mnras},
    pages = {staf411},
    year = {2025},
    month = {03},
    issn = {0035-8711},
    doi = {10.1093/mnras/staf411},
    url = {https://doi.org/10.1093/mnras/staf411},
    eprint = {https://academic.oup.com/mnras/advance-article-pdf/doi/10.1093/mnras/staf411/62388066/staf411.pdf},
}

@article{Casoli_1982,
  title={Can giant molecular clouds form in spiral arms},
  author={Casoli, F and Combes, F},
  journal={\aap},
  volume={110},
  pages={287--294},
  year={1982}
}

@article{Dobbs_2008,
    author = {Dobbs, C. L.},
    title = {GMC formation by agglomeration and self gravity},
    journal = {\mnras},
    volume = {391},
    number = {2},
    pages = {844-858},
    year = {2008},
    month = {11},
    issn = {0035-8711},
    doi = {10.1111/j.1365-2966.2008.13939.x},
    url = {https://doi.org/10.1111/j.1365-2966.2008.13939.x},
    eprint = {https://academic.oup.com/mnras/article-pdf/391/2/844/5774001/mnras0391-0844.pdf},
}

@ARTICLE{Toomre_1964,
       author = {{Toomre}, A.},
        title = "{On the gravitational stability of a disk of stars.}",
      journal = {\apj},
         year = 1964,
        month = may,
       volume = {139},
        pages = {1217-1238},
          doi = {10.1086/147861},
       adsurl = {https://ui.adsabs.harvard.edu/abs/1964ApJ...139.1217T}
}

@ARTICLE{Pathak_2024,
       author = {{Pathak}, Debosmita and {Leroy}, Adam K. and {Thompson}, Todd A. and {Lopez}, Laura A. and {Belfiore}, Francesco and {Boquien}, M{\'e}d{\'e}ric and {Dale}, Daniel A. and {Glover}, Simon C.~O. and {Klessen}, Ralf S. and {Koch}, Eric W. and {Rosolowsky}, Erik and {Sandstrom}, Karin M. and {Schinnerer}, Eva and {Smith}, Rowan and {Sun}, Jiayi and {Sutter}, Jessica and {Williams}, Thomas G. and {Bigiel}, Frank and {Cao}, Yixian and {Chastenet}, J{\'e}r{\'e}my and {Chevance}, M{\'e}lanie and {Chown}, Ryan and {Emsellem}, Eric and {Faesi}, Christopher M. and {Larson}, Kirsten L. and {Lee}, Janice C. and {Meidt}, Sharon and {Ostriker}, Eve C. and {Ramambason}, Lise and {Sarbadhicary}, Sumit K. and {Thilker}, David A.},
        title = "{A Two-Component Probability Distribution Function Describes the Mid-IR Emission from the Disks of Star-forming Galaxies}",
      journal = {\aj},
         year = 2024,
        month = jan,
       volume = {167},
       number = {1},
          eid = {39},
        pages = {39},
          doi = {10.3847/1538-3881/ad110d},
archivePrefix = {arXiv},
       eprint = {2311.18067},
 primaryClass = {astro-ph.GA},
       adsurl = {https://ui.adsabs.harvard.edu/abs/2024AJ....167...39P}
}

@article{Koda_2009,
doi = {10.1088/0004-637X/700/2/L132},
url = {https://dx.doi.org/10.1088/0004-637X/700/2/L132},
year = {2009},
month = {jul},
publisher = {The American Astronomical Society},
volume = {700},
number = {2},
pages = {L132},
author = {Koda, Jin and Scoville, Nick and Sawada, Tsuyoshi and La Vigne, Misty A. and Vogel, Stuart N. and Potts, Ashley E. and Carpenter, John M. and Corder, Stuartt A. and Wright, Melvyn C. H. and White, Stephen M. and Zauderer, B. Ashley and Patience, Jenny and Sargent, Anneila I. and Bock, Douglas C. J. and Hawkins, David and Hodges, Mark and Kemball, Athol and Lamb, James W. and Plambeck, Richard L. and Pound, Marc W. and Scott, Stephen L. and Teuben, Peter and Woody, David P.},
title = {DYNAMICALLY DRIVEN EVOLUTION OF THE INTERSTELLAR MEDIUM IN M51},
journal = {\apj}
}

@article{Ragan_2014,
	author = {Ragan, S. E. and Henning, Th. and Tackenberg, J.and Beuther, H. and Johnston, K. G. and Kainulainen, J. and Linz, H.},
	title = {Giant molecular filaments in the Milky Way⋆},
	DOI= "10.1051/0004-6361/201423401",
	url= "https://doi.org/10.1051/0004-6361/201423401",
	journal = {A\&A},
	year = 2014,
	volume = 568,
	pages = "A73",
	month = "",
}

@article{Dobbs_2011,
    author = {Dobbs, C. L. and Burkert, A. and Pringle, J. E.},
    title = {The properties of the interstellar medium in disc galaxies with stellar feedback},
    journal = {\mnras},
    volume = {417},
    number = {2},
    pages = {1318-1334},
    year = {2011},
    month = {10},
    issn = {0035-8711},
    doi = {10.1111/j.1365-2966.2011.19346.x},
    url = {https://doi.org/10.1111/j.1365-2966.2011.19346.x},
    eprint = {https://academic.oup.com/mnras/article-pdf/417/2/1318/3070165/mnras0417-1318.pdf},
}

@article{Querejeta_2024,
	author = {{Querejeta, Miguel} and {Leroy, Adam K.} and {Meidt, Sharon E.} and {Schinnerer, Eva} and {Belfiore, Francesco} and {Emsellem, Eric} and {Klessen, Ralf S.} and {Sun, Jiayi} and {Sormani, Mattia} and {Bešlić, Ivana} and {Cao, Yixian} and {Chevance, Mélanie} and {Colombo, Dario} and {Dale, Daniel A.} and {García-Burillo, Santiago} and {Glover, Simon C. O.} and {Grasha, Kathryn} and {Groves, Brent} and {Koch, Eric. W.} and {Neumann, Lukas} and {Pan, Hsi-An} and {Pessa, Ismael} and {Pety, Jérôme} and {Pinna, Francesca} and {Ramambason, Lise} and {Razza, Alessandro} and {Romanelli, Andrea} and {Rosolowsky, Erik} and {Ruiz-García, Marina} and {Sánchez-Blázquez, Patricia} and {Smith, Rowan} and {Stuber, Sophia} and {Ubeda, Leonardo} and {Usero, Antonio} and {Williams, Thomas G.}},
	title = {Do spiral arms enhance star formation efficiency?},
	DOI= "10.1051/0004-6361/202449733",
	url= "https://doi.org/10.1051/0004-6361/202449733",
	journal = {A\&A},
	year = 2024,
	volume = 687,
	pages = "A293",
}

@article{Faesi_2018,
doi = {10.3847/1538-4357/aaad60},
url = {https://dx.doi.org/10.3847/1538-4357/aaad60},
year = {2018},
month = {apr},
publisher = {The American Astronomical Society},
volume = {857},
number = {1},
pages = {19},
author = {Faesi, Christopher M. and Lada, Charles J. and Forbrich, Jan},
title = {The ALMA View of GMCs in NGC 300: Physical Properties and Scaling Relations at 10 pc Resolution},
journal = {\apj}
}

@article{Meidt_2021,
doi = {10.3847/1538-4357/abf35b},
url = {https://dx.doi.org/10.3847/1538-4357/abf35b},
year = {2021},
month = {jun},
publisher = {The American Astronomical Society},
volume = {913},
number = {2},
pages = {113},
author = {Meidt, Sharon E. and Leroy, Adam K. and Querejeta, Miguel and Schinnerer, Eva and Sun, Jiayi and van der Wel, Arjen and Emsellem, Eric and Henshaw, Jonathan and Hughes, Annie and Kruijssen, J. M. Diederik and Rosolowsky, Erik and Schruba, Andreas and Barnes, Ashley and Bigiel, Frank and Blanc, Guillermo A. and Chevance, Melanie and Cao, Yixian and Dale, Daniel A. and Faesi, Christopher and Glover, Simon C. O. and Grasha, Kathryn and Groves, Brent and Herrera, Cynthia and Klessen, Ralf S. and Kreckel, Kathryn and Liu, Daizhong and Pan, Hsi-An and Pety, Jerome and Saito, Toshiki and Usero, Antonio and Watkins, Elizabeth and Williams, Thomas G.},
title = {The Organization of Cloud-scale Gas Density Structure: High-resolution CO versus 3.6 μm Brightness Contrasts in Nearby Galaxies},
journal = {\apj}
}

@article{Sormani_2019,
    author = {Sormani, Mattia C and Barnes, Ashley T},
    title = {Mass inflow rate into the Central Molecular Zone: observational determination and evidence of episodic accretion},
    journal = {\mnras},
    volume = {484},
    number = {1},
    pages = {1213-1219},
    year = {2019},
    month = {01},
    issn = {0035-8711},
    doi = {10.1093/mnras/stz046},
    url = {https://doi.org/10.1093/mnras/stz046},
    eprint = {https://academic.oup.com/mnras/article-pdf/484/1/1213/27577311/stz046.pdf},
}

@article{Longmore_2012,
    author = {Longmore, S. N. and Bally, J. and Testi, L. and Purcell, C. R. and Walsh, A. J. and Bressert, E. and Pestalozzi, M. and Molinari, S. and Ott, J. and Cortese, L. and Battersby, C. and Murray, N. and Lee, E. and Kruijssen, J. M. D. and Schisano, E. and Elia, D.},
    title = {Variations in the Galactic star formation rate and density thresholds for star formation},
    journal = {\mnras},
    volume = {429},
    number = {2},
    pages = {987-1000},
    year = {2012},
    month = {12},
    issn = {0035-8711},
    doi = {10.1093/mnras/sts376},
    url = {https://doi.org/10.1093/mnras/sts376},
    eprint = {https://academic.oup.com/mnras/article-pdf/429/2/987/18448189/sts376.pdf},
}

@ARTICLE{Tress_2021,
       author = {{Tre{\ss}}, Robin G. and {Sormani}, Mattia C. and {Smith}, Rowan J. and {Glover}, Simon C.~O. and {Klessen}, Ralf S. and {Mac Low}, Mordecai-Mark and {Clark}, Paul and {Duarte-Cabral}, Ana},
        title = "{Simulations of the star-forming molecular gas in an interacting M51-like galaxy: cloud population statistics}",
      journal = {\mnras},
         year = 2021,
        month = aug,
       volume = {505},
       number = {4},
        pages = {5438-5459},
          doi = {10.1093/mnras/stab1683},
archivePrefix = {arXiv},
       eprint = {2012.05919},
 primaryClass = {astro-ph.GA},
       adsurl = {https://ui.adsabs.harvard.edu/abs/2021MNRAS.505.5438T}
}

@article{Sandstrom_2023b,
doi = {10.3847/2041-8213/aca972},
url = {https://dx.doi.org/10.3847/2041-8213/aca972},
year = {2023},
month = {feb},
publisher = {The American Astronomical Society},
volume = {944},
number = {2},
pages = {L8},
author = {Sandstrom, Karin M. and Koch, Eric W. and Leroy, Adam K. and Rosolowsky, Erik and Emsellem, Eric and Smith, Rowan J. and Egorov, Oleg V. and Williams, Thomas G. and Larson, Kirsten L. and Lee, Janice C. and Schinnerer, Eva and Thilker, David A. and Barnes, Ashley T. and Belfiore, Francesco and Bigiel, F. and Blanc, Guillermo A. and Bolatto, Alberto D. and Boquien, Médéric and Cao, Yixian and Chastenet, Jérémy and Chevance, Mélanie and Chiang, I-Da and Dale, Daniel A. and Faesi, Christopher M. and Glover, Simon C. O. and Grasha, Kathryn and Groves, Brent and Hassani, Hamid and Henshaw, Jonathan D. and Hughes, Annie and Kim, Jaeyeon and Klessen, Ralf S. and Kreckel, Kathryn and Kruijssen, J. M. Diederik and Lopez, Laura A. and Liu, Daizhong and Meidt, Sharon E. and Murphy, Eric J. and Pan, Hsi-An and Querejeta, Miguel and Saito, Toshiki and Sardone, Amy and Sormani, Mattia C. and Sutter, Jessica and Usero, Antonio and Watkins, Elizabeth J.},
title = {PHANGS–JWST First Results: Tracing the Diffuse Interstellar Medium with JWST Imaging of Polycyclic Aromatic Hydrocarbon Emission in Nearby Galaxies},
journal = {\apjl}
}

@ARTICLE{Saintonge_2022,
       author = {{Saintonge}, Am{\'e}lie and {Catinella}, Barbara},
        title = "{The Cold Interstellar Medium of Galaxies in the Local Universe}",
      journal = {\araa},
         year = 2022,
        month = aug,
       volume = {60},
        pages = {319-361},
          doi = {10.1146/annurev-astro-021022-043545},
archivePrefix = {arXiv},
       eprint = {2202.00690},
 primaryClass = {astro-ph.GA},
       adsurl = {https://ui.adsabs.harvard.edu/abs/2022ARA&A..60..319S}
}

@misc{Mills_2017,
      title={The Milky Way's Central Molecular Zone}, 
      author={E. A. C. Mills},
      year={2017},
      eprint={1705.05332},
      archivePrefix={arXiv},
      primaryClass={astro-ph.GA},
      url={https://arxiv.org/abs/1705.05332}, 
}

@ARTICLE{Leroy_2025,
       author = {{Leroy}, Adam K. and {Sun}, Jiayi and {Meidt}, Sharon and {Agertz}, Oscar and {Chiang}, I. -Da and {Gensior}, Jindra and {Glover}, Simon C.~O. and {Gnedin}, Oleg Y. and {Hughes}, Annie and {Schinnerer}, Eva and {Barnes}, Ashley T. and {Bigiel}, Frank and {Bolatto}, Alberto D. and {Colombo}, Dario and {den Brok}, Jakob and {Chevance}, M{\'e}lanie and {Chown}, Ryan and {Eibensteiner}, Cosima and {Gleis}, Damian R. and {Grasha}, Kathryn and {Henshaw}, Jonathan D. and {Klessen}, Ralf S. and {Koch}, Eric W. and {Oakes}, Elias K. and {Pan}, Hsi-An and {Querejeta}, Miguel and {Rosolowsky}, Erik and {Saito}, Toshiki and {Sandstrom}, Karin and {Sarbadhicary}, Sumit K. and {Teng}, Yu-Hsuan and {Usero}, Antonio and {Utomo}, Dyas and {Williams}, Thomas G.},
        title = "{Cloud-scale Gas Properties, Depletion Times, and Star Formation Efficiency per Freefall Time in PHANGS{\textendash}ALMA}",
      journal = {\apj},
         year = 2025,
        month = may,
       volume = {985},
       number = {1},
          eid = {14},
        pages = {14},
          doi = {10.3847/1538-4357/adbcab},
archivePrefix = {arXiv},
       eprint = {2502.04481},
 primaryClass = {astro-ph.GA},
       adsurl = {https://ui.adsabs.harvard.edu/abs/2025ApJ...985...14L}
}

@ARTICLE{Savage_1977,
       author = {{Savage}, B.~D. and {Bohlin}, R.~C. and {Drake}, J.~F. and {Budich}, W.},
        title = "{A survey of interstellar molecular hydrogen. I.}",
      journal = {\apj},
         year = 1977,
        month = aug,
       volume = {216},
        pages = {291-307},
          doi = {10.1086/155471},
       adsurl = {https://ui.adsabs.harvard.edu/abs/1977ApJ...216..291S}
}

@article{Elmegreen_2011,
   title={Star Formation in Spiral Arms},
   volume={51},
   ISSN={1638-1963},
   url={http://dx.doi.org/10.1051/eas/1151002},
   DOI={10.1051/eas/1151002},
   journal={EAS Publications Series},
   publisher={EDP Sciences},
   author={Elmegreen, B.G.},
   year={2011},
   pages={19–30} }

@ARTICLE{Smith_2014,
       author = {{Smith}, Rowan J. and {Glover}, Simon C.~O. and {Clark}, Paul C. and {Klessen}, Ralf S. and {Springel}, Volker},
        title = "{CO-dark gas and molecular filaments in Milky Way-type galaxies}",
      journal = {\mnras},
         year = 2014,
        month = jun,
       volume = {441},
       number = {2},
        pages = {1628-1645},
          doi = {10.1093/mnras/stu616},
archivePrefix = {arXiv},
       eprint = {1403.1589},
 primaryClass = {astro-ph.GA},
       adsurl = {https://ui.adsabs.harvard.edu/abs/2014MNRAS.441.1628S}
}

@ARTICLE{Glover_2016,
       author = {{Glover}, Simon C.~O. and {Smith}, Rowan J.},
        title = "{CO-dark gas and molecular filaments in Milky Way-type galaxies - II. The temperature distribution of the gas}",
      journal = {\mnras},
         year = 2016,
        month = nov,
       volume = {462},
       number = {3},
        pages = {3011-3025},
          doi = {10.1093/mnras/stw1879},
archivePrefix = {arXiv},
       eprint = {1607.08253},
 primaryClass = {astro-ph.GA},
       adsurl = {https://ui.adsabs.harvard.edu/abs/2016MNRAS.462.3011G}
}

@article{Hu_2022,
  title={Dependence of X CO on Metallicity, Intensity, and Spatial Scale in a Self-regulated Interstellar Medium},
  author={Hu, Chia-Yu and Schruba, Andreas and Sternberg, Amiel and van Dishoeck, Ewine F},
  journal={\apj},
  volume={931},
  number={1},
  pages={28},
  year={2022},
  publisher={IOP Publishing}
}

@article{Leroy_2011,
doi = {10.1088/0004-637X/737/1/12},
url = {https://dx.doi.org/10.1088/0004-637X/737/1/12},
year = {2011},
month = {jul},
publisher = {The American Astronomical Society},
volume = {737},
number = {1},
pages = {12},
author = {Leroy, Adam K. and Bolatto, Alberto and Gordon, Karl and Sandstrom, Karin and Gratier, Pierre and Rosolowsky, Erik and Engelbracht, Charles W. and Mizuno, Norikazu and Corbelli, Edvige and Fukui, Yasuo and Kawamura, Akiko},
title = {THE CO-TO-H2 CONVERSION FACTOR FROM INFRARED DUST EMISSION ACROSS THE LOCAL GROUP},
journal = {\apj}
}

@article{Glover_2011,
  title={On the relationship between molecular hydrogen and carbon monoxide abundances in molecular clouds},
  author={Glover, SCO and Low, M-M Mac},
  journal={\mnras},
  volume={412},
  number={1},
  pages={337--350},
  year={2011},
  publisher={Blackwell Publishing Ltd Oxford, UK}
}

@ARTICLE{Wolfire_2010,
       author = {{Wolfire}, Mark G. and {Hollenbach}, David and {McKee}, Christopher F.},
        title = "{The Dark Molecular Gas}",
      journal = {\apj},
         year = 2010,
        month = jun,
       volume = {716},
       number = {2},
        pages = {1191-1207},
          doi = {10.1088/0004-637X/716/2/1191},
archivePrefix = {arXiv},
       eprint = {1004.5401},
 primaryClass = {astro-ph.GA},
       adsurl = {https://ui.adsabs.harvard.edu/abs/2010ApJ...716.1191W}
}

@ARTICLE{Van_1988,
       author = {{van Dishoeck}, Ewine F. and {Black}, John H.},
        title = "{The Photodissociation and Chemistry of Interstellar CO}",
      journal = {\apj},
         year = 1988,
        month = nov,
       volume = {334},
        pages = {771},
          doi = {10.1086/166877},
       adsurl = {https://ui.adsabs.harvard.edu/abs/1988ApJ...334..771V}
}

@article{Jameson_2016,
doi = {10.3847/0004-637X/825/1/12},
url = {https://dx.doi.org/10.3847/0004-637X/825/1/12},
year = {2016},
month = {jun},
publisher = {The American Astronomical Society},
volume = {825},
number = {1},
pages = {12},
author = {Jameson, Katherine E. and Bolatto, Alberto D. and Leroy, Adam K. and Meixner, Margaret and Roman-Duval, Julia and Gordon, Karl and Hughes, Annie and Israel, Frank P. and Rubio, Monica and Indebetouw, Remy and Madden, Suzanne C. and Bot, Caroline and Hony, Sacha and Cormier, Diane and Pellegrini, Eric W. and Galametz, Maud and Sonneborn, George},
title = {THE RELATIONSHIP BETWEEN MOLECULAR GAS, H i, AND STAR FORMATION IN THE LOW-MASS, LOW-METALLICITY MAGELLANIC CLOUDS},
journal = {\apj}
}

@ARTICLE{Houk_2004,
       author = {{Houck}, J.~R. and {Roellig}, T.~L. and {van Cleve}, J. and {Forrest}, W.~J. and {Herter}, T. and {Lawrence}, C.~R. and {Matthews}, K. and {Reitsema}, H.~J. and {Soifer}, B.~T. and {Watson}, D.~M. and {Weedman}, D. and {Huisjen}, M. and {Troeltzsch}, J. and {Barry}, D.~J. and {Bernard-Salas}, J. and {Blacken}, C.~E. and {Brandl}, B.~R. and {Charmandaris}, V. and {Devost}, D. and {Gull}, G.~E. and {Hall}, P. and {Henderson}, C.~P. and {Higdon}, S.~J.~U. and {Pirger}, B.~E. and {Schoenwald}, J. and {Sloan}, G.~C. and {Uchida}, K.~I. and {Appleton}, P.~N. and {Armus}, L. and {Burgdorf}, M.~J. and {Fajardo-Acosta}, S.~B. and {Grillmair}, C.~J. and {Ingalls}, J.~G. and {Morris}, P.~W. and {Teplitz}, H.~I.},
        title = "{The Infrared Spectrograph (IRS) on the Spitzer Space Telescope}",
      journal = {\apjs},
         year = 2004,
        month = sep,
       volume = {154},
       number = {1},
        pages = {18-24},
          doi = {10.1086/423134},
archivePrefix = {arXiv},
       eprint = {astro-ph/0406167},
 primaryClass = {astro-ph},
       adsurl = {https://ui.adsabs.harvard.edu/abs/2004ApJS..154...18H}
}

@article{Cortzen_2018,
    author = {Cortzen, I and Garrett, J and Magdis, G and Rigopoulou, D and Valentino, F and Pereira-Santaella, M and Combes, F and Alonso-Herrero, A and Toft, S and Daddi, E and Elbaz, D and Gómez-Guijarro, C and Stockmann, M and Huang, J and Kramer, C},
    title = {PAHs as tracers of the molecular gas in star-forming galaxies},
    journal = {\mnras},
    volume = {482},
    number = {2},
    pages = {1618-1633},
    year = {2018},
    month = {10},
    issn = {0035-8711},
    doi = {10.1093/mnras/sty2777},
    url = {https://doi.org/10.1093/mnras/sty2777},
    eprint = {https://academic.oup.com/mnras/article-pdf/482/2/1618/26288803/sty2777.pdf},
}

@article{Regan_2006,
doi = {10.1086/505382},
url = {https://dx.doi.org/10.1086/505382},
year = {2006},
month = {dec},
publisher = {},
volume = {652},
number = {2},
pages = {1112},
author = {Regan, Michael W. and Thornley, Michele D. and Vogel, Stuart N. and Sheth, Kartik and Draine, Bruce T. and Hollenbach, David J. and Meyer, Martin and Dale, Daniel A. and Engelbracht, Charles W. and Kennicutt, Robert C. and Armus, Lee and Buckalew, Brent and Calzetti, Daniela and Gordon, Karl D. and Helou, George and Leitherer, Claus and Malhotra, Sangeeta and Murphy, Eric and Rieke, George H. and Rieke, Marcia J. and Smith, J. D.},
title = {The Radial Distribution of the Interstellar Medium in Disk Galaxies: Evidence for Secular Evolution},
journal = {\apj}
}

@ARTICLE{Wright_2010,
       author = {{Wright}, Edward L. and {Eisenhardt}, Peter R.~M. and {Mainzer}, Amy K. and {Ressler}, Michael E. and {Cutri}, Roc M. and {Jarrett}, Thomas and {Kirkpatrick}, J. Davy and {Padgett}, Deborah and {McMillan}, Robert S. and {Skrutskie}, Michael and {Stanford}, S.~A. and {Cohen}, Martin and {Walker}, Russell G. and {Mather}, John C. and {Leisawitz}, David and {Gautier}, III, Thomas N. and {McLean}, Ian and {Benford}, Dominic and {Lonsdale}, Carol J. and {Blain}, Andrew and {Mendez}, Bryan and {Irace}, William R. and {Duval}, Valerie and {Liu}, Fengchuan and {Royer}, Don and {Heinrichsen}, Ingolf and {Howard}, Joan and {Shannon}, Mark and {Kendall}, Martha and {Walsh}, Amy L. and {Larsen}, Mark and {Cardon}, Joel G. and {Schick}, Scott and {Schwalm}, Mark and {Abid}, Mohamed and {Fabinsky}, Beth and {Naes}, Larry and {Tsai}, Chao-Wei},
        title = "{The Wide-field Infrared Survey Explorer (WISE): Mission Description and Initial On-orbit Performance}",
      journal = {\aj},
         year = 2010,
        month = dec,
       volume = {140},
       number = {6},
        pages = {1868-1881},
          doi = {10.1088/0004-6256/140/6/1868},
archivePrefix = {arXiv},
       eprint = {1008.0031},
 primaryClass = {astro-ph.IM},
       adsurl = {https://ui.adsabs.harvard.edu/abs/2010AJ....140.1868W}
}

@ARTICLE{Savage_1979,
       author = {{Savage}, B.~D. and {Mathis}, J.~S.},
        title = "{Observed properties of interstellar dust.}",
      journal = {\araa},
         year = 1979,
        month = jan,
       volume = {17},
        pages = {73-111},
          doi = {10.1146/annurev.aa.17.090179.000445},
       adsurl = {https://ui.adsabs.harvard.edu/abs/1979ARA&A..17...73S}
}

@ARTICLE{Bohlin_1978,
       author = {{Bohlin}, R.~C. and {Savage}, B.~D. and {Drake}, J.~F.},
        title = "{A survey of interstellar H I from Lalpha absorption measurements. II.}",
      journal = {\apj},
         year = 1978,
        month = aug,
       volume = {224},
        pages = {132-142},
          doi = {10.1086/156357},
       adsurl = {https://ui.adsabs.harvard.edu/abs/1978ApJ...224..132B}
}

@article{Cabral_2017,
    author = {Duarte-Cabral, Ana and Dobbs, C. L.},
    title = {The evolution of giant molecular filaments},
    journal = {\mnras},
    volume = {470},
    number = {4},
    pages = {4261-4273},
    year = {2017},
    month = {06},
    issn = {0035-8711},
    doi = {10.1093/mnras/stx1524},
    url = {https://doi.org/10.1093/mnras/stx1524},
    eprint = {https://academic.oup.com/mnras/article-pdf/470/4/4261/49203620/mnras\_470\_4\_4261.pdf},
}

@article{Sun_2020,
doi = {10.3847/2041-8213/abb3be},
url = {https://dx.doi.org/10.3847/2041-8213/abb3be},
year = {2020},
month = {sep},
publisher = {The American Astronomical Society},
volume = {901},
number = {1},
pages = {L8},
author = {Sun, Jiayi and Leroy, Adam K. and Schinnerer, Eva and Hughes, Annie and Rosolowsky, Erik and Querejeta, Miguel and Schruba, Andreas and Liu, Daizhong and Saito, Toshiki and Herrera, Cinthya N. and Faesi, Christopher and Usero, Antonio and Pety, Jérôme and Kruijssen, J. M. Diederik and Ostriker, Eve C. and Bigiel, Frank and Blanc, Guillermo A. and Bolatto, Alberto D. and Boquien, Médéric and Chevance, Mélanie and Dale, Daniel A. and Deger, Sinan and Emsellem, Eric and Glover, Simon C. O. and Grasha, Kathryn and Groves, Brent and Henshaw, Jonathan and Jimenez-Donaire, Maria J. and Kim, Jenny J. and Klessen, Ralf S. and Kreckel, Kathryn and Lee, Janice C. and Meidt, Sharon and Sandstrom, Karin and Sardone, Amy E. and Utomo, Dyas and Williams, Thomas G.},
title = {Molecular Gas Properties on Cloud Scales across the Local Star-forming Galaxy Population},
journal = {\apjl}
}

@article{Sun_2020a,
doi = {10.3847/1538-4357/ab781c},
url = {https://dx.doi.org/10.3847/1538-4357/ab781c},
year = {2020},
month = {apr},
publisher = {The American Astronomical Society},
volume = {892},
number = {2},
pages = {148},
author = {Sun, Jiayi and Leroy, Adam K. and Ostriker, Eve C. and Hughes, Annie and Rosolowsky, Erik and Schruba, Andreas and Schinnerer, Eva and Blanc, Guillermo A. and Faesi, Christopher and Kruijssen, J. M. Diederik and Meidt, Sharon and Utomo, Dyas and Bigiel, Frank and Bolatto, Alberto D. and Chevance, Mélanie and Chiang, I-Da and Dale, Daniel and Emsellem, Eric and Glover, Simon C. O. and Grasha, Kathryn and Henshaw, Jonathan and Herrera, Cinthya N. and Jimenez-Donaire, Maria Jesus and Lee, Janice C. and Pety, Jérôme and Querejeta, Miguel and Saito, Toshiki and Sandstrom, Karin and Usero, Antonio},
title = {Dynamical Equilibrium in the Molecular ISM in 28 Nearby Star-forming Galaxies},
journal = {\apj}
}

@article{Leroy_2022,
doi = {10.3847/1538-4357/ac3490},
url = {https://dx.doi.org/10.3847/1538-4357/ac3490},
year = {2022},
month = {mar},
publisher = {The American Astronomical Society},
volume = {927},
number = {2},
pages = {149},
author = {Leroy, Adam K. and Rosolowsky, Erik and Usero, Antonio and Sandstrom, Karin and Schinnerer, Eva and Schruba, Andreas and Bolatto, Alberto D. and Sun, Jiayi and Barnes, Ashley. T. and Belfiore, Francesco and Bigiel, Frank and den Brok, Jakob S. and Cao, Yixian and Chiang, I-Da and Chevance, Mélanie and Dale, Daniel A. and Eibensteiner, Cosima and Faesi, Christopher M. and Glover, Simon C. O. and Hughes, Annie and Jiménez Donaire, María J. and Klessen, Ralf S. and Koch, Eric W. and Kruijssen, J. M. Diederik and Liu, Daizhong and Meidt, Sharon E. and Pan, Hsi-An and Pety, Jérôme and Puschnig, Johannes and Querejeta, Miguel and Saito, Toshiki and Sardone, Amy and Watkins, Elizabeth J. and Weiss, Axel and Williams, Thomas G.},
title = {Low-J CO Line Ratios from Single-dish CO Mapping Surveys and PHANGS-ALMA},
journal = {\apj}
}

@ARTICLE{Demachi_2024,
       author = {{Demachi}, Fumika and {Fukui}, Yasuo and {Yamada}, Rin I. and {Tachihara}, Kengo and {Hayakawa}, Takahiro and {Tokuda}, Kazuki and {Fujita}, Shinji and {Kobayashi}, Masato I.~N. and {Muraoka}, Kazuyuki and {Konishi}, Ayu and {Tsuge}, Kisetsu and {Onishi}, Toshikazu and {Kawamura}, Akiko},
        title = "{Giant molecular clouds and their type classification in M 74: Toward understanding star formation and cloud evolution}",
      journal = {\pasj},
         year = 2024,
        month = oct,
       volume = {76},
       number = {5},
        pages = {1059-1083},
          doi = {10.1093/pasj/psae071},
archivePrefix = {arXiv},
       eprint = {2305.19192},
 primaryClass = {astro-ph.GA},
       adsurl = {https://ui.adsabs.harvard.edu/abs/2024PASJ...76.1059D}
}

@ARTICLE{Regan_2004,
       author = {{Regan}, Michael W. and {Thornley}, Michele D. and {Bendo}, George J. and {Draine}, Bruce T. and {Li}, Aigen and {Dale}, Daniel A. and {Engelbracht}, Charles W. and {Kennicutt}, Robert C., Jr. and {Armus}, Lee and {Calzetti}, Daniela and {Gordon}, Karl D. and {Helou}, George and {Hollenbach}, David J. and {Jarrett}, Thomas H. and {Kewley}, Lisa J. and {Leitherer}, Claus and {Malhotra}, Sangeeta and {Meyer}, Martin and {Misselt}, Karl A. and {Morrison}, Jane E. and {Murphy}, Eric J. and {Muzerolle}, James and {Rieke}, George H. and {Rieke}, Marcia J. and {Roussel}, H{\'e}l{\`e}ne and {Smith}, John-David T. and {Walter}, Fabian},
        title = "{Spitzer Infrared Nearby Galaxies Survey (SINGS) Imaging of NGC 7331: A Panchromatic View of a Ringed Galaxy}",
      journal = {\apjs},
         year = 2004,
        month = sep,
       volume = {154},
       number = {1},
        pages = {204-210},
          doi = {10.1086/423204},
       adsurl = {https://ui.adsabs.harvard.edu/abs/2004ApJS..154..204R}
}

@article{Gao_2019,
doi = {10.3847/1538-4357/ab557c},
url = {https://dx.doi.org/10.3847/1538-4357/ab557c},
year = {2019},
month = {dec},
publisher = {The American Astronomical Society},
volume = {887},
number = {2},
pages = {172},
author = {Yang Gao and Ting Xiao and Cheng Li and Xue-Jian Jiang and Qing-Hua Tan and Yu Gao and Christine D. Wilson and Martin Bureau and Amélie Saintonge and José R. Sánchez-Gallego and Toby Brown and Christopher J. R. Clark and Ho Seong Hwang and Isabella Lamperti and Lin Lin and Lijie Liu and Dengrong Lu and Hsi-An Pan and Jixian Sun and Thomas G. Williams},
title = {Estimating the Molecular Gas Mass of Low-redshift Galaxies from a Combination of Mid-infrared Luminosity and Optical Properties},
journal = {\apj}
}

@article{Chown_2021,
    author = {Chown, Ryan and Li, Cheng and Parker, Laura and Wilson, Christine D and Li, Niu and Gao, Yang},
    title = {A new estimator of resolved molecular gas in nearby galaxies},
    journal = {\mnras},
    volume = {500},
    number = {1},
    pages = {1261-1278},
    year = {2020},
    month = {10},
    issn = {0035-8711},
    doi = {10.1093/mnras/staa3288},
    url = {https://doi.org/10.1093/mnras/staa3288},
    eprint = {https://academic.oup.com/mnras/article-pdf/500/1/1261/34370066/staa3288.pdf},
}

@article{Leroy_2021,
   title={PHANGS–ALMA: Arcsecond CO(2–1) Imaging of Nearby Star-forming Galaxies},
   volume={257},
   ISSN={1538-4365},
   url={http://dx.doi.org/10.3847/1538-4365/ac17f3},
   DOI={10.3847/1538-4365/ac17f3},
   number={2},
   journal={\apjs},
   publisher={American Astronomical Society},
   author={Leroy, Adam K. and Schinnerer, Eva and Hughes, Annie and Rosolowsky, Erik and Pety, Jérôme and Schruba, Andreas and Usero, Antonio and Blanc, Guillermo A. and Chevance, Mélanie and Emsellem, Eric and Faesi, Christopher M. and Herrera, Cinthya N. and Liu, Daizhong and Meidt, Sharon E. and Querejeta, Miguel and Saito, Toshiki and Sandstrom, Karin M. and Sun 孙, Jiayi 嘉 懿 and Williams, Thomas G. and Anand, Gagandeep S. and Barnes, Ashley T. and Behrens, Erica A. and Belfiore, Francesco and Benincasa, Samantha M. and Bešlić, Ivana and Bigiel, Frank and Bolatto, Alberto D. and den Brok, Jakob S. and Cao, Yixian and Chandar, Rupali and Chastenet, Jérémy and Chiang 江, I-Da 宜 達 and Congiu, Enrico and Dale, Daniel A. and Deger, Sinan and Eibensteiner, Cosima and Egorov, Oleg V. and García-Rodríguez, Axel and Glover, Simon C. O. and Grasha, Kathryn and Henshaw, Jonathan D. and Ho, I-Ting and Kepley, Amanda A. and Kim, Jaeyeon and Klessen, Ralf S. and Kreckel, Kathryn and Koch, Eric W. and Kruijssen, J. M. Diederik and Larson, Kirsten L. and Lee, Janice C. and Lopez, Laura A. and Machado, Josh and Mayker, Ness and McElroy, Rebecca and Murphy, Eric J. and Ostriker, Eve C. and Pan, Hsi-An and Pessa, Ismael and Puschnig, Johannes and Razza, Alessandro and Sánchez-Blázquez, Patricia and Santoro, Francesco and Sardone, Amy and Scheuermann, Fabian and Sliwa, Kazimierz and Sormani, Mattia C. and Stuber, Sophia K. and Thilker, David A. and Turner, Jordan A. and Utomo, Dyas and Watkins, Elizabeth J. and Whitmore, Bradley},
   year={2021},
   month=nov, pages={43} }

@ARTICLE{Williams_1994,
       author = {{Williams}, Jonathan P. and {de Geus}, Eugene J. and {Blitz}, Leo},
        title = "{Determining Structure in Molecular Clouds}",
      journal = {\apj},
         year = 1994,
        month = jun,
       volume = {428},
        pages = {693},
          doi = {10.1086/174279},
       adsurl = {https://ui.adsabs.harvard.edu/abs/1994ApJ...428..693W}
}

@article{Heyer_2001,
doi = {10.1086/320218},
url = {https://dx.doi.org/10.1086/320218},
year = {2001},
month = {apr},
publisher = {},
volume = {551},
number = {2},
pages = {852},
author = {Mark H. Heyer and John M. Carpenter and Ronald L. Snell},
title = {The Equilibrium State of Molecular Regions in the Outer
Galaxy},
journal = {\apj}
}

@article{denbrok_2020,
    author = {den Brok, J S and Cantalupo, S and Mackenzie, R and Marino, R A and Pezzulli, G and Matthee, J and Johnson, S D and Krumpe, M and Urrutia, T and Kollatschny, W},
    title = "{Probing the AGN unification model at redshift z ∼ 3 with MUSE observations of giant Ly α nebulae}",
    journal = {\mnras},
    volume = {495},
    number = {2},
    pages = {1874-1887},
    year = {2020},
    month = {05},
    issn = {0035-8711},
    doi = {10.1093/mnras/staa1269},
    url = {https://doi.org/10.1093/mnras/staa1269},
    eprint = {https://academic.oup.com/mnras/article-pdf/495/2/1874/33315491/staa1269.pdf},
}

@article{Sanchez_2019,
    author = {Sánchez, S F and Barrera-Ballesteros, J K and López-Cobá, C and Brough, S and Bryant, J J and Bland-Hawthorn, J and Croom, S M and van de Sande, J and Cortese, L and Goodwin, M and Lawrence, J S and López-Sánchez, A R and Sweet, S M and Owers, M S and Richards, S N and Walcher, C J and SAMI Team},
    title = "{The SAMI galaxy survey: exploring the gas-phase mass–metallicity relation}",
    journal = {\mnras},
    volume = {484},
    number = {3},
    pages = {3042-3070},
    year = {2019},
    month = {01},
    issn = {0035-8711},
    doi = {10.1093/mnras/stz019},
    url = {https://doi.org/10.1093/mnras/stz019},
    eprint = {https://academic.oup.com/mnras/article-pdf/484/3/3042/27711745/stz019.pdf},
}

@article{Sanchez_2014,
	author = {{Sánchez}, S. F. and {Rosales-Ortega, F. F.} and {Iglesias-Páramo, J.} and {Mollá, M.} and {Barrera-Ballesteros, J.} and {Marino, R. A.} and {Pérez, E.} and {Sánchez-Blazquez, P.} and {González Delgado, R.} and {Cid Fernandes, R.} and {de Lorenzo-Cáceres, A.} and {Mendez-Abreu, J.} and {Galbany, L.} and {Falcon-Barroso, J.} and {Miralles-Caballero, D.} and {Husemann, B.} and {García-Benito, R.} and {Mast, D.} and {Walcher, C. J.} and {Gil de Paz, A.} and {García-Lorenzo, B.} and {Jungwiert, B.} and {Vílchez, J. M.} and {Jílková, Lucie} and {Lyubenova, M.} and {Cortijo-Ferrero, C.} and {Díaz, A. I.} and {Wisotzki, L.} and {Márquez, I.} and {Bland-Hawthorn, J.} and {Ellis, S.} and {van de Ven, G.} and {Jahnke, K.} and {Papaderos, P.} and {Gomes, J. M.} and {Mendoza, M. A.} and {López-Sánchez, Á. R.} and {The CALIFA collaboration}},
	title = {A characteristic oxygen abundance gradient in galaxy disks
          unveiled with CALIFA⋆⋆⋆},
	DOI= "10.1051/0004-6361/201322343",
	url= "https://doi.org/10.1051/0004-6361/201322343",
	journal = {A\&A},
	year = 2014,
	volume = 563,
	pages = "A49",
	month = "",
}

@ARTICLE{astropy2013,
       author = {{Astropy Collaboration} and {Robitaille}, Thomas P. and
         {Tollerud}, Erik J. and {Greenfield}, Perry and {Droettboom}, Michael and
         {Bray}, Erik and {Aldcroft}, Tom and {Davis}, Matt and
         {Ginsburg}, Adam and {Price-Whelan}, Adrian M. and
         {Kerzendorf}, Wolfgang E. and {Conley}, Alexander and {Crighton}, Neil and
         {Barbary}, Kyle and {Muna}, Demitri and {Ferguson}, Henry and
         {Grollier}, Fr{\'e}d{\'e}ric and {Parikh}, Madhura M. and
         {Nair}, Prasanth H. and {Unther}, Hans M. and {Deil}, Christoph and
         {Woillez}, Julien and {Conseil}, Simon and {Kramer}, Roban and
         {Turner}, James E.~H. and {Singer}, Leo and {Fox}, Ryan and
         {Weaver}, Benjamin A. and {Zabalza}, Victor and {Edwards}, Zachary I. and
         {Azalee Bostroem}, K. and {Burke}, D.~J. and {Casey}, Andrew R. and
         {Crawford}, Steven M. and {Dencheva}, Nadia and {Ely}, Justin and
         {Jenness}, Tim and {Labrie}, Kathleen and {Lim}, Pey Lian and
         {Pierfederici}, Francesco and {Pontzen}, Andrew and {Ptak}, Andy and
         {Refsdal}, Brian and {Servillat}, Mathieu and {Streicher}, Ole},
        title = "{Astropy: A community Python package for astronomy}",
      journal = {\aap},
         year = 2013,
        month = oct,
       volume = {558},
          eid = {A33},
        pages = {A33},
          doi = {10.1051/0004-6361/201322068},
archivePrefix = {arXiv},
       eprint = {1307.6212},
 primaryClass = {astro-ph.IM},
       adsurl = {https://ui.adsabs.harvard.edu/abs/2013A\&A...558A..33A}
}

@ARTICLE{astropy2018,
       author = {{Astropy Collaboration} and {Price-Whelan}, A.~M. and
         {Sip{\H{o}}cz}, B.~M. and {G{\"u}nther}, H.~M. and {Lim}, P.~L. and
         {Crawford}, S.~M. and {Conseil}, S. and {Shupe}, D.~L. and
         {Craig}, M.~W. and {Dencheva}, N. and {Ginsburg}, A. and {Vand
        erPlas}, J.~T. and {Bradley}, L.~D. and {P{\'e}rez-Su{\'a}rez}, D. and
         {de Val-Borro}, M. and {Aldcroft}, T.~L. and {Cruz}, K.~L. and
         {Robitaille}, T.~P. and {Tollerud}, E.~J. and {Ardelean}, C. and
         {Babej}, T. and {Bach}, Y.~P. and {Bachetti}, M. and {Bakanov}, A.~V. and
         {Bamford}, S.~P. and {Barentsen}, G. and {Barmby}, P. and
         {Baumbach}, A. and {Berry}, K.~L. and {Biscani}, F. and {Boquien}, M. and
         {Bostroem}, K.~A. and {Bouma}, L.~G. and {Brammer}, G.~B. and
         {Bray}, E.~M. and {Breytenbach}, H. and {Buddelmeijer}, H. and
         {Burke}, D.~J. and {Calderone}, G. and {Cano Rodr{\'\i}guez}, J.~L. and
         {Cara}, M. and {Cardoso}, J.~V.~M. and {Cheedella}, S. and {Copin}, Y. and
         {Corrales}, L. and {Crichton}, D. and {D'Avella}, D. and {Deil}, C. and
         {Depagne}, {\'E}. and {Dietrich}, J.~P. and {Donath}, A. and
         {Droettboom}, M. and {Earl}, N. and {Erben}, T. and {Fabbro}, S. and
         {Ferreira}, L.~A. and {Finethy}, T. and {Fox}, R.~T. and
         {Garrison}, L.~H. and {Gibbons}, S.~L.~J. and {Goldstein}, D.~A. and
         {Gommers}, R. and {Greco}, J.~P. and {Greenfield}, P. and
         {Groener}, A.~M. and {Grollier}, F. and {Hagen}, A. and {Hirst}, P. and
         {Homeier}, D. and {Horton}, A.~J. and {Hosseinzadeh}, G. and {Hu}, L. and
         {Hunkeler}, J.~S. and {Ivezi{\'c}}, {\v{Z}}. and {Jain}, A. and
         {Jenness}, T. and {Kanarek}, G. and {Kendrew}, S. and {Kern}, N.~S. and
         {Kerzendorf}, W.~E. and {Khvalko}, A. and {King}, J. and {Kirkby}, D. and
         {Kulkarni}, A.~M. and {Kumar}, A. and {Lee}, A. and {Lenz}, D. and
         {Littlefair}, S.~P. and {Ma}, Z. and {Macleod}, D.~M. and
         {Mastropietro}, M. and {McCully}, C. and {Montagnac}, S. and
         {Morris}, B.~M. and {Mueller}, M. and {Mumford}, S.~J. and {Muna}, D. and
         {Murphy}, N.~A. and {Nelson}, S. and {Nguyen}, G.~H. and
         {Ninan}, J.~P. and {N{\"o}the}, M. and {Ogaz}, S. and {Oh}, S. and
         {Parejko}, J.~K. and {Parley}, N. and {Pascual}, S. and {Patil}, R. and
         {Patil}, A.~A. and {Plunkett}, A.~L. and {Prochaska}, J.~X. and
         {Rastogi}, T. and {Reddy Janga}, V. and {Sabater}, J. and
         {Sakurikar}, P. and {Seifert}, M. and {Sherbert}, L.~E. and
         {Sherwood-Taylor}, H. and {Shih}, A.~Y. and {Sick}, J. and
         {Silbiger}, M.~T. and {Singanamalla}, S. and {Singer}, L.~P. and
         {Sladen}, P.~H. and {Sooley}, K.~A. and {Sornarajah}, S. and
         {Streicher}, O. and {Teuben}, P. and {Thomas}, S.~W. and
         {Tremblay}, G.~R. and {Turner}, J.~E.~H. and {Terr{\'o}n}, V. and
         {van Kerkwijk}, M.~H. and {de la Vega}, A. and {Watkins}, L.~L. and
         {Weaver}, B.~A. and {Whitmore}, J.~B. and {Woillez}, J. and
         {Zabalza}, V. and {Astropy Contributors}},
        title = "{The Astropy Project: Building an Open-science Project and Status of the v2.0 Core Package}",
      journal = {\aj},
         year = 2018,
        month = sep,
       volume = {156},
       number = {3},
          eid = {123},
        pages = {123},
          doi = {10.3847/1538-3881/aabc4f},
archivePrefix = {arXiv},
       eprint = {1801.02634},
 primaryClass = {astro-ph.IM},
       adsurl = {https://ui.adsabs.harvard.edu/abs/2018AJ....156..123A}
}

@ARTICLE{scipy2020,
       author = {{Virtanen}, Pauli and {Gommers}, Ralf and {Oliphant},
         Travis E. and {Haberland}, Matt and {Reddy}, Tyler and
         {Cournapeau}, David and {Burovski}, Evgeni and {Peterson}, Pearu
         and {Weckesser}, Warren and {Bright}, Jonathan and {van der Walt},
         St{\'e}fan J.  and {Brett}, Matthew and {Wilson}, Joshua and
         {Jarrod Millman}, K.  and {Mayorov}, Nikolay and {Nelson}, Andrew
         R.~J. and {Jones}, Eric and {Kern}, Robert and {Larson}, Eric and
         {Carey}, CJ and {Polat}, {\.I}lhan and {Feng}, Yu and {Moore},
         Eric W. and {Vand erPlas}, Jake and {Laxalde}, Denis and
         {Perktold}, Josef and {Cimrman}, Robert and {Henriksen}, Ian and
         {Quintero}, E.~A. and {Harris}, Charles R and {Archibald}, Anne M.
         and {Ribeiro}, Ant{\^o}nio H. and {Pedregosa}, Fabian and
         {van Mulbregt}, Paul and {Contributors}, SciPy 1. 0},
        title = "{SciPy 1.0: Fundamental Algorithms for Scientific
                  Computing in Python}",
      journal = {\natm},
      year = "2020",
      volume={17},
      pages={261--272},
      adsurl = {https://rdcu.be/b08Wh},
      doi = {https://doi.org/10.1038/s41592-019-0686-2},
}

@article{Mazumdar_2021,
   title={High resolution LAsMA 12CO and 13CO observation of the G305 giant molecular cloud complex: II. Effect of feedback on clump properties},
   volume={656},
   ISSN={1432-0746},
   url={http://dx.doi.org/10.1051/0004-6361/202142036},
   DOI={10.1051/0004-6361/202142036},
   journal={\aap},
   publisher={EDP Sciences},
   author={Mazumdar, P. and Wyrowski, F. and Urquhart, J. S. and Colombo, D. and Menten, K. M. and Neupane, S. and Thompson, M. A.},
   year={2021},
   month=dec, pages={A101} }

@ARTICLE{Lin_1964,
       author = {{Lin}, C.~C. and {Shu}, Frank H.},
        title = "{On the Spiral Structure of Disk Galaxies.}",
      journal = {\apj},
         year = 1964,
        month = aug,
       volume = {140},
        pages = {646},
          doi = {10.1086/147955},
       adsurl = {https://ui.adsabs.harvard.edu/abs/1964ApJ...140..646L}
}

@article{Yu_2022,
   title={The EDGE-CALIFA survey: The role of spiral arms and bars in driving central molecular gas concentrations},
   volume={666},
   ISSN={1432-0746},
   url={http://dx.doi.org/10.1051/0004-6361/202244306},
   DOI={10.1051/0004-6361/202244306},
   journal={\aap},
   publisher={EDP Sciences},
   author={Yu, Si-Yue and Kalinova, Veselina and Colombo, Dario and Bolatto, Alberto D. and Wong, Tony and Levy, Rebecca C. and Villanueva, Vicente and Sánchez, Sebastián F. and Ho, Luis C. and Vogel, Stuart N. and Teuben, Peter and Rubio, Mónica},
   year={2022},
   month=oct, pages={A175} }

@ARTICLE{Roberts_1979,
       author = {{Roberts}, Jr., W.~W. and {Huntley}, J.~M. and {van Albada}, G.~D.},
        title = "{Gas dynamics in barred spirals: gaseous density waves and galactic shocks.}",
      journal = {\apj},
         year = 1979,
        month = oct,
       volume = {233},
        pages = {67-84},
          doi = {10.1086/157367},
       adsurl = {https://ui.adsabs.harvard.edu/abs/1979ApJ...233...67R}
}

@ARTICLE{matplotlib2007,
       author = {{Hunter}, John D.},
        title = "{Matplotlib: A 2D Graphics Environment}",
      journal = {Computing in Science and Engineering},
         year = 2007,
        month = may,
       volume = {9},
       number = {3},
        pages = {90-95},
          doi = {10.1109/MCSE.2007.55},
       adsurl = {https://ui.adsabs.harvard.edu/abs/2007CSE.....9...90H}
}

@ARTICLE{Asplund_2009,
       author = {{Asplund}, Martin and {Grevesse}, Nicolas and {Sauval}, A. Jacques and {Scott}, Pat},
        title = "{The Chemical Composition of the Sun}",
      journal = {\araa},
         year = 2009,
        month = sep,
       volume = {47},
       number = {1},
        pages = {481-522},
          doi = {10.1146/annurev.astro.46.060407.145222},
archivePrefix = {arXiv},
       eprint = {0909.0948},
 primaryClass = {astro-ph.SR},
       adsurl = {https://ui.adsabs.harvard.edu/abs/2009ARA\&A..47..481A}
}

@article{Teng_2024,
doi = {10.3847/1538-4357/ad10ae},
url = {https://dx.doi.org/10.3847/1538-4357/ad10ae},
year = {2024},
month = {jan},
publisher = {The American Astronomical Society},
volume = {961},
number = {1},
pages = {42},
author = {Teng, Yu-Hsuan and Chiang, I-Da and Sandstrom, Karin M. and Sun, Jiayi and Leroy, Adam K. and Bolatto, Alberto D. and Usero, Antonio and Ostriker, Eve C. and Querejeta, Miguel and Chastenet, Jérémy and Bigiel, Frank and Boquien, Médéric and den Brok, Jakob and Cao, Yixian and Chevance, Mélanie and Chown, Ryan and Colombo, Dario and Eibensteiner, Cosima and Glover, Simon C. O. and Grasha, Kathryn and Henshaw, Jonathan D. and Jiménez-Donaire, María J. and Liu, Daizhong and Murphy, Eric J. and Pan, Hsi-An and Stuber, Sophia K. and Williams, Thomas G.},
title = {Star Formation Efficiency in Nearby Galaxies Revealed with a New CO-to-H2 Conversion Factor Prescription},
journal = {\apj}
}

@ARTICLE{Hughes_2013,
       author = {{Hughes}, Annie and {Meidt}, Sharon E. and {Colombo}, Dario and {Schinnerer}, Eva and {Pety}, Jer{\^o}me and {Leroy}, Adam K. and {Dobbs}, Clare L. and {Garc{\'\i}a-Burillo}, Santiago and {Thompson}, Todd A. and {Dumas}, Ga{\"e}lle and {Schuster}, Karl F. and {Kramer}, Carsten},
        title = "{A Comparative Study of Giant Molecular Clouds in M51, M33, and the Large Magellanic Cloud}",
      journal = {\apj},
         year = 2013,
        month = dec,
       volume = {779},
       number = {1},
          eid = {46},
        pages = {46},
          doi = {10.1088/0004-637X/779/1/46},
archivePrefix = {arXiv},
       eprint = {1309.3453},
 primaryClass = {astro-ph.GA},
       adsurl = {https://ui.adsabs.harvard.edu/abs/2013ApJ...779...46H}
}

@ARTICLE{Leroy_2016,
       author = {{Leroy}, Adam K. and {Hughes}, Annie and {Schruba}, Andreas and {Rosolowsky}, Erik and {Blanc}, Guillermo A. and {Bolatto}, Alberto D. and {Colombo}, Dario and {Escala}, Andres and {Kramer}, Carsten and {Kruijssen}, J.~M. Diederik and {Meidt}, Sharon and {Pety}, Jerome and {Querejeta}, Miguel and {Sandstrom}, Karin and {Schinnerer}, Eva and {Sliwa}, Kazimierz and {Usero}, Antonio},
        title = "{A Portrait of Cold Gas in Galaxies at 60 pc Resolution and a Simple Method to Test Hypotheses That Link Small-scale ISM Structure to Galaxy-scale Processes}",
      journal = {\apj},
         year = 2016,
        month = nov,
       volume = {831},
       number = {1},
          eid = {16},
        pages = {16},
          doi = {10.3847/0004-637X/831/1/16},
archivePrefix = {arXiv},
       eprint = {1606.07077},
 primaryClass = {astro-ph.GA},
       adsurl = {https://ui.adsabs.harvard.edu/abs/2016ApJ...831...16L}
}

@ARTICLE{Colombo_2019,
       author = {{Colombo}, D. and {Rosolowsky}, E. and {Duarte-Cabral}, A. and {Ginsburg}, A. and {Glenn}, J. and {Zetterlund}, E. and {Hernandez}, A.~K. and {Dempsey}, J. and {Currie}, M.~J.},
        title = "{The integrated properties of the molecular clouds from the JCMT CO(3-2) High-Resolution Survey}",
      journal = {\mnras},
         year = 2019,
        month = mar,
       volume = {483},
       number = {4},
        pages = {4291-4340},
          doi = {10.1093/mnras/sty3283},
archivePrefix = {arXiv},
       eprint = {1812.04688},
 primaryClass = {astro-ph.GA},
       adsurl = {https://ui.adsabs.harvard.edu/abs/2019MNRAS.483.4291C}
}

@ARTICLE{Sanders_1985,
       author = {{Sanders}, D.~B. and {Mirabel}, I.~F.},
        title = "{CO detections and IRAS observations of bright radio spiral galaxies at cz<9000 kilometers per second.}",
      journal = {\apjl},
         year = 1985,
        month = nov,
       volume = {298},
        pages = {L31-L35},
          doi = {10.1086/184561},
       adsurl = {https://ui.adsabs.harvard.edu/abs/1985ApJ...298L..31S}
}

@ARTICLE{Stutzki_1990,
       author = {{Stutzki}, J. and {Guesten}, R.},
        title = "{High Spatial Resolution Isotopic CO and CS Observations of M17 SW: The Clumpy Structure of the Molecular Cloud Core}",
      journal = {\apj},
         year = 1990,
        month = jun,
       volume = {356},
        pages = {513},
          doi = {10.1086/168859},
       adsurl = {https://ui.adsabs.harvard.edu/abs/1990ApJ...356..513S}
}

@ARTICLE{Dame_2001,
       author = {{Dame}, T.~M. and {Hartmann}, Dap and {Thaddeus}, P.},
        title = "{The Milky Way in Molecular Clouds: A New Complete CO Survey}",
      journal = {\apj},
         year = 2001,
        month = feb,
       volume = {547},
       number = {2},
        pages = {792-813},
          doi = {10.1086/318388},
archivePrefix = {arXiv},
       eprint = {astro-ph/0009217},
 primaryClass = {astro-ph},
       adsurl = {https://ui.adsabs.harvard.edu/abs/2001ApJ...547..792D}
}

@INPROCEEDINGS{Blitz_2007,
       author = {{Blitz}, L. and {Fukui}, Y. and {Kawamura}, A. and {Leroy}, A. and {Mizuno}, N. and {Rosolowsky}, E.},
        title = "{Giant Molecular Clouds in Local Group Galaxies}",
    booktitle = {Protostars and Planets V},
         year = 2007,
       editor = {{Reipurth}, Bo and {Jewitt}, David and {Keil}, Klaus},
        month = jan,
        pages = {81},
          doi = {10.48550/arXiv.astro-ph/0602600},
archivePrefix = {arXiv},
       eprint = {astro-ph/0602600},
 primaryClass = {astro-ph},
       adsurl = {https://ui.adsabs.harvard.edu/abs/2007prpl.conf...81B}
}

@article{Mok_2020,
doi = {10.3847/1538-4357/ab7a14},
url = {https://dx.doi.org/10.3847/1538-4357/ab7a14},
year = {2020},
month = {apr},
publisher = {The American Astronomical Society},
volume = {893},
number = {2},
pages = {135},
author = {Mok, Angus and Chandar, Rupali and Fall, S. Michael},
title = {Mass Functions of Giant Molecular Clouds and Young Star Clusters in Six Nearby Galaxies},
journal = {\apj}
}

@article{Sun_2018,
doi = {10.3847/1538-4357/aac326},
url = {https://dx.doi.org/10.3847/1538-4357/aac326},
year = {2018},
month = {jun},
publisher = {The American Astronomical Society},
volume = {860},
number = {2},
pages = {172},
author = {Sun, Jiayi and Leroy, Adam K. and Schruba, Andreas and Rosolowsky, Erik and Hughes, Annie and Kruijssen, J. M. Diederik and Meidt, Sharon and Schinnerer, Eva and Blanc, Guillermo A. and Bigiel, Frank and Bolatto, Alberto D. and Chevance, Mélanie and Groves, Brent and Herrera, Cinthya N. and Hygate, Alexander P. S. and Pety, Jérôme and Querejeta, Miguel and Usero, Antonio and Utomo, Dyas},
title = {Cloud-scale Molecular Gas Properties in 15 Nearby Galaxies},
journal = {\apj}
}

@ARTICLE{Sun_2022,
       author = {Sun, Jiayi and {Leroy}, Adam K. and {Rosolowsky}, Erik and {Hughes}, Annie and {Schinnerer}, Eva and {Schruba}, Andreas and {Koch}, Eric W. and {Blanc}, Guillermo A. and {Chiang}, I-Da and {Groves}, Brent and {Liu}, Daizhong and {Meidt}, Sharon and {Pan}, Hsi-An and {Pety}, J{\'e}r{\^o}me and {Querejeta}, Miguel and {Saito}, Toshiki and {Sandstrom}, Karin and {Sardone}, Amy and {Usero}, Antonio and {Utomo}, Dyas and {Williams}, Thomas G. and {Barnes}, Ashley T. and {Benincasa}, Samantha M. and {Bigiel}, Frank and {Bolatto}, Alberto D. and {Boquien}, M{\'e}d{\'e}ric and {Chevance}, M{\'e}lanie and {Dale}, Daniel A. and {Deger}, Sinan and {Emsellem}, Eric and {Glover}, Simon C.~O. and {Grasha}, Kathryn and {Henshaw}, Jonathan D. and {Klessen}, Ralf S. and {Kreckel}, Kathryn and {Kruijssen}, J.~M. Diederik and {Ostriker}, Eve C. and {Thilker}, David A.},
        title = "{Molecular Cloud Populations in the Context of Their Host Galaxy Environments: A Multiwavelength Perspective}",
      journal = {\aj},
         year = 2022,
        month = aug,
       volume = {164},
       number = {2},
          eid = {43},
        pages = {43},
          doi = {10.3847/1538-3881/ac74bd},
archivePrefix = {arXiv},
       eprint = {2206.07055},
 primaryClass = {astro-ph.GA},
       adsurl = {https://ui.adsabs.harvard.edu/abs/2022AJ....164...43S}
}

@ARTICLE{Houlan_1992,
       author = {{Houlahan}, Padraig and {Scalo}, John},
        title = "{Recognition and Characterization of Hierarchical Interstellar Structure. II. Structure Tree Statistics}",
      journal = {\apj},
         year = 1992,
        month = jul,
       volume = {393},
        pages = {172},
          doi = {10.1086/171495},
       adsurl = {https://ui.adsabs.harvard.edu/abs/1992ApJ...393..172H}
}

@ARTICLE{Draine_2007,
       author = {{Draine}, B.~T. and {Li}, Aigen},
        title = "{Infrared Emission from Interstellar Dust. IV. The Silicate-Graphite-PAH Model in the Post-Spitzer Era}",
      journal = {\apj},
         year = 2007,
        month = mar,
       volume = {657},
       number = {2},
        pages = {810-837},
          doi = {10.1086/511055},
archivePrefix = {arXiv},
       eprint = {astro-ph/0608003},
 primaryClass = {astro-ph},
       adsurl = {https://ui.adsabs.harvard.edu/abs/2007ApJ...657..810D}
}

@article{Querejeta_2021,
   title={Stellar structures, molecular gas, and star formation across the PHANGS sample of nearby galaxies},
   volume={656},
   ISSN={1432-0746},
   url={http://dx.doi.org/10.1051/0004-6361/202140695},
   DOI={10.1051/0004-6361/202140695},
   journal={\aap},
   publisher={EDP Sciences},
   author={Querejeta, M. and Schinnerer, E. and Meidt, S. and Sun, J. and Leroy, A. K. and Emsellem, E. and Klessen, R. S. and Muñoz-Mateos, J. C. and Salo, H. and Laurikainen, E. and Bešlić, I. and Blanc, G. A. and Chevance, M. and Dale, D. A. and Eibensteiner, C. and Faesi, C. and García-Rodríguez, A. and Glover, S. C. O. and Grasha, K. and Henshaw, J. and Herrera, C. and Hughes, A. and Kreckel, K. and Kruijssen, J. M. D. and Liu, D. and Murphy, E. J. and Pan, H.-A. and Pety, J. and Razza, A. and Rosolowsky, E. and Saito, T. and Schruba, A. and Usero, A. and Watkins, E. J. and Williams, T. G.},
   year={2021},
   month=dec, pages={A133} }

@ARTICLE{Sheth_2010,
       author = {{Sheth}, Kartik and {Regan}, Michael and {Hinz}, Joannah L. and {Gil de Paz}, Armando and {Men{\'e}ndez-Delmestre}, Kar{\'\i}n and {Mu{\~n}oz-Mateos}, Juan-Carlos and {Seibert}, Mark and {Kim}, Taehyun and {Laurikainen}, Eija and {Salo}, Heikki and {Gadotti}, Dimitri A. and {Laine}, Jarkko and {Mizusawa}, Trisha and {Armus}, Lee and {Athanassoula}, E. and {Bosma}, Albert and {Buta}, Ronald J. and {Capak}, Peter and {Jarrett}, Thomas H. and {Elmegreen}, Debra M. and {Elmegreen}, Bruce G. and {Knapen}, Johan H. and {Koda}, Jin and {Helou}, George and {Ho}, Luis C. and {Madore}, Barry F. and {Masters}, Karen L. and {Mobasher}, Bahram and {Ogle}, Patrick and {Peng}, Chien Y. and {Schinnerer}, Eva and {Surace}, Jason A. and {Zaritsky}, Dennis and {Comer{\'o}n}, S{\'e}bastien and {de Swardt}, Bonita and {Meidt}, Sharon E. and {Kasliwal}, Mansi and {Aravena}, Manuel},
        title = "{The Spitzer Survey of Stellar Structure in Galaxies (S4G)}",
      journal = {\pasp},
         year = 2010,
        month = dec,
       volume = {122},
       number = {898},
        pages = {1397-1414},
          doi = {10.1086/657638},
archivePrefix = {arXiv},
       eprint = {1010.1592},
 primaryClass = {astro-ph.CO},
       adsurl = {https://ui.adsabs.harvard.edu/abs/2010PASP..122.1397S}
}

@article{Spoon_2007,
doi = {10.1086/511268},
url = {https://dx.doi.org/10.1086/511268},
year = {2006},
month = {dec},
publisher = {},
volume = {654},
number = {1},
pages = {L49},
author = {H. W. W. Spoon and J. A. Marshall and J. R. Houck and M. Elitzur and L. Hao and L. Armus and B. R. Brandl and V. Charmandaris},
title = {Mid-Infrared Galaxy Classification Based on Silicate Obscuration and PAH Equivalent Width},
journal = {\apj}
}

@article{Smith_2007,
doi = {10.1086/510549},
url = {https://dx.doi.org/10.1086/510549},
year = {2007},
month = {feb},
publisher = {},
volume = {656},
number = {2},
pages = {770},
author = {J. D. T. Smith and B. T. Draine and D. A. Dale and J. Moustakas and R. C. Kennicutt, Jr. and G. Helou and L. Armus and H. Roussel and K. Sheth and G. J. Bendo and B. A. Buckalew and D. Calzetti and C. W. Engelbracht and K. D. Gordon and D. J. Hollenbach and A. Li and S. Malhotra and E. J. Murphy and F. Walter},
title = {The Mid-Infrared Spectrum of Star-forming Galaxies: Global Properties of Polycyclic Aromatic Hydrocarbon Emission},
journal = {\apj}
}

@article{Lee_2023,
doi = {10.3847/2041-8213/acaaae},
url = {https://dx.doi.org/10.3847/2041-8213/acaaae},
year = {2023},
month = {feb},
publisher = {The American Astronomical Society},
volume = {944},
number = {2},
pages = {L17},
author = {Janice C. Lee and Karin M. Sandstrom and Adam K. Leroy and David A. Thilker and Eva Schinnerer and Erik Rosolowsky and Kirsten L. Larson and Oleg V. Egorov and Thomas G. Williams and Judy Schmidt and Eric Emsellem and Gagandeep S. Anand and Ashley T. Barnes and Francesco Belfiore and Ivana Bešlić and Frank Bigiel and Guillermo A. Blanc and Alberto D. Bolatto and Médéric Boquien and Jakob den Brok and Yixian Cao and Rupali Chandar and Jérémy Chastenet and Mélanie Chevance and I-Da Chiang and Enrico Congiu and Daniel A. Dale and Sinan Deger and Cosima Eibensteiner and Christopher M. Faesi and Simon C. O. Glover and Kathryn Grasha and Brent Groves and Hamid Hassani and Kiana F. Henny and Jonathan D. Henshaw and Nils Hoyer and Annie Hughes and Sarah Jeffreson and María J. Jiménez-Donaire and Jaeyeon Kim and Hwihyun Kim and Ralf S. Klessen and Eric W. Koch and Kathryn Kreckel and J. M. Diederik Kruijssen and Jing Li and Daizhong Liu and Laura A. Lopez and Daniel Maschmann and Ness Mayker Chen and Sharon E. Meidt and Eric J. Murphy and Justus Neumann and Nadine Neumayer and Hsi-An Pan and Ismael Pessa and Jérôme Pety and Miguel Querejeta and Francesca Pinna and M. Jimena Rodríguez and Toshiki Saito and Patricia Sánchez-Blázquez and Francesco Santoro and Amy Sardone and Rowan J. Smith and Mattia C. Sormani and Fabian Scheuermann and Sophia K. Stuber and Jessica Sutter and Jiayi Sun and Yu-Hsuan Teng and Robin G. Treß and Antonio Usero and Elizabeth J. Watkins and Bradley C. Whitmore and Alessandro Razza},
title = {The PHANGS–JWST Treasury Survey: Star Formation, Feedback, and Dust Physics at High Angular Resolution in Nearby GalaxieS},
journal = {\apjl}
}

@article{Meidt_2023,
doi = {10.3847/2041-8213/acaaa8},
url = {https://dx.doi.org/10.3847/2041-8213/acaaa8},
year = {2023},
month = {feb},
publisher = {The American Astronomical Society},
volume = {944},
number = {2},
pages = {L18},
author = {Meidt, Sharon E. and Rosolowsky, Erik and Sun, Jiayi and Koch, Eric W. and Klessen, Ralf S. and Leroy, Adam K. and Schinnerer, Eva and Barnes, Ashley. T. and Glover, Simon C. O. and Lee, Janice C. and van der Wel, Arjen and Watkins, Elizabeth J. and Williams, Thomas G. and Bigiel, F. and Boquien, Médéric and Blanc, Guillermo A. and Cao, Yixian and Chevance, Mélanie and Dale, Daniel A. and Egorov, Oleg V. and Emsellem, Eric and Grasha, Kathryn and Henshaw, Jonathan D. and Kruijssen, J. M. Diederik and Larson, Kirsten L. and Liu, Daizhong and Murphy, Eric J. and Pety, Jérôme and Querejeta, Miguel and Saito, Toshiki and Sandstrom, Karin M. and Smith, Rowan J. and Sormani, Mattia C. and Thilker, David A.},
title = {PHANGS–JWST First Results: Interstellar Medium Structure on the Turbulent Jeans Scale in Four Disk Galaxies Observed by JWST and the Atacama Large Millimeter/submillimeter Array},
journal = {\apjl}
}

@ARTICLE{Fukui_2001,
       author = {{Fukui}, Yasuo and {Mizuno}, Norikazu and {Yamaguchi}, Reiko and {Mizuno}, Akira and {Onishi}, Toshikazu},
        title = "{On the Mass Spectrum of Giant Molecular Clouds in the Large Magellanic Cloud}",
      journal = {\pasj},
         year = 2001,
        month = dec,
       volume = {53},
       number = {6},
        pages = {L41-L44},
          doi = {10.1093/pasj/53.6.L41},
       adsurl = {https://ui.adsabs.harvard.edu/abs/2001PASJ...53L..41F}
}

@ARTICLE{Fukui_2010,
       author = {{Fukui}, Yasuo and {Kawamura}, Akiko},
        title = "{Molecular Clouds in Nearby Galaxies}",
      journal = {\araa},
         year = 2010,
        month = sep,
       volume = {48},
        pages = {547-580},
          doi = {10.1146/annurev-astro-081309-130854},
       adsurl = {https://ui.adsabs.harvard.edu/abs/2010ARA&A..48..547F}
}

@article{Thilker_2023,
doi = {10.3847/2041-8213/acaeac},
url = {https://dx.doi.org/10.3847/2041-8213/acaeac},
year = {2023},
month = {feb},
publisher = {The American Astronomical Society},
volume = {944},
number = {2},
pages = {L13},
author = {Thilker, David A. and Lee, Janice C. and Deger, Sinan and Barnes, Ashley T. and Bigiel, Frank and Boquien, Médéric and Cao, Yixian and Chevance, Mélanie and Dale, Daniel A. and Egorov, Oleg V. and Glover, Simon C. O. and Grasha, Kathryn and Henshaw, Jonathan D. and Klessen, Ralf S. and Koch, Eric and Kruijssen, J. M. Diederik and Leroy, Adam K. and Lessing, Ryan A. and Meidt, Sharon E. and Pinna, Francesca and Querejeta, Miguel and Rosolowsky, Erik and Sandstrom, Karin M. and Schinnerer, Eva and Smith, Rowan J. and Watkins, Elizabeth J. and Williams, Thomas G. and Anand, Gagandeep S. and Belfiore, Francesco and Blanc, Guillermo A. and Chandar, Rupali and Congiu, Enrico and Emsellem, Eric and Groves, Brent and Kreckel, Kathryn and Larson, Kirsten L. and Liu, Daizhong and Pessa, Ismael and Whitmore, Bradley C.},
title = {PHANGS–JWST First Results: The Dust Filament Network of NGC 628 and Its Relation to Star Formation Activity},
journal = {\apjl}
}

@article{Bolatto_2008,
doi = {10.1086/591513},
url = {https://dx.doi.org/10.1086/591513},
year = {2008},
month = {oct},
publisher = {},
volume = {686},
number = {2},
pages = {948},
author = {Bolatto, Alberto D. and Leroy, Adam K. and Rosolowsky, Erik and Walter, Fabian and Blitz, Leo},
title = {The Resolved Properties of Extragalactic Giant Molecular Clouds},
journal = {\apj}
}

@ARTICLE{Schruba_2011,
       author = {{Schruba}, Andreas and {Leroy}, Adam K. and {Walter}, Fabian and {Bigiel}, Frank and {Brinks}, Elias and {de Blok}, W.~J.~G. and {Dumas}, Gaelle and {Kramer}, Carsten and {Rosolowsky}, Erik and {Sandstrom}, Karin and {Schuster}, Karl and {Usero}, Antonio and {Weiss}, Axel and {Wiesemeyer}, Helmut},
        title = "{A Molecular Star Formation Law in the Atomic-gas-dominated Regime in Nearby Galaxies}",
      journal = {\aj},
         year = 2011,
        month = aug,
       volume = {142},
       number = {2},
          eid = {37},
        pages = {37},
          doi = {10.1088/0004-6256/142/2/37},
archivePrefix = {arXiv},
       eprint = {1105.4605},
 primaryClass = {astro-ph.CO},
       adsurl = {https://ui.adsabs.harvard.edu/abs/2011AJ....142...37S}
}

@ARTICLE{Leroy_2023a,
       author = {Adam K. Leroy and {Bolatto}, Alberto D. and {Sandstrom}, Karin and {Rosolowsky}, Erik and {Barnes}, Ashley. T. and {Bigiel}, F. and {Boquien}, M{\'e}d{\'e}ric and {den Brok}, Jakob S. and {Cao}, Yixian and {Chastenet}, J{\'e}r{\'e}my and {Chevance}, M{\'e}lanie and {Chiang}, I-Da and {Chown}, Ryan and {Colombo}, Dario and {Ellison}, Sara L. and {Emsellem}, Eric and {Grasha}, Kathryn and {Henshaw}, Jonathan D. and {Hughes}, Annie and {Klessen}, Ralf S. and {Koch}, Eric W. and {Kim}, Jaeyeon and {Kreckel}, Kathryn and {Kruijssen}, J.~M. Diederik and {Larson}, Kirsten L. and {Lee}, Janice C. and {Levy}, Rebecca C. and {Lin}, Lihwai and {Liu}, Daizhong and {Meidt}, Sharon E. and {Pety}, J{\'e}r{\^o}me and {Querejeta}, Miguel and {Rubio}, M{\'o}nica and {Saito}, Toshiki and {Salim}, Samir and {Schinnerer}, Eva and {Sormani}, Mattia C. and {Sun}, Jiayi and {Thilker}, David A. and {Usero}, Antonio and {Vogel}, Stuart N. and {Watkins}, Elizabeth J. and {Whitcomb}, Cory M. and {Williams}, Thomas G. and {Wilson}, Christine D.},
        title = "{PHANGS-JWST First Results: A Global and Moderately Resolved View of Mid-infrared and CO Line Emission from Galaxies at the Start of the JWST Era}",
      journal = {\apjl},
         year = 2023,
        month = feb,
       volume = {944},
       number = {2},
          eid = {L10},
        pages = {L10},
          doi = {10.3847/2041-8213/acab01},
archivePrefix = {arXiv},
       eprint = {2212.09774},
 primaryClass = {astro-ph.GA},
       adsurl = {https://ui.adsabs.harvard.edu/abs/2023ApJ...944L..10L}
}

@article{Vieira_2023,
    author = {Faustino Vieira, Helena and Duarte-Cabral, Ana and Davis, Timothy A and Peretto, Nicolas and Smith, Matthew W L and Querejeta, Miguel and Colombo, Dario and Anderson, Michael},
    title = "{A high-resolution extinction mapping technique for face-on disc galaxies}",
    journal = {\mnras},
    volume = {524},
    number = {1},
    pages = {161-175},
    year = {2023},
    month = {06},
    issn = {0035-8711},
    doi = {10.1093/mnras/stad1876},
    url = {https://doi.org/10.1093/mnras/stad1876},
    eprint = {https://academic.oup.com/mnras/article-pdf/524/1/161/56132502/stad1876.pdf},
}

@ARTICLE{Vieira_2024,
       author = {Faustino Vieira, Helena and Duarte-Cabral, Ana and Davis, Timothy A. and Peretto, Nicolas and Smith, Matthew W.~L. and Querejeta, Miguel and Colombo, Dario and Anderson, Michael},
        title = "{Molecular clouds in M51 from high-resolution extinction mapping}",
      journal = {\mnras},
         year = 2024,
        month = jan,
       volume = {527},
       number = {2},
        pages = {3639-3658},
          doi = {10.1093/mnras/stad3327},
archivePrefix = {arXiv},
       eprint = {2310.18210},
 primaryClass = {astro-ph.GA},
       adsurl = {https://ui.adsabs.harvard.edu/abs/2024MNRAS.527.3639F}
}

@article{Leroy_2013,
doi = {10.1088/0004-6256/146/2/19},
url = {https://dx.doi.org/10.1088/0004-6256/146/2/19},
year = {2013},
month = {jun},
publisher = {The American Astronomical Society},
volume = {146},
number = {2},
pages = {19},
author = {Adam K. Leroy and Fabian Walter and Karin Sandstrom and Andreas Schruba and Juan-Carlos Munoz-Mateos and Frank Bigiel and Alberto Bolatto and Elias Brinks and W. J. G. de Blok and Sharon Meidt and Hans-Walter Rix and Erik Rosolowsky and Eva Schinnerer and Karl-Friedrich Schuster and Antonio Usero},
title = {MOLECULAR GAS AND STAR FORMATION IN NEARBY DISK GALAXIES},
journal = {\aj}
}

@article{Bigiel_2008,
doi = {10.1088/0004-6256/136/6/2846},
url = {https://dx.doi.org/10.1088/0004-6256/136/6/2846},
year = {2008},
month = {nov},
publisher = {The American Astronomical Society},
volume = {136},
number = {6},
pages = {2846},
author = {F. Bigiel and A. Leroy and F. Walter and E. Brinks and W. J. G. de Blok and B. Madore and M. D. Thornley},
title = {THE STAR FORMATION LAW IN NEARBY GALAXIES ON SUB-KPC SCALES},
journal = {\aj}
}

@ARTICLE{Rosolowsky_2006,
       author = {{Rosolowsky}, Erik and {Leroy}, Adam},
        title = "{Bias-free Measurement of Giant Molecular Cloud Properties}",
      journal = {\pasp},
         year = 2006,
        month = apr,
       volume = {118},
       number = {842},
        pages = {590-610},
          doi = {10.1086/502982},
archivePrefix = {arXiv},
       eprint = {astro-ph/0601706},
 primaryClass = {astro-ph},
       adsurl = {https://ui.adsabs.harvard.edu/abs/2006PASP..118..590R}
}

@article{Rosolowsky_2021,
    author = {Rosolowsky, Erik and Hughes, Annie and Leroy, Adam K and Sun, Jiayi and Querejeta, Miguel and Schruba, Andreas and Usero, Antonio and Herrera, Cinthya N and Liu, Daizhong and Pety, Jérôme and Saito, Toshiki and Bešlić, Ivana and Bigiel, Frank and Blanc, Guillermo and Chevance, Mélanie and Dale, Daniel A and Deger, Sinan and Faesi, Christopher M and Glover, Simon C O and Henshaw, Jonathan D and Klessen, Ralf S and Kruijssen, J M Diederik and Larson, Kirsten and Lee, Janice and Meidt, Sharon and Mok, Angus and Schinnerer, Eva and Thilker, David A and Williams, Thomas G},
    title = "{Giant molecular cloud catalogues for PHANGS-ALMA: methods and initial results}",
    journal = {\mnras},
    volume = {502},
    number = {1},
    pages = {1218-1245},
    year = {2021},
    month = {01},
    issn = {0035-8711},
    doi = {10.1093/mnras/stab085},
    url = {https://doi.org/10.1093/mnras/stab085},
    eprint = {https://academic.oup.com/mnras/article-pdf/502/1/1218/36179849/stab085.pdf},
}

@ARTICLE{Rosolowsky_2005,
       author = {{Rosolowsky}, E.},
        title = "{The Mass Spectra of Giant Molecular Clouds in the Local Group}",
      journal = {\pasp},
         year = 2005,
        month = dec,
       volume = {117},
       number = {838},
        pages = {1403-1410},
          doi = {10.1086/497582},
archivePrefix = {arXiv},
       eprint = {astro-ph/0508679},
 primaryClass = {astro-ph},
       adsurl = {https://ui.adsabs.harvard.edu/abs/2005PASP..117.1403R}
}

@article{Dobbs_2013,
    author = {Dobbs, C. L. and Pringle, J. E.},
    title = {The exciting lives of giant molecular clouds},
    journal = {\mnras},
    volume = {432},
    number = {1},
    pages = {653-667},
    year = {2013},
    month = {04},
    issn = {0035-8711},
    doi = {10.1093/mnras/stt508},
    url = {https://doi.org/10.1093/mnras/stt508},
    eprint = {https://academic.oup.com/mnras/article-pdf/432/1/653/3936468/stt508.pdf},
}

@article{Heyer_2009,
doi = {10.1088/0004-637X/699/2/1092},
url = {https://dx.doi.org/10.1088/0004-637X/699/2/1092},
year = {2009},
month = {jun},
publisher = {The American Astronomical Society},
volume = {699},
number = {2},
pages = {1092},
author = {Heyer, Mark and Krawczyk, Coleman and Duval, Julia and Jackson, James M.},
title = {RE-EXAMINING LARSON'S SCALING RELATIONSHIPS IN GALACTIC MOLECULAR CLOUDS},
journal = {\apj}
}

@article{Colombo_2014,
doi = {10.1088/0004-637X/784/1/3},
url = {https://dx.doi.org/10.1088/0004-637X/784/1/3},
year = {2014},
month = {feb},
publisher = {The American Astronomical Society},
volume = {784},
number = {1},
pages = {3},
author = {Colombo, Dario and Hughes, Annie and Schinnerer, Eva and Meidt, Sharon E. and Leroy, Adam K. and Pety, Jérôme and Dobbs, Clare L. and García-Burillo, Santiago and Dumas, Gaëlle and Thompson, Todd A. and Schuster, Karl F. and Kramer, Carsten},
title = {THE PdBI ARCSECOND WHIRLPOOL SURVEY (PAWS): ENVIRONMENTAL DEPENDENCE OF GIANT MOLECULAR CLOUD PROPERTIES IN M51*},
journal = {\apj}
}

@ARTICLE{Colombo_2015,
       author = {{Colombo}, D. and {Rosolowsky}, E. and {Ginsburg}, A. and {Duarte-Cabral}, A. and {Hughes}, A.},
        title = "{Graph-based interpretation of the molecular interstellar medium segmentation}",
      journal = {\mnras},
         year = 2015,
        month = dec,
       volume = {454},
       number = {2},
        pages = {2067-2091},
          doi = {10.1093/mnras/stv2063},
archivePrefix = {arXiv},
       eprint = {1510.04253},
 primaryClass = {astro-ph.GA},
       adsurl = {https://ui.adsabs.harvard.edu/abs/2015MNRAS.454.2067C},
    shortlabel = {C15},
}

@ARTICLE{Bolatto_2013,
       author = {{Bolatto}, Alberto D. and {Wolfire}, Mark and {Leroy}, Adam K.},
        title = "{The CO-to-H$_{2}$ Conversion Factor}",
      journal = {\araa},
         year = 2013,
        month = aug,
       volume = {51},
       number = {1},
        pages = {207-268},
          doi = {10.1146/annurev-astro-082812-140944},
archivePrefix = {arXiv},
       eprint = {1301.3498},
 primaryClass = {astro-ph.GA},
       adsurl = {https://ui.adsabs.harvard.edu/abs/2013ARA\&A..51..207B}
}

@article{chown2025,
doi = {10.3847/1538-4357/adbd40},
url = {https://dx.doi.org/10.3847/1538-4357/adbd40},
year = {2025},
month = {apr},
publisher = {The American Astronomical Society},
volume = {983},
number = {1},
pages = {64},
author = {Chown, Ryan and Leroy, Adam K. and Sandstrom, Karin and Chastenet, Jérémy and Sutter, Jessica and Koch, Eric W. and Koziol, Hannah B. and Neumann, Lukas and Sun, Jiayi and Williams, Thomas G. and Baron, Dalya and Anand, Gagandeep S. and Barnes, Ashley. T. and Bazzi, Zein and Belfiore, Francesco and Bigiel, Frank and Bolatto, Alberto and Boquien, Médéric and Cao, Yixian and Chevance, Mélanie and Colombo, Dario and Dale, Daniel A. and den Brok, Jakob and Egorov, Oleg V. and Eibensteiner, Cosima and Emsellem, Eric and Hassani, Hamid and Henshaw, Jonathan D. and He, Hao and Kim, Jaeyeon and Klessen, Ralf S. and Kreckel, Kathryn and Larson, Kirsten L. and Lee, Janice C. and Meidt, Sharon E. and Murphy, Eric J. and Oakes, Elias K. and Ostriker, Eve C. and Pan, Hsi-An and Pathak, Debosmita and Rosolowsky, Erik and Sarbadhicary, Sumit K. and Schinnerer, Eva and Teng, Yu-Hsuan and Thilker, David A. and Weinbeck, Tony D. and Watkins, Elizabeth J.},
title = {Polycyclic Aromatic Hydrocarbon and CO(2–1) Emission at 50–150 pc Scales in 70 Nearby Galaxies},
journal = {\apj}
}

@article{Williams_2024,
doi = {10.3847/1538-4365/ad4be5},
url = {https://dx.doi.org/10.3847/1538-4365/ad4be5},
year = {2024},
month = {jul},
publisher = {The American Astronomical Society},
volume = {273},
number = {1},
pages = {13},
author = {Thomas G. Williams and Janice C. Lee and Kirsten L. Larson and Adam K. Leroy and Karin Sandstrom and Eva Schinnerer and David A. Thilker and Francesco Belfiore and Oleg V. Egorov and Erik Rosolowsky and Jessica Sutter and Joseph DePasquale and Alyssa Pagan and Travis A. Berger and Gagandeep S. Anand and Ashley T. Barnes and Frank Bigiel and Médéric Boquien and Yixian Cao and Jérémy Chastenet and Mélanie Chevance and Ryan Chown and Daniel A. Dale and Sinan Deger and Cosima Eibensteiner and Eric Emsellem and Christopher M. Faesi and Simon C. O. Glover and Kathryn Grasha and Stephen Hannon and Hamid Hassani and Jonathan D. Henshaw and María J. Jiménez-Donaire and Jaeyeon Kim and Ralf S. Klessen and Eric W. Koch and Jing Li and Daizhong Liu and Sharon E. Meidt and J. Eduardo Méndez-Delgado and Eric J. Murphy and Justus Neumann and Lukas Neumann and Nadine Neumayer and Elias K. Oakes and Debosmita Pathak and Jérôme Pety and Francesca Pinna and Miguel Querejeta and Lise Ramambason and Andrea Romanelli and Mattia C. Sormani and Sophia K. Stuber and Jiayi Sun and Yu-Hsuan Teng and Antonio Usero and Elizabeth J. Watkins and Tony D. Weinbeck},
title = {PHANGS-JWST: Data-processing Pipeline and First Full Public Data Release},
journal = {\apjs}
}

@article{tielens2008,
	adsurl = {https://ui.adsabs.harvard.edu/abs/2008ARA\&A..46..289T},
	author = {{Tielens}, A.~G.~G.~M.},
	doi = {10.1146/annurev.astro.46.060407.145211},
	journal = {\araa},
	month = sep,
	pages = {289-337},
	title = {{Interstellar polycyclic aromatic hydrocarbon molecules.}},
	volume = {46},
	year = 2008}

@article{sandstrom2023,
	adsurl = {https://ui.adsabs.harvard.edu/abs/2023ApJ...944L...7S},
	archiveprefix = {arXiv},
	author = {{Sandstrom}, Karin M. and {Chastenet}, J{\'e}r{\'e}my and {Sutter}, Jessica and {Leroy}, Adam K. and {Egorov}, Oleg V. and {Williams}, Thomas G. and {Bolatto}, Alberto D. and {Boquien}, M{\'e}d{\'e}ric and {Cao}, Yixian and {Dale}, Daniel A. and {Lee}, Janice C. and {Rosolowsky}, Erik and {Schinnerer}, Eva and {Barnes}, Ashley. T. and {Belfiore}, Francesco and {Bigiel}, F. and {Chevance}, M{\'e}lanie and {Grasha}, Kathryn and {Groves}, Brent and {Hassani}, Hamid and {Hughes}, Annie and {Klessen}, Ralf S. and {Kruijssen}, J.~M. Diederik and {Larson}, Kirsten L. and {Liu}, Daizhong and {Lopez}, Laura A. and {Meidt}, Sharon E. and {Murphy}, Eric J. and {Sormani}, Mattia C. and {Thilker}, David A. and {Watkins}, Elizabeth J.},
	doi = {10.3847/2041-8213/acb0cf},
	eid = {L7},
	eprint = {2301.00854},
	journal = {\apjl},
	month = feb,
	number = {2},
	pages = {L7},
	primaryclass = {astro-ph.GA},
	title = {{PHANGS-JWST First Results: Mapping the 3.3 {\ensuremath{\mu}}m Polycyclic Aromatic Hydrocarbon Vibrational Band in Nearby Galaxies with NIRCam Medium Bands}},
	volume = {944},
	year = 2023}

@ARTICLE{sutter2024,
       author = {{Sutter}, Jessica and {Sandstrom}, Karin and {Chastenet}, J{\'e}r{\'e}my and {Leroy}, Adam K. and {Koch}, Eric W. and {Williams}, Thomas G. and {Chown}, Ryan and {Belfiore}, Francesco and {Bigiel}, Frank and {Boquien}, M{\'e}d{\'e}ric and {Cao}, Yixian and {Chevance}, M{\'e}lanie and {Dale}, Daniel A. and {Egorov}, Oleg V. and {Glover}, Simon C.~O. and {Groves}, Brent and {Klessen}, Ralf S. and {Kreckel}, Kathryn and {Larson}, Kirsten L. and {Oakes}, Elias K. and {Pathak}, Debosmita and {Ramambason}, Lise and {Rosolowsky}, Erik and {Watkins}, Elizabeth J.},
        title = "{The Fraction of Dust Mass in the Form of Polycyclic Aromatic Hydrocarbons on 10{\textendash}50 pc Scales in Nearby Galaxies}",
      journal = {\apj},
         year = 2024,
        month = aug,
       volume = {971},
       number = {2},
          eid = {178},
        pages = {178},
          doi = {10.3847/1538-4357/ad54bd},
archivePrefix = {arXiv},
       eprint = {2405.15102},
 primaryClass = {astro-ph.GA},
       adsurl = {https://ui.adsabs.harvard.edu/abs/2024ApJ...971..178S}
}

@ARTICLE{Gao_2022,
       author = {{Gao}, Yang and {Tan}, Qing-Hua and {Gao}, Yu and {Fang}, Min and {Chown}, Ryan and {Jiao}, Qian and {Luo}, Chun-Sheng},
        title = "{The Correlation between WISE 12 {\ensuremath{\mu}}m Emission and Molecular Gas Tracers on Subkiloparsec Scales in Nearby Star-forming Galaxies}",
      journal = {\apj},
         year = 2022,
        month = dec,
       volume = {940},
       number = {2},
          eid = {133},
        pages = {133},
          doi = {10.3847/1538-4357/ac9af1},
archivePrefix = {arXiv},
       eprint = {2210.01982},
 primaryClass = {astro-ph.GA},
       adsurl = {https://ui.adsabs.harvard.edu/abs/2022ApJ...940..133G}
}

@article{Leroy_2023,
doi = {10.3847/2041-8213/acaf85},
url = {https://dx.doi.org/10.3847/2041-8213/acaf85},
year = {2023},
month = {feb},
publisher = {The American Astronomical Society},
volume = {944},
number = {2},
pages = {L9},
author = {Adam K. Leroy and Karin Sandstrom and Erik Rosolowsky and Francesco Belfiore and Alberto D. Bolatto and Yixian Cao and Eric W. Koch and Eva Schinnerer and Ashley. T. Barnes and Ivana Bešlić and F. Bigiel and Guillermo A. Blanc and Jérémy Chastenet and Ness Mayker Chen and Mélanie Chevance and Ryan Chown and Enrico Congiu and Daniel A. Dale and Oleg V. Egorov and Eric Emsellem and Cosima Eibensteiner and Christopher M. Faesi and Simon C. O. Glover and Kathryn Grasha and Brent Groves and Hamid Hassani and Jonathan D. Henshaw and Annie Hughes and María J. Jiménez-Donaire and Jaeyeon Kim and Ralf S. Klessen and Kathryn Kreckel and J. M. Diederik Kruijssen and Kirsten L. Larson and Janice C. Lee and Rebecca C. Levy and Daizhong Liu and Laura A. Lopez and Sharon E. Meidt and Eric J. Murphy and Justus Neumann and Ismael Pessa and Jérôme Pety and Toshiki Saito and Amy Sardone and Jiayi Sun and David A. Thilker and Antonio Usero and Elizabeth J. Watkins and Cory M. Whitcomb and Thomas G. Williams},
title = {PHANGS–JWST First Results: Mid-infrared Emission Traces Both Gas Column Density and Heating at 100 pc Scales},
journal = {\apjl}
}
}

\begin{appendix}
\label{S:appendix_summary}

\onecolumn

\section{SCIMES version update} \label{a:scimes_update}

In this study, we implemented a modification to SCIMES to enhance computational efficiency. The primary modification involved implementing parallel processing using the $\mathtt{Parallel}$\footnote{https://joblib.readthedocs.io/en/latest/generated/joblib.Parallel.html} function from the $\mathtt{joblib}$\footnote{https://joblib.readthedocs.io/en/latest/index.html} library in $\mathtt{Python}$. This optimization was applied to the spectral embedding step, where dimensionality reduction is performed.

To make use of the parallel processing feature, an additional parameter was added ($\mathtt{n\_jobs = -1}$). The $\mathtt{n\_jobs}$ parameter controls the number of concurrent tasks executed in parallel. When set to a positive integer, it specifies the exact number of worker processes or threads used for computation. If $\mathtt{n\_jobs = -1}$, the system utilizes all available CPU cores, while $\mathtt{n\_jobs = -2}$ reserves one core for other tasks. If $\mathtt{n\_jobs = 1}$, the code runs sequentially, similar to a standard Python loop. This improvement significantly reduces processing time, particularly when handling large molecular cloud datasets. 

This modified version of SCIMES was used throughout this analysis and is available on the following $\mathtt{GitHub}$\footnote{\url{https://github.com/Astroua/SCIMES/tree/new_version}} page.

\section{Tables of properties}

\begin{longtable}{lcccccccccccc}
\caption{The global and cloud properties of 66 galaxies in the PHANGS-JWST sample.}\\
\hline
\label{T:sample}
Galaxy & RA & Dec & $i$ & $D$ &$\log M_*$ & $\log \mathrm{SFR}/M_*$ & $f_{all}$ & $f_{flg}$ & cycle & $\sigmol$ & $\Reff$ & $\Mmol$ \\
 & [deg] & [deg] & [deg] & [Mpc] & $\log[\msun]$ & $\log [1/yr]$ & [$\%$] & [$\%$] & & $[\msunperpcsq]$ & [pc] & $\log [ \msun]$ \\
\hline
\endfirsthead
\caption{continued.}\\
\hline
Galaxy & RA & Dec & $i$ & $D$ & $\log M_*$ & $\log \mathrm{SFR}/M_*$ & $f_{all}$ & $f_{flg}$ & cycle & $\sigmol$ & $\Reff$ & $\Mmol$ \\

& [deg] & [deg] & [deg] & [Mpc] & $\log[\msun]$ & $\log [1/yr]$ & [$\%$] & [$\%$] & & $[\msunperpcsq]$ & [pc] & $\log [ \msun]$ \\
\hline
\endhead
\hline
\endfoot
IC5273 & 344.86 & -37.70 & 52.00 & 14.18 & 9.72 & -9.99 & 44.10 & 35.06 & 2 & $35.09^{+34.04}_{-15.04}$ & $41.07^{+47.98}_{-22.95}$ & $5.19^{+0.77}_{-0.48}$ \\
IC5332 & 353.61 & -36.10 & 26.90 & 9.01 & 9.67 & -10.05 & 48.96 & 29.94 & 2 & $5.93^{+5.30}_{-1.45}$ & $33.66^{+58.04}_{-20.57}$ & $4.33^{+0.81}_{-0.61}$ \\
NGC0628 & 24.17 & 15.78 & 8.90 & 9.84 & 10.34 & -10.10 & 49.24 & 44.51 & 1 & $13.85^{+17.57}_{-6.98}$ & $29.70^{+43.45}_{-16.30}$ & $4.51^{+0.81}_{-0.46}$ \\
NGC1087 & 41.60 & -0.50 & 42.90 & 15.85 & 9.94 & -9.83 & 43.38 & 23.50 & 1 & $41.42^{+61.91}_{-26.08}$ & $36.28^{+41.56}_{-18.92}$ & $5.16^{+0.65}_{-0.45}$ \\
NGC1097 & 41.58 & -30.28 & 48.60 & 13.58 & 10.76 & -10.08 & 38.20 & 22.11 & 2 & $9.08^{+13.11}_{-3.91}$ & $38.15^{+47.88}_{-20.34}$ & $4.60^{+0.71}_{-0.49}$ \\
NGC1300 & 49.92 & -19.41 & 31.80 & 18.99 & 10.62 & -10.55 & 40.65 & 30.58 & 1 & $10.05^{+14.46}_{-4.73}$ & $40.74^{+45.19}_{-21.02}$ & $4.70^{+0.65}_{-0.49}$ \\
NGC1365 & 53.40 & -36.14 & 55.40 & 19.57 & 11.00 & -9.76 & 33.11 & 18.93 & 1 & $6.16^{+7.88}_{-2.26}$ & $37.22^{+48.46}_{-20.27}$ & $4.44^{+0.67}_{-0.50}$ \\
NGC1385 & 54.37 & -24.50 & 44.00 & 17.22 & 9.98 & -9.66 & 42.30 & 14.05 & 1 & $37.15^{+57.66}_{-22.02}$ & $33.41^{+35.06}_{-17.18}$ & $5.04^{+0.59}_{-0.41}$ \\
NGC1433 & 55.51 & -47.22 & 28.60 & 18.63 & 10.87 & -10.82 & 40.85 & 29.39 & 1 & $6.98^{+7.53}_{-2.25}$ & $36.94^{+52.16}_{-19.98}$ & $4.52^{+0.75}_{-0.60}$ \\
NGC1511 & 59.91 & -67.64 & 72.70 & 15.28 & 9.91 & -9.55 & 33.69 & 6.28 & 2 & $27.61^{+43.60}_{-14.33}$ & $34.61^{+31.06}_{-17.14}$ & $5.02^{+0.48}_{-0.41}$ \\
NGC1512 & 60.98 & -43.35 & 42.50 & 18.83 & 10.72 & -10.61 & 39.53 & 31.77 & 1 & $7.23^{+7.66}_{-2.38}$ & $42.17^{+49.79}_{-22.93}$ & $4.63^{+0.70}_{-0.59}$ \\
NGC1546 & 63.65 & -56.06 & 70.30 & 17.69 & 10.35 & -10.43 & 34.75 & 3.25 & 2 & $11.45^{+21.89}_{-5.10}$ & $32.18^{+39.70}_{-18.45}$ & $4.59^{+0.57}_{-0.39}$ \\
NGC1559 & 64.40 & -62.78 & 65.40 & 19.44 & 10.36 & -9.79 & 37.85 & 15.83 & 2 & $39.00^{+51.84}_{-20.14}$ & $35.46^{+38.96}_{-18.78}$ & $5.13^{+0.59}_{-0.40}$ \\
NGC1566 & 65.00 & -54.94 & 29.50 & 17.69 & 10.79 & -10.13 & 41.41 & 23.02 & 1 & $14.72^{+23.01}_{-7.22}$ & $36.34^{+42.95}_{-18.83}$ & $4.76^{+0.67}_{-0.47}$ \\
NGC1637 & 70.37 & -2.86 & 31.10 & 11.70 & 9.95 & -10.14 & 46.20 & 40.10 & 2 & $15.53^{+15.76}_{-7.03}$ & $40.74^{+47.57}_{-22.86}$ & $4.84^{+0.77}_{-0.49}$ \\
NGC1672 & 71.43 & -59.25 & 42.60 & 19.40 & 10.73 & -9.85 & 34.40 & 22.94 & 1 & $21.22^{+26.80}_{-8.84}$ & $36.33^{+38.42}_{-18.51}$ & $4.92^{+0.62}_{-0.42}$ \\
NGC1792 & 76.31 & -37.98 & 65.10 & 16.20 & 10.61 & -10.04 & 36.03 & 6.28 & 2 & $34.46^{+49.02}_{-18.81}$ & $35.79^{+43.02}_{-19.68}$ & $5.10^{+0.59}_{-0.43}$ \\
NGC1809 & 75.52 & -69.57 & 57.60 & 19.95 & 9.77 & -9.01 & 38.02 & 30.04 & 2 & $18.50^{+22.66}_{-9.77}$ & $33.81^{+52.02}_{-17.30}$ & $4.78^{+0.74}_{-0.40}$ \\
NGC2090 & 86.76 & -34.25 & 64.50 & 11.75 & 10.04 & -10.43 & 44.24 & 38.11 & 2 & $10.54^{+10.65}_{-4.32}$ & $36.90^{+39.21}_{-19.69}$ & $4.60^{+0.64}_{-0.47}$ \\
NGC2283 & 101.47 & -18.21 & 43.70 & 13.68 & 9.89 & -10.17 & 45.01 & 40.23 & 2 & $22.66^{+35.73}_{-11.83}$ & $40.10^{+45.67}_{-22.60}$ & $4.99^{+0.70}_{-0.50}$ \\
NGC2566 & 124.69 & -25.50 & 48.50 & 23.44 & 10.71 & -9.77 & 31.02 & 20.57 & 2 & $10.47^{+15.14}_{-4.92}$ & $40.93^{+45.69}_{-19.76}$ & $4.76^{+0.60}_{-0.48}$ \\
NGC2775 & 137.58 & 7.04 & 41.20 & 23.15 & 11.07 & -11.13 & 40.63 & 32.35 & 2 & $6.16^{+3.46}_{-1.59}$ & $43.58^{+54.40}_{-22.86}$ & $4.53^{+0.72}_{-0.49}$ \\
NGC2835 & 139.47 & -22.35 & 41.30 & 12.22 & 10.00 & -9.90 & 47.96 & 45.65 & 1 & $13.57^{+15.96}_{-6.32}$ & $36.25^{+48.42}_{-18.63}$ & $4.68^{+0.74}_{-0.44}$ \\
NGC2903 & 143.04 & 21.50 & 66.80 & 10.00 & 10.63 & -10.15 & 37.75 & 21.16 & 2 & $16.76^{+23.37}_{-8.25}$ & $37.87^{+47.25}_{-22.32}$ & $4.82^{+0.67}_{-0.51}$ \\
NGC2997 & 146.41 & -31.19 & 33.00 & 14.06 & 10.73 & -10.09 & 38.81 & 30.90 & 2 & $17.65^{+27.22}_{-8.92}$ & $41.13^{+51.03}_{-22.22}$ & $4.93^{+0.67}_{-0.45}$ \\
NGC3059 & 147.53 & -73.92 & 29.40 & 20.23 & 10.38 & -10.00 & 42.67 & 27.21 & 2 & $22.36^{+27.70}_{-12.44}$ & $35.99^{+43.62}_{-18.47}$ & $4.91^{+0.67}_{-0.47}$ \\
NGC3137 & 152.28 & -29.06 & 70.30 & 16.37 & 9.88 & -10.19 & 34.74 & 28.63 & 2 & $8.78^{+7.06}_{-2.97}$ & $47.97^{+55.26}_{-26.71}$ & $4.81^{+0.64}_{-0.55}$ \\
NGC3239 & 156.27 & 17.16 & 60.30 & 10.86 & 9.17 & -9.58 & 44.55 & 41.91 & 2 & $10.32^{+22.08}_{-3.66}$ & $44.07^{+51.67}_{-27.87}$ & $4.83^{+0.65}_{-0.64}$ \\
NGC3351 & 160.99 & 11.70 & 45.10 & 9.96 & 10.37 & -10.25 & 45.81 & 36.52 & 1 & $7.18^{+6.01}_{-2.27}$ & $42.10^{+56.53}_{-24.40}$ & $4.56^{+0.80}_{-0.58}$ \\
NGC3507 & 165.86 & 18.14 & 21.70 & 23.55 & 10.40 & -10.40 & 43.13 & 39.70 & 2 & $10.10^{+11.16}_{-4.44}$ & $37.51^{+47.30}_{-18.48}$ & $4.62^{+0.73}_{-0.48}$ \\
NGC3511 & 165.85 & -23.09 & 75.10 & 13.94 & 10.03 & -10.12 & 38.13 & 17.62 & 2 & $22.54^{+18.27}_{-10.04}$ & $48.07^{+48.32}_{-25.45}$ & $5.14^{+0.64}_{-0.49}$ \\
NGC3521 & 166.45 & -0.03 & 68.80 & 13.24 & 11.02 & -10.45 & 37.83 & 13.25 & 2 & $34.04^{+46.42}_{-19.58}$ & $37.55^{+46.81}_{-21.47}$ & $5.11^{+0.62}_{-0.44}$ \\
NGC3596 & 168.78 & 14.79 & 25.10 & 11.30 & 9.66 & -10.18 & 47.76 & 41.95 & 2 & $16.55^{+17.15}_{-7.88}$ & $38.93^{+37.03}_{-22.58}$ & $4.77^{+0.69}_{-0.50}$ \\
NGC3621 & 169.57 & -32.81 & 65.80 & 7.06 & 10.06 & -10.06 & 44.10 & 30.34 & 2 & $45.31^{+62.52}_{-24.74}$ & $35.19^{+45.35}_{-19.62}$ & $5.14^{+0.69}_{-0.39}$ \\
NGC3626 & 170.02 & 18.36 & 46.60 & 20.05 & 10.46 & -11.13 & 32.00 & 18.98 & 2 & $8.92^{+8.23}_{-3.96}$ & $30.00^{+36.90}_{-14.36}$ & $4.33^{+0.70}_{-0.32}$ \\
NGC3627 & 170.06 & 12.99 & 57.30 & 11.32 & 10.84 & -10.25 & 41.68 & 11.95 & 1 & $24.73^{+38.33}_{-12.36}$ & $28.30^{+33.24}_{-15.23}$ & $4.75^{+0.57}_{-0.38}$ \\
NGC4254 & 184.71 & 14.42 & 34.40 & 13.10 & 10.42 & -9.93 & 43.56 & 17.81 & 1 & $48.20^{+67.27}_{-25.61}$ & $30.63^{+34.84}_{-16.07}$ & $5.07^{+0.66}_{-0.41}$ \\
NGC4298 & 185.39 & 14.61 & 59.20 & 14.92 & 10.02 & -10.36 & 40.42 & 18.23 & 2 & $13.87^{+12.43}_{-5.31}$ & $41.37^{+42.98}_{-21.91}$ & $4.81^{+0.65}_{-0.45}$ \\
NGC4303 & 185.48 & 4.47 & 23.50 & 16.99 & 10.51 & -9.78 & 42.06 & 24.39 & 1 & $26.55^{+44.62}_{-16.26}$ & $31.69^{+39.42}_{-16.51}$ & $4.86^{+0.66}_{-0.46}$ \\
NGC4321 & 185.73 & 15.82 & 38.50 & 15.21 & 10.75 & -10.20 & 43.28 & 30.23 & 1 & $13.69^{+16.42}_{-6.34}$ & $35.20^{+45.54}_{-18.96}$ & $4.66^{+0.76}_{-0.45}$ \\
NGC4424 & 186.80 & 9.42 & 58.20 & 16.20 & 9.91 & -10.43 & 30.47 & 5.39 & 2 & $11.47^{+31.44}_{-5.19}$ & $26.07^{+24.99}_{-12.71}$ & $4.52^{+0.40}_{-0.53}$ \\
NGC4457 & 187.25 & 3.57 & 17.40 & 15.10 & 10.42 & -10.93 & 38.03 & 22.90 & 2 & $6.70^{+24.07}_{-3.15}$ & $21.59^{+38.89}_{-10.20}$ & $4.17^{+0.66}_{-0.49}$ \\
NGC4496A & 187.91 & 3.94 & 53.80 & 14.86 & 9.53 & -9.74 & 45.60 & 43.93 & 2 & $13.42^{+28.66}_{-6.36}$ & $43.38^{+48.53}_{-23.57}$ & $4.92^{+0.69}_{-0.54}$ \\
NGC4535 & 188.58 & 8.20 & 44.70 & 15.77 & 10.54 & -10.20 & 44.42 & 25.84 & 1 & $10.16^{+15.37}_{-4.36}$ & $39.34^{+48.02}_{-20.93}$ & $4.69^{+0.63}_{-0.46}$ \\
NGC4536 & 188.61 & 2.19 & 66.00 & 16.25 & 10.40 & -9.86 & 25.31 & 17.55 & 2 & $11.22^{+13.89}_{-4.54}$ & $52.24^{+54.49}_{-30.30}$ & $4.94^{+0.63}_{-0.50}$ \\
NGC4540 & 188.71 & 15.55 & 28.70 & 15.76 & 9.79 & -10.56 & 41.38 & 28.57 & 2 & $13.90^{+14.69}_{-7.23}$ & $33.15^{+45.51}_{-17.52}$ & $4.65^{+0.73}_{-0.48}$ \\
NGC4548 & 188.86 & 14.50 & 38.30 & 16.22 & 10.69 & -10.97 & 39.96 & 25.24 & 2 & $6.78^{+7.18}_{-2.08}$ & $46.58^{+58.90}_{-25.84}$ & $4.67^{+0.70}_{-0.51}$ \\
NGC4569 & 189.21 & 13.16 & 70.00 & 15.76 & 10.81 & -10.68 & 29.04 & 4.39 & 2 & $6.62^{+7.26}_{-1.79}$ & $28.06^{+28.44}_{-12.40}$ & $4.27^{+0.67}_{-0.49}$ \\
NGC4571 & 189.23 & 14.22 & 32.70 & 14.90 & 10.09 & -10.63 & 47.16 & 40.20 & 2 & $6.61^{+5.19}_{-1.89}$ & $48.69^{+52.52}_{-27.23}$ & $4.68^{+0.64}_{-0.58}$ \\
NGC4579 & 189.43 & 11.82 & 40.20 & 21.00 & 11.15 & -10.81 & 41.47 & 21.33 & 2 & $7.01^{+6.28}_{-2.31}$ & $34.79^{+48.22}_{-18.12}$ & $4.40^{+0.75}_{-0.50}$ \\
NGC4654 & 190.99 & 13.13 & 55.60 & 21.98 & 10.57 & -9.99 & 38.93 & 12.82 & 2 & $15.39^{+17.60}_{-7.44}$ & $41.55^{+43.96}_{-20.99}$ & $4.86^{+0.62}_{-0.42}$ \\
NGC4689 & 191.94 & 13.76 & 38.70 & 15.00 & 10.22 & -10.61 & 44.43 & 39.62 & 2 & $9.34^{+8.61}_{-3.60}$ & $40.21^{+55.75}_{-22.89}$ & $4.63^{+0.78}_{-0.52}$ \\
NGC4694 & 192.06 & 10.98 & 60.70 & 15.76 & 9.86 & -10.66 & 36.91 & 7.00 & 2 & $11.22^{+19.92}_{-5.35}$ & $27.54^{+34.68}_{-15.00}$ & $4.41^{+0.64}_{-0.37}$ \\
NGC4731 & 192.76 & -6.39 & 64.00 & 13.28 & 9.48 & -9.70 & 35.02 & 22.12 & 2 & $19.81^{+33.59}_{-9.68}$ & $43.55^{+48.54}_{-26.72}$ & $5.01^{+0.67}_{-0.41}$ \\
NGC4781 & 193.60 & -10.54 & 59.00 & 11.31 & 9.64 & -9.96 & 40.82 & 29.42 & 2 & $48.89^{+52.62}_{-25.69}$ & $38.76^{+36.97}_{-20.93}$ & $5.27^{+0.64}_{-0.49}$ \\
NGC4826 & 194.18 & 21.68 & 59.10 & 4.41 & 10.24 & -10.93 & 32.90 & 5.89 & 2 & $32.82^{+28.26}_{-23.74}$ & $16.76^{+30.58}_{-11.76}$ & $4.40^{+0.76}_{-0.87}$ \\
NGC4941 & 196.05 & -5.55 & 53.40 & 15.00 & 10.17 & -10.53 & 38.22 & 25.92 & 2 & $5.83^{+3.55}_{-1.34}$ & $53.40^{+52.25}_{-28.85}$ & $4.71^{+0.61}_{-0.58}$ \\
NGC4951 & 196.28 & -6.49 & 70.20 & 15.00 & 9.79 & -10.24 & 34.74 & 23.28 & 2 & $16.61^{+15.02}_{-7.28}$ & $44.39^{+45.12}_{-24.80}$ & $4.94^{+0.56}_{-0.49}$ \\
NGC5042 & 198.88 & -23.98 & 49.40 & 16.78 & 9.90 & -10.12 & 38.56 & 35.86 & 2 & $10.62^{+11.23}_{-4.67}$ & $46.55^{+53.74}_{-26.12}$ & $4.78^{+0.74}_{-0.50}$ \\
NGC5068 & 199.73 & -21.04 & 35.70 & 5.20 & 9.41 & -9.97 & 52.12 & 48.66 & 1 & $16.82^{+27.90}_{-8.48}$ & $31.84^{+47.06}_{-17.81}$ & $4.69^{+0.85}_{-0.51}$ \\
NGC5134 & 201.33 & -21.13 & 22.70 & 19.92 & 10.41 & -10.75 & 39.91 & 35.33 & 2 & $10.26^{+13.21}_{-4.65}$ & $35.85^{+44.62}_{-18.38}$ & $4.59^{+0.77}_{-0.51}$ \\
NGC5248 & 204.38 & 8.89 & 47.40 & 14.87 & 10.41 & -10.05 & 39.72 & 24.09 & 2 & $17.23^{+20.92}_{-8.56}$ & $39.05^{+43.44}_{-21.28}$ & $4.85^{+0.66}_{-0.45}$ \\
NGC5643 & 218.17 & -44.17 & 29.90 & 12.68 & 10.34 & -9.92 & 45.26 & 39.21 & 2 & $21.51^{+32.02}_{-11.57}$ & $34.45^{+46.73}_{-19.12}$ & $4.83^{+0.78}_{-0.49}$ \\
NGC6300 & 259.25 & -62.82 & 49.60 & 11.58 & 10.47 & -10.19 & 45.77 & 36.80 & 2 & $15.31^{+19.55}_{-7.92}$ & $36.28^{+46.67}_{-20.70}$ & $4.75^{+0.66}_{-0.53}$ \\
NGC7456 & 345.54 & -39.57 & 67.30 & 15.70 & 9.64 & -10.08 & 33.11 & 29.65 & 2 & $7.62^{+6.09}_{-2.69}$ & $51.98^{+52.75}_{-28.99}$ & $4.76^{+0.62}_{-0.54}$ \\
NGC7496 & 347.45 & -43.43 & 35.90 & 18.72 & 10.00 & -9.65 & 37.34 & 28.87 & 1 & $19.43^{+25.74}_{-8.90}$ & $41.81^{+46.65}_{-22.04}$ & $5.00^{+0.65}_{-0.48}$ \\

\end{longtable}

\tablefoot{\textit{Global properties of the galaxies:} Right Ascension (RA), Declination (Dec), $i$, $D$, M${_\star}$, sSFR, and the PHANGS-JWST cycle of the galaxy. \textit{Cloud properties:} the fraction of flux in all the clouds with respect to the total flux of the galaxy $f_{all}$, the fraction of flux in the clouds after flagging and used in the analysis $f_{flg}$, median $\sigmol$, $\Reff$, and $\Mmol$. The 84th - 50th and 50th - 16th percentiles are shown in superscript and subscript, respectively.}

\vspace{1em}
\clearpage

\section{Catalog information}\label{s:catalog}

\renewcommand{\arraystretch}{1.2} 

\begin{table*}[h]
\caption{PAH cloud catalog columns and their descriptions.}
\begin{tabular}{|p{0.18\textwidth} p{0.18\textwidth} p{0.6\textwidth}|}
\hline
\textbf{Catalog Column} & \textbf{Variable} & \textbf{Description} \\ 
\hline
\textit{ID} &  & ID of the cloud in a specific galaxy\\ 
\textit{galaxy} &   & The galaxy of a specific cloud \\ 
\textit{Env} & & Galactic environment of the cloud (1 = Center, 2 + 3 = Bar,  5 + 6 = Spiral Arm, 9 + 10 = Disk, 4 + 7 + 8 = Interarm) \\
\textit{pos$\_$x} &  & x position of the cloud (pixels) \\ 
\textit{pos$\_$y} &  & y position of the cloud (pixels) \\
\textit{pos$\_$ra} &  & Right Ascension of the cloud (degrees) \\ 
\textit{pos$\_$dec} &  & Declination of the cloud (degrees) \\
\textit{Lpah} &   & PAH luminosity of the cloud ($\mathrm{MJy~sr^{-1} ~pc^{2}}$) \\
\textit{Lco} &   & PAH-to-CO converted luminosity of the cloud ($\mathrm{K~km~s^{-1} ~pc^{2}}$) \\
\textit{mass$\_$cst} & \Mmol  & Molecular mass of the cloud using a constant MW \aco\ prescription (M$_{\odot}$) \\ 
\textit{mass$\_$sl} & \Mmol  & Molecular mass of the cloud using \cite{schinnerer2024} \aco\ prescription (M$_{\odot}$) \\ 
\textit{mass$\_$s} & \Mmol  & Molecular mass of the cloud using \cite{Sun_2020a} \aco\ prescription (M$_{\odot}$) \\ 
\textit{mass$\_$b} & \Mmol  & Molecular mass of the cloud using \cite{Bolatto_2013} \aco\ prescription (M$_{\odot}$) \\ 
\textit{mass$\_$t} & \Mmol  & Molecular mass of the cloud using \cite{Teng_2024} \aco\ prescription (M$_{\odot}$) \\
\textit{rad$\_$eq} &  & Radius of the cloud using the area of the cloud (pc)\\
\textit{rad$\_$eq$\_$dec} & \Reff & Beam-deconvolved radius of the cloud using the area of the cloud (pc)\\
\textit{rad} & R & Radius of the cloud using a HWHM factor of 1.18 (pc)\\
\textit{SD$\_$cst} & \sigmol  & Molecular mass surface density of the cloud using a constant MW \aco\ prescription (M$_{\odot}\,\mathrm{pc}^{-2}$) \\
\textit{SD$\_$sl} & \sigmol  & Molecular mass surface density of the cloud using the \cite{schinnerer2024} \aco\ prescription  (M$_{\odot}\,\mathrm{pc}^{-2}$) \\
\textit{SD$\_$s} & \sigmol  & Molecular mass surface density of the cloud using the \cite{Sun_2020a} \aco\ prescription  (M$_{\odot}\,\mathrm{pc}^{-2}$) \\
\textit{SD$\_$b} & \sigmol  & Molecular mass surface density of the cloud using the \cite{Bolatto_2013} \aco\ prescription  (M$_{\odot}\,\mathrm{pc}^{-2}$) \\
\textit{SD$\_$t} & \sigmol  & Molecular mass surface density of the cloud using the \cite{Teng_2024} \aco\ prescription  (M$_{\odot}\,\mathrm{pc}^{-2}$) \\
\textit{Distance} & R$_{\mathrm{gal}}$ & Distance from the cloud to the center of the galaxy (kpc)\\
\textit{Distance$\_$Re} & R$_{\mathrm{gal}}$ & Distance from the cloud to the center of the galaxy (R$_{e}$)\\
\textit{f$\_$all} & $f_{all}$ & Flag to remove clouds according to our flagging method (set = True to remove) \\
\textit{overlap$\_$ratio$\_$all} & & The percentage overlap of the cloud in velocity space \\
\textit{edge$\_$clouds} & & Flag to check if the cloud is on the edge of the FOV (1 = edge, 0 = non-edge) \\
\textit{err$\_$Lpah} &   & Error on the PAH luminosity of the cloud ($\mathrm{MJy~sr^{-1} ~pc^{2}}$) \\
\textit{err$\_$Lco} &   & Error on the PAH-to-CO converted luminosity of the cloud ($\mathrm{K~km~s^{-1} ~pc^{2}}$) \\
\textit{rad$\_$dec$\_$err$\_$fin} & \Reff  & Error on the Beam-deconvolved radius of the cloud using the area of the cloud (pc)\\
\textit{rad$\_$err$\_$fin} &  & Error on the radius of the cloud using the area of the cloud (pc)\\
\textit{mass$\_$err$\_$fin} &   & Error on the mass of the cloud. Add $\_cst$ in the end or the other \aco\ prescription notations (e.g., sl, s, b, t) to specify the error on the corresponding \aco\ prescription mass (M$_{\odot}$)\\
\textit{SD$\_$err$\_$fin} &   & Error on the molecular mass surface density of the cloud. Add $\_cst$ in the end or the other \aco\ prescription notations (e.g., sl, s, b, t) to specify the error on the corresponding \aco\ prescription mass (\msunperpcsq)\\
\hline
\end{tabular}
\label{T:property_description}
\end{table*}

We publish two catalogs, one at the native resolution and sensitivity of each galaxy and another at the homogenized resolution of $30$ pc and a common sensitivity of $0.19$ \mjysr . The \Mmol\ estimates using the different prescriptions are calculated for all the clouds in the different galaxies, except for two galaxies (NGC 4424 and NGC 4694) using the \cite{Teng_2024} prescription.

The native resolution sample comprises 146,040 PAH clouds, and 108,019 clouds after flagging using $f_{all}$. The $\mathtt{ min\_npix}$ is set to be $ 3\times\Omega_{\rm beam}/\Omega_{\rm pix}$, instead of $1\times\Omega_{\rm beam}/\Omega_{\rm pix}$ for this sample. This measure was taken to decrease the segmentation error on the smallest structures. Additionally, the scaling parameter implementation is similar to the homogenized sample.

\section{Additional plots}

\subsection{Galaxy-by-galaxy property plots}
\begin{figure}[h]
    \centering
    \caption{Summary of galaxy-by-galaxy properties. The plots are available at \href{https://zenodo.org/records/15428261}{https://zenodo.org/records/15428261}. \textit{Top left:} Continuum-subtracted images of the galaxy. \textit{Top right:} PAH clouds identified using SCIMES. The flagged clouds are in gray, unflagged in green, and PAH clouds with CO cloud counterparts in red. \textit{Middle left:} \sigmol\ violin plots per galactic environment for the specific galaxy (transparent), and the full sample (colored). The median values and number of PAH clouds per environment are also represented in the plot. \textit{Middle center:} log(\Mmol )-log(\Reff ) scaling relation for the PAH clouds in the galaxy (blue), clouds per galactic environment in the galaxy (see colors in plot), and for the full sample (black). The median values are also represented in the plot per galactic environment. \textit{Middle right:} Global mass spectra (black), galaxy-specific mass spectra (blue), and per galactic environment in the galaxy. The fit values for the galaxy-specific mass spectra are displayed in the plot. \textit{Bottom:}  \sigmol\ (\textit{left}), \Mmol\ (\textit{center}), and \Reff\ (\textit{right}) as a function of R$_{\rm gal}$. A scatterplot of the PAH clouds, color-coded by the density of clouds, is also represented. The running galaxy median (filled blue circles) is plotted for a bin width of 0.1 R$_{e}$. The gray-shaded region represents the interquartile range of the medians per galaxy. The error bars on the median are the standard errors (1.253$\sigma$/$\sqrt{N}$), where N is the number of clouds contributing to a specific bin. The blue and black dashed lines represent the median property of the galaxy and the full sample, respectively.} 
    \label{fig:appendix}
\end{figure}

\subsection{Galaxy-by-galaxy cloud property distribution as a function of galactocentric radius}

\begin{figure*}[h]
    \centering
    \includegraphics[width=0.33\textwidth]{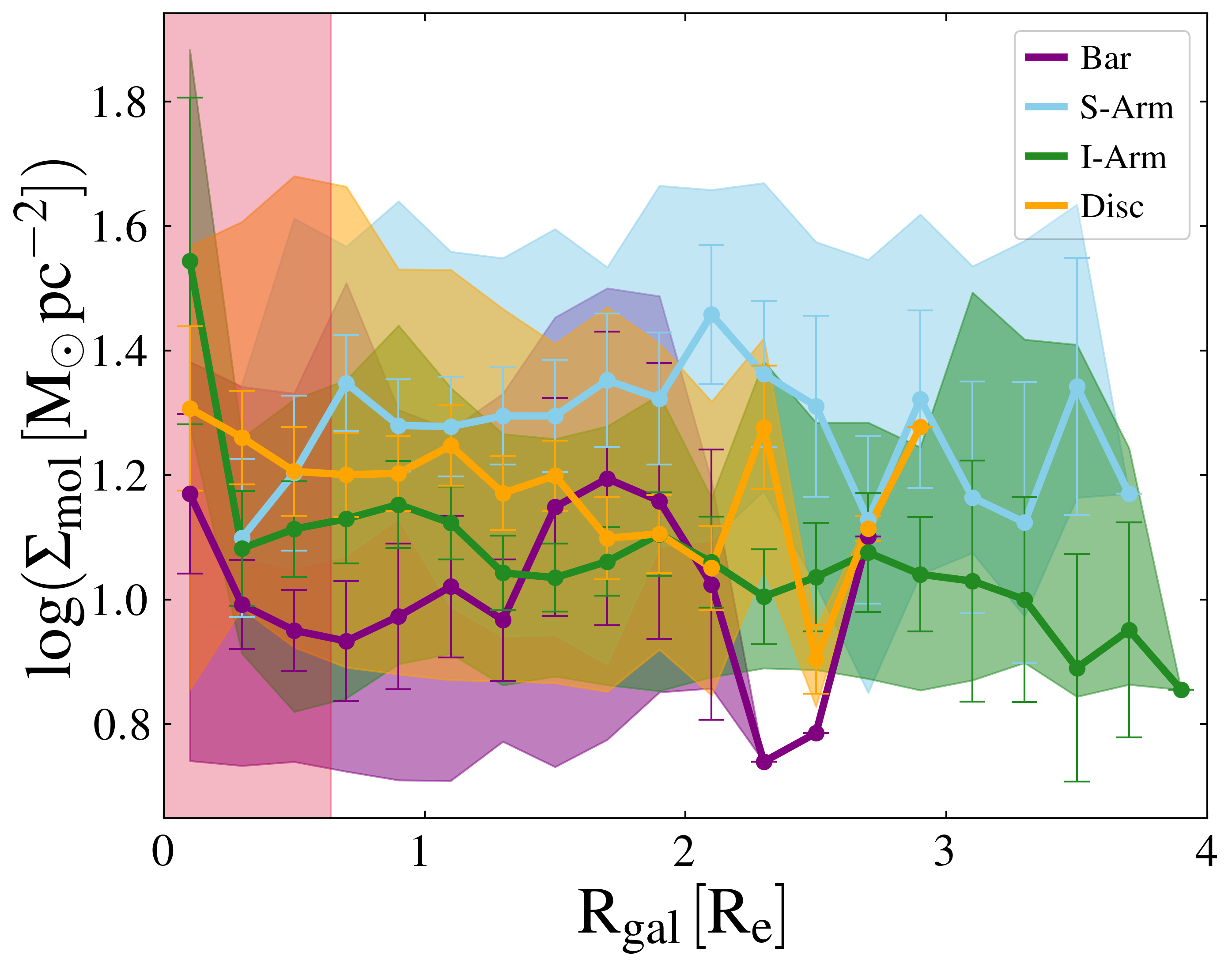}
    \includegraphics[width=0.32\textwidth]{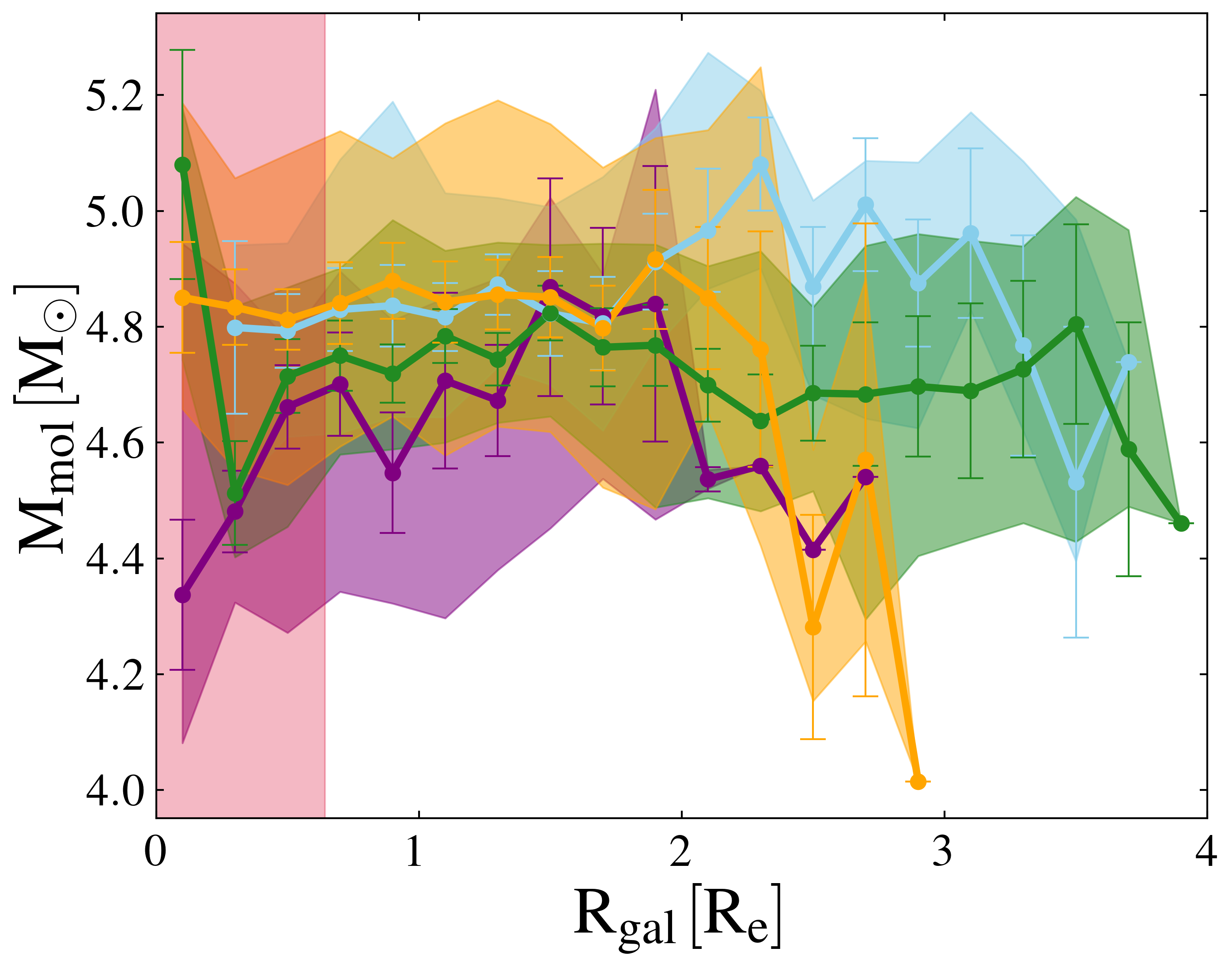}
    \includegraphics[width=0.33\textwidth]{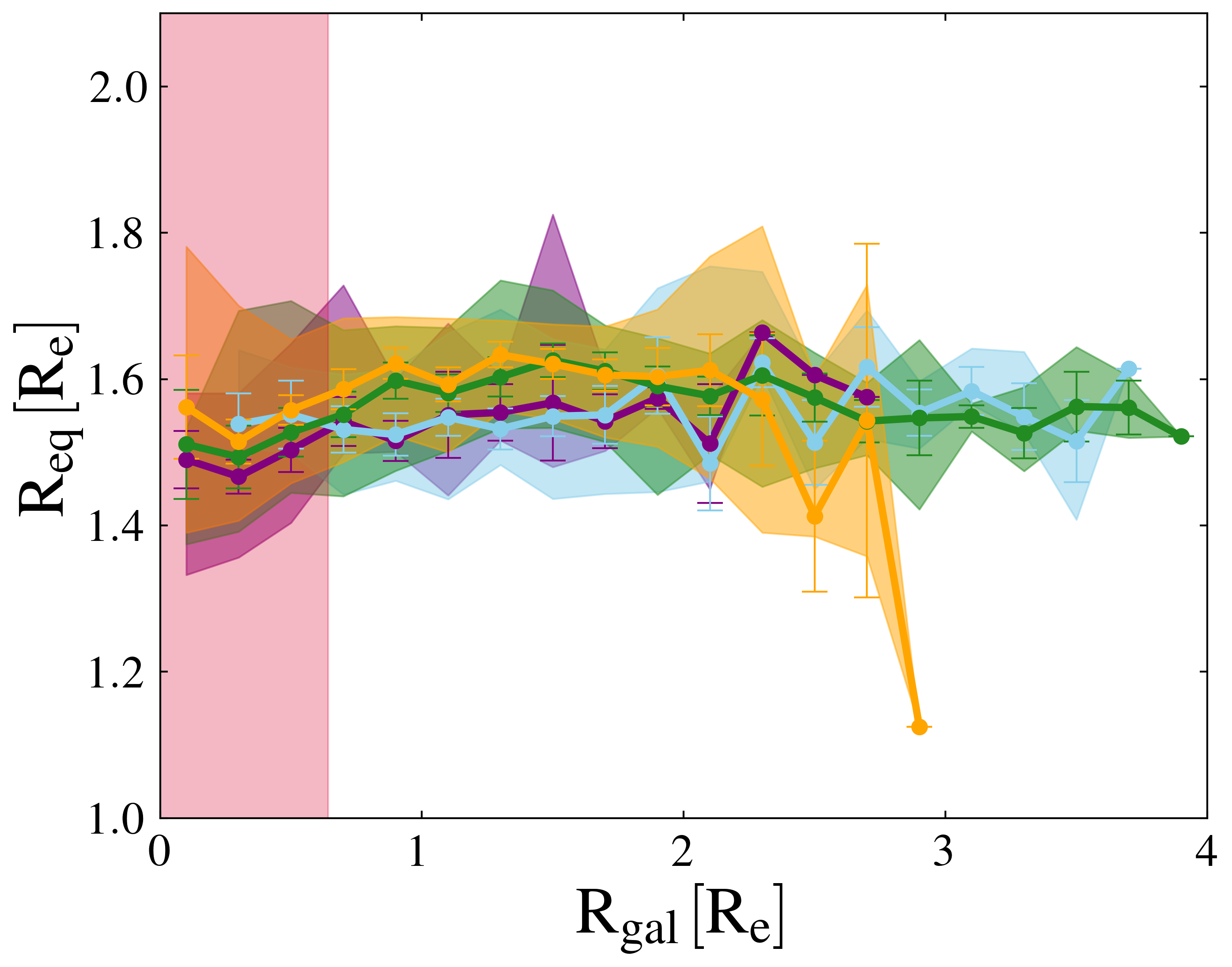}
    \caption{Properties of the PAH clouds vs galactocentric radius per environment for all the clouds in the 66 galaxies: \sigmol\ (\textit{left}), \Mmol\ (\textit{middle}), and \Reff\ (\textit{right}). The running median property per galaxy median (dashed line) is plotted for a bin width of 0.2 R$_{e}$. The shaded region represents the 84-50th and 50-16th percentiles of the medians per galaxy. The error bars on the median are the standard errors (1.253$\sigma$/$\sqrt{N}$), where N is the number of galaxies contributing to a specific bin.}
    \label{fig:radial_envs}
\end{figure*}

\begin{figure*}[h]
    \centering
    \includegraphics[width=0.33\textwidth]{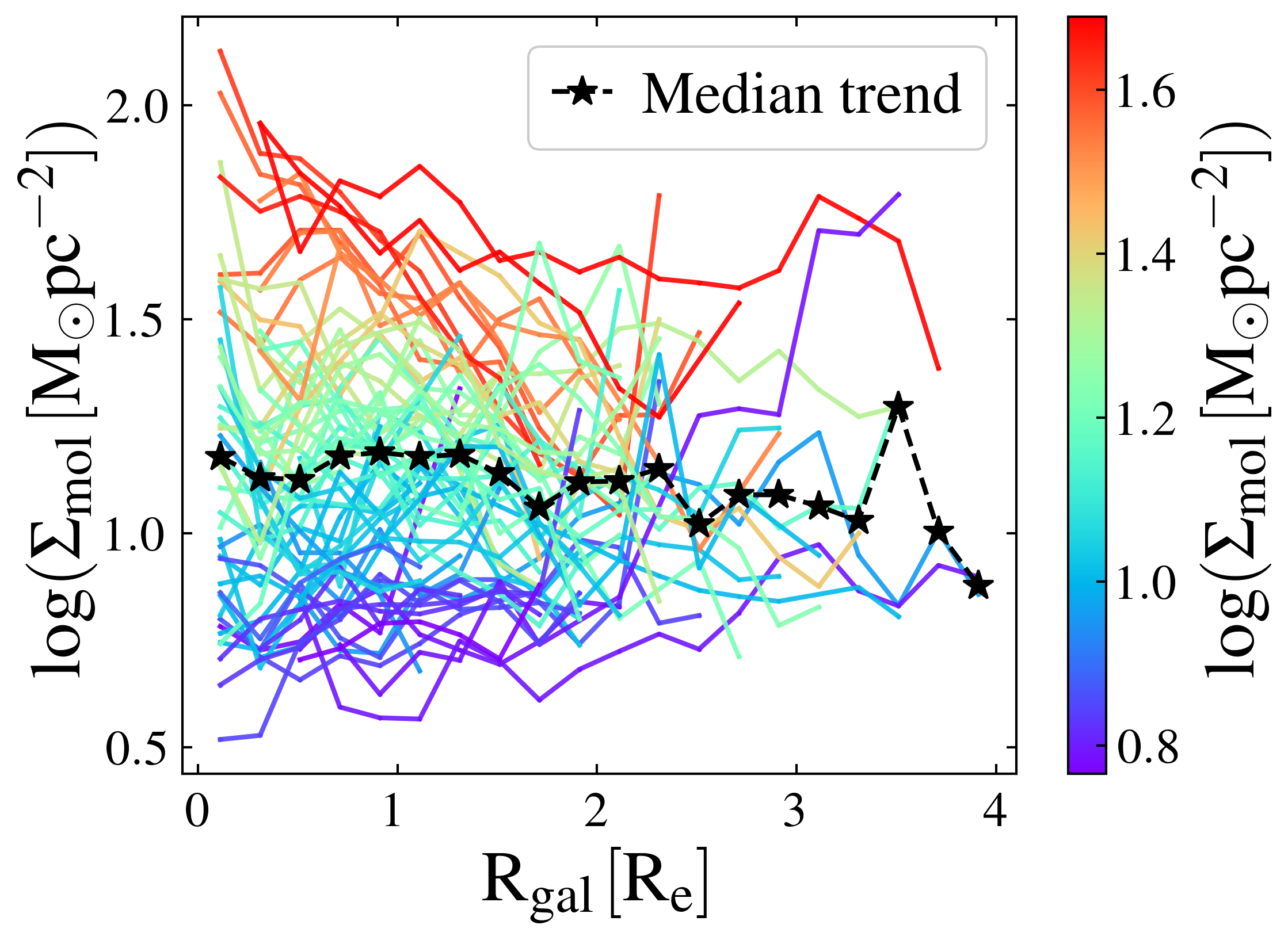}
    \includegraphics[width=0.33\textwidth]{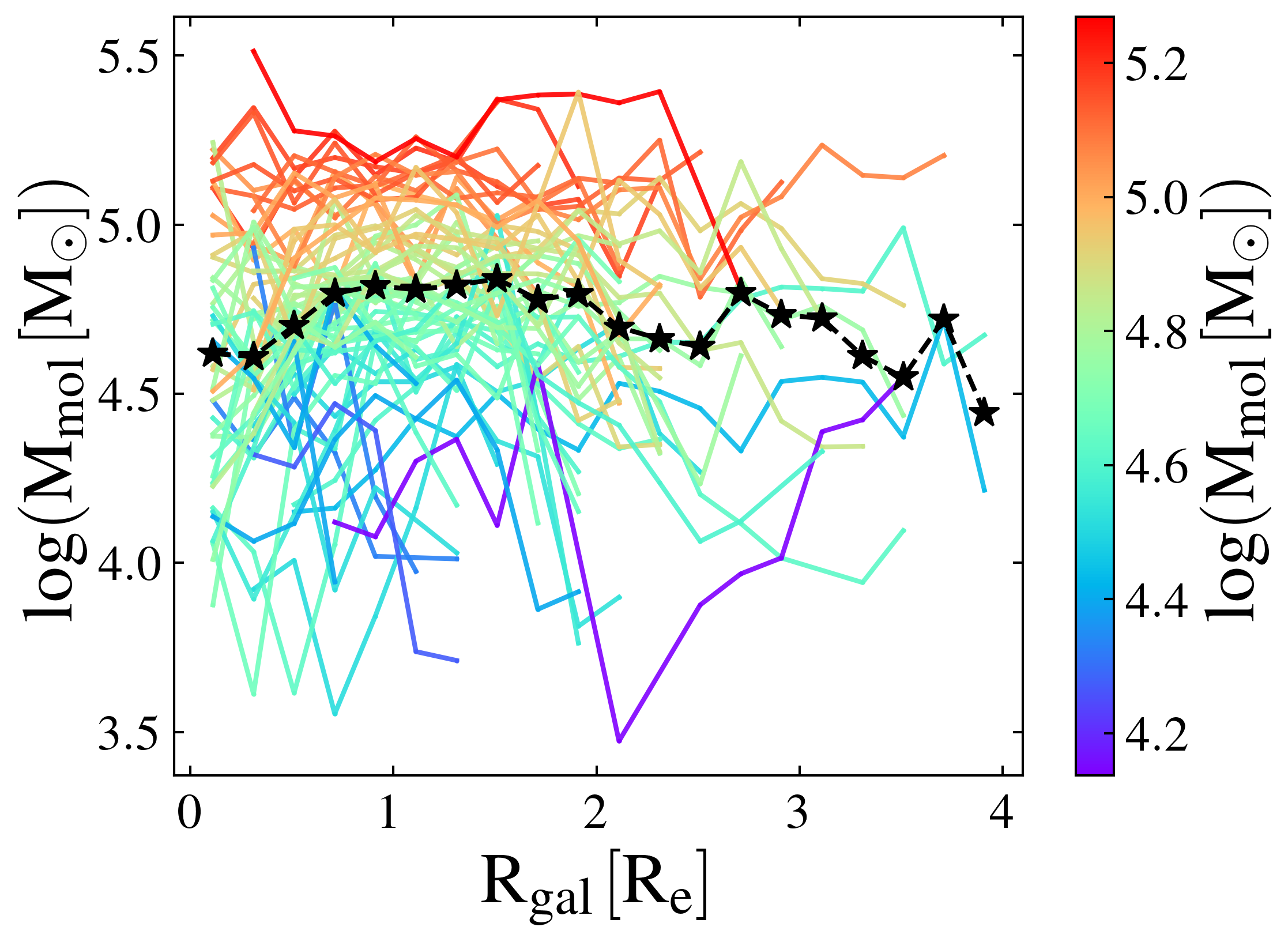}
    \includegraphics[width=0.33\textwidth]{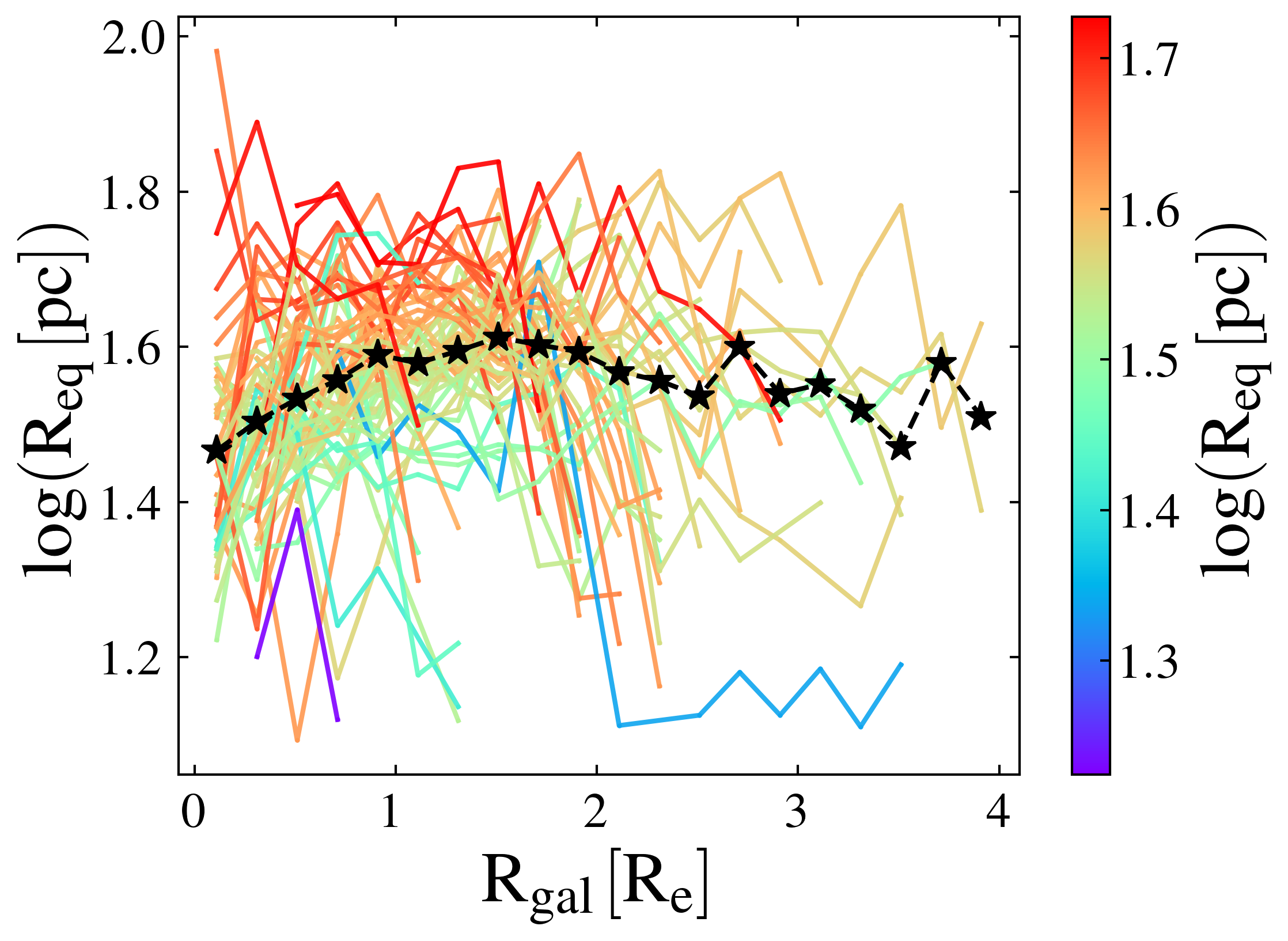}
    \caption{Properties of the PAH clouds vs galactocentric radius per environment for all the clouds in the 66 galaxies: \sigmol\ (\textit{left}), \Mmol\ (\textit{middle}), and \Reff\ (\textit{right}). The running median property per galaxy median (dashed line) is plotted for a bin width of 0.2 R$_{e}$. In each plot,  the individual galaxy trends are color-coded by the median of the property.}
    \label{fig:radial_all}
\end{figure*}

Here, we present the radial profiles per environment. In Sect.~\ref{SS:rad_prof}, we presented the global trends and showed that \Reff\ shows a flat profile and the \sigmol\ radial profile is slightly decreasing. Fig.~\ref{fig:radial_envs} further shows that \Reff\ is also flat per environment with a slight decrease after 2  R$_{e}$ for interarm clouds. The \Mmol\ and \sigmol\ profiles generally show a decreasing trend per environment after 0.5 R$_{e}$. 

The individual-galaxy radial profiles are shown in Fig.~\ref{fig:radial_all}. The trends largely vary per galaxy. However, the consensus is a decreasing \Mmol\ and \sigmol\ profile after 0.5 R$_{e}$. Upon adding the flagged clouds or adopting another \aco\ prescription (e.g., only metallicity dependent, or a constant \aco), we notice a bump in the \Mmol\ and \sigmol\ profiles toward central regions due to higher \aco\ values and the addition of overlapping structure, which mostly affects central regions. This confirms a general declining \Mmol\ and \sigmol\ radial profiles.

\onecolumn
\subsection{Galaxy-by-galaxy mass spectra}

\begin{figure*}[h]
    \centering
    \includegraphics[width=0.90
    \paperwidth]{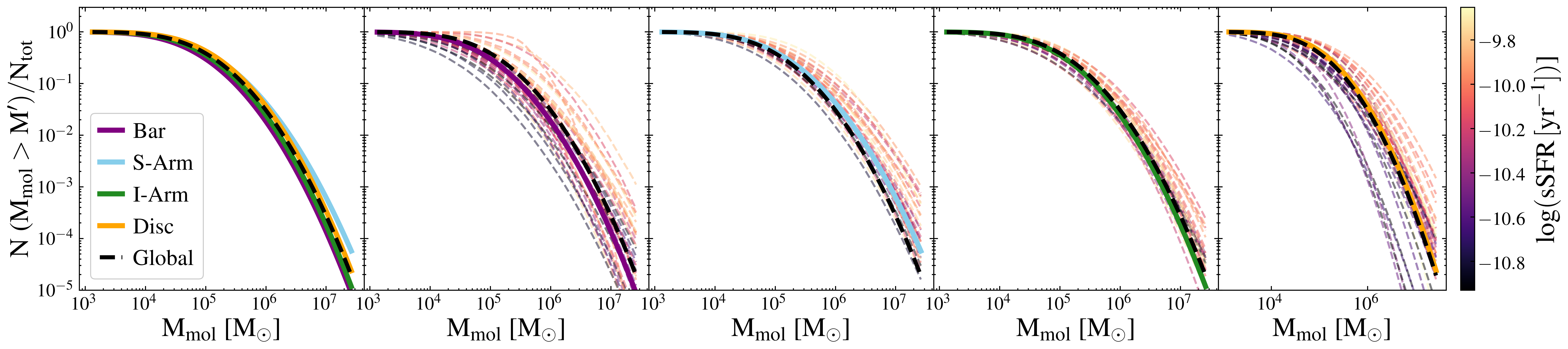}
    \caption{Normalized survival function fits for PAH clouds in different environments. The dashed black block curve represents the fit for all the clouds, and the other block curves are for all clouds in a specific environment, as labeled (i.e., bar in purple, spiral arm in light blue, interarm in green, and disk in yellow). The dashed curves, color-coded by sSFR, are the fits to the galaxy-by-galaxy per environment (depending on the block curves).}
    \label{fig:CDF_Norm}
\end{figure*}

\subsection{Mass-radius relationship}

\begin{figure*}[h]
    \centering
    \includegraphics[width=0.92
    \paperwidth]{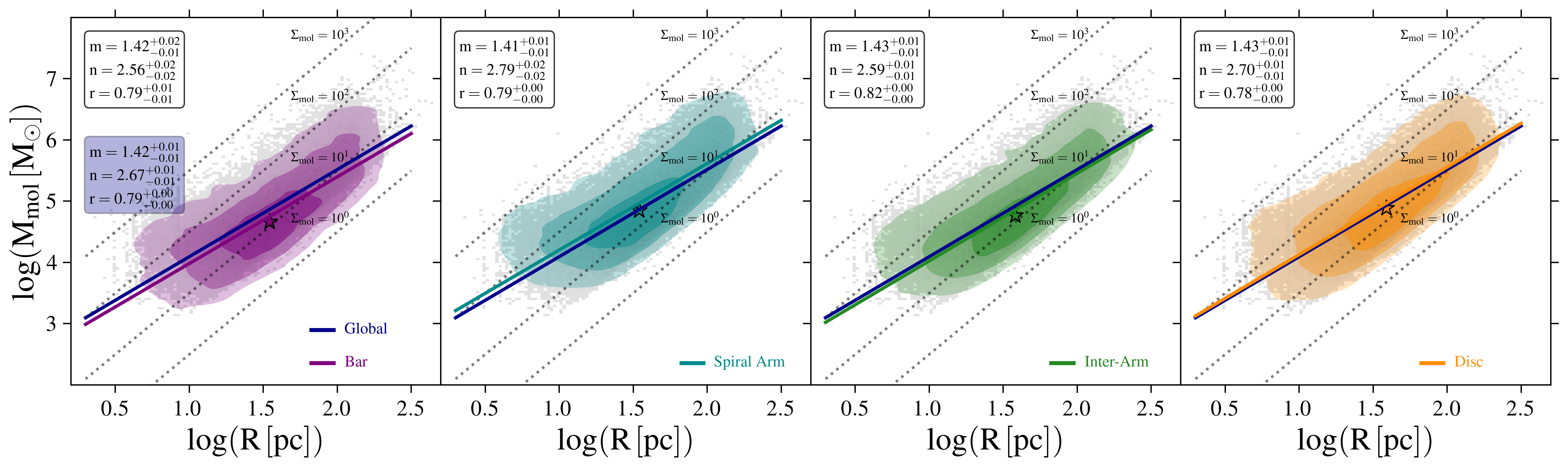}
    \caption{log(\Mmol)-log(\Reff) scaling relation for the clouds in the different environments. Colored density contours represent 1, 2, and 3 $\sigma$ contours in their corresponding galactic environments, and the stars show the median values. The solid line in each plot is a linear regression for the clouds in a specific environment. The slope and intercept are given by m and n, respectively, and the correlation coefficient is given by r. The 2D histogram of the full sample is shown in gray. The blue line represents the linear regression for the clouds regardless of environment, and its fit parameters are represented in the blue box in the leftmost plot. The diagonal dotted lines represent constant \sigmol\ lines at $10^{0},10^{1},10^{2}$, and $10^{3}$ \msunperpcsq .}
    \label{fig:Scaling}
\end{figure*}

The mass-radius relationship is shown in Fig.~\ref{fig:Scaling} for the full Global cloud sample and the clouds per galactic environment. We compare the distribution to constant \sigmol\ lines plotted and notice that a significant number of clouds exist at the typical observed $\sigmol \leq 10\, \msunperpcsq$, which corresponds to the peak of the lognormal column density distribution in the F770W band \citep[$\sim 10^{21}$ cm$^{-2}$ ;][]{Pathak_2024}. However, these clouds appear to span over all \Reff , similar to the clouds in regions above $10\,\msunperpcsq$. 

We fit a linear regression using $linmix\footnote{https://linmix.readthedocs.io/en/latest/}$ and take the median values of the fit parameters with errors as 84th - 50th and 50th - 16th percentiles. Generally, the slopes are similar in each galactic environment, and the spiral arm clouds are located at higher \Mmol and \sigmol\ values compared to the other environments (see Fig.~\ref{fig:Scaling}).

\onecolumn

\subsection{Correlations}

\begin{figure*}[h]
    \centering
    \includegraphics[width=0.9
    \paperwidth]{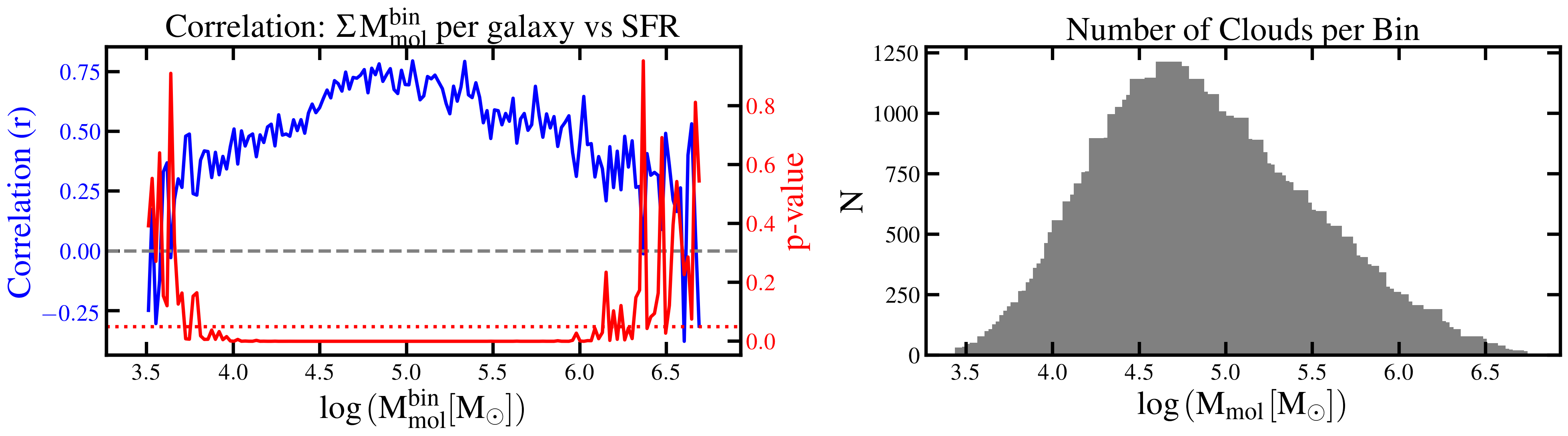}
    \caption{\textit{Left:} Spearman correlation coefficient (solid blue) and the corresponding probability values (solid red) between the sum of the binned cloud \Mmol\ ($\rm M^{bin}_{mol}$) and global SFR per galaxy. The red dotted line indicates a probability value of 0.05, while the gray dashed line marks a correlation coefficient of zero. \textit{Right:} Number of clouds per bin.}
    \label{fig:sfr_corr}
\end{figure*}

\subsection{The scaling parameter}\label{ss:scaling_appendix}

\begin{figure*}[h]
    \centering
    \includegraphics[width=0.9
    \paperwidth]{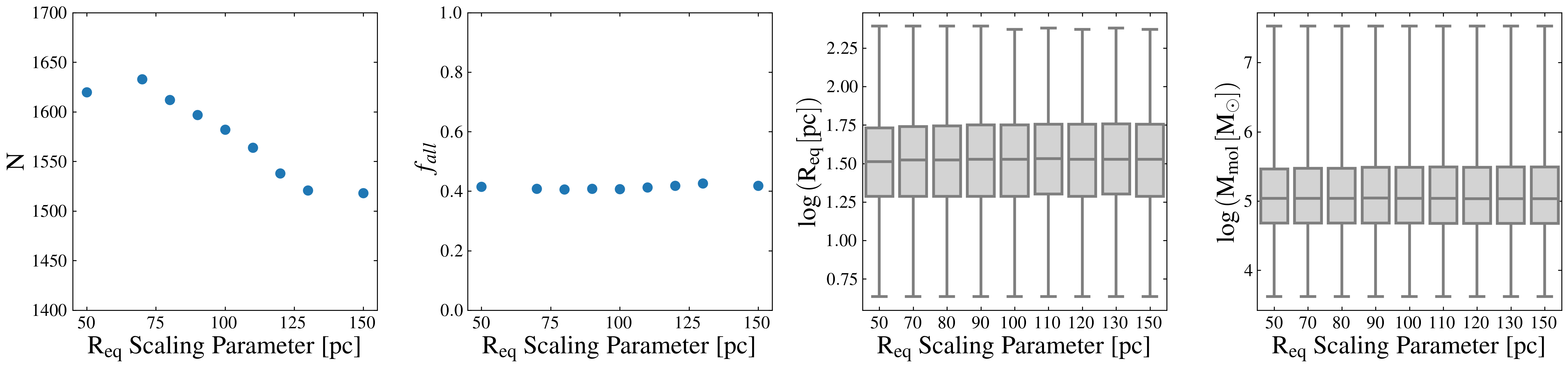}
    \includegraphics[width=0.9
    \paperwidth]{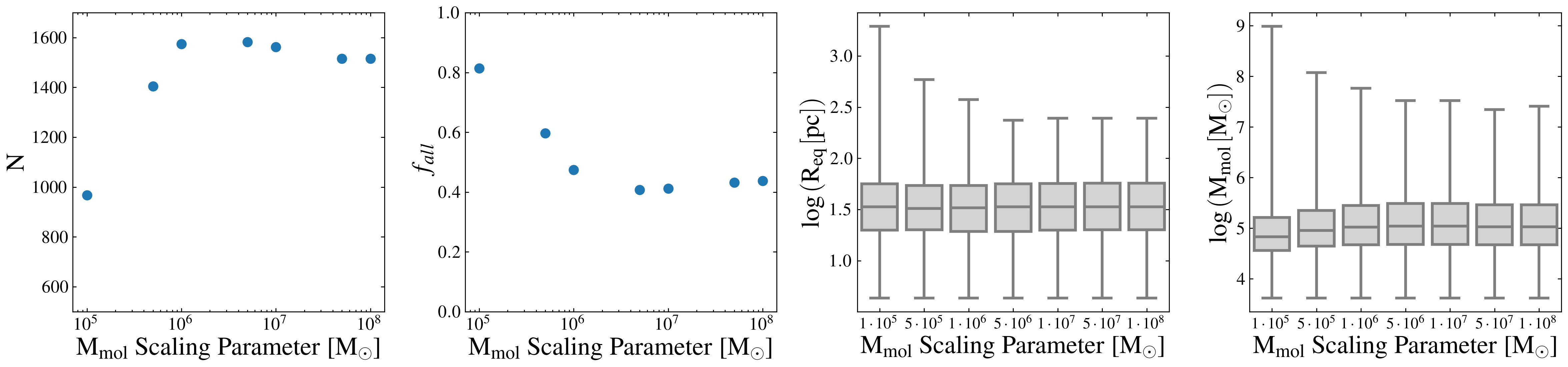}
    \caption{
    \textit{Top row:} Variation in the total number of clouds ($N$),  flux within the clouds relative to the total flux ($f_{all}$),  equivalent radius of the clouds (\Reff), and molecular mass of the clouds (\Mmol) as a function of varying radius scaling parameter (from 50 to 150~pc), assuming a constant \Mmol\ scaling parameter of $5\times 10^{6}~\msun$. 
    \textit{Bottom row:} Same as the top row, but with a varying \Mmol\ scaling parameter between $10^{5}$ and $10^{8}~\msun$, assuming a constant \Reff\ scaling parameter of 100~pc. The box plots represent the median and the interquartile range of their distributions, and the whiskers cover the rest of the distribution. All results are shown for NGC 1385.
    }
    \label{fig:scaling_effect}
\end{figure*}

We tested how changing the scaling parameters in SCIMES for both \Reff\ and \Mmol\ affects the way clouds are grouped, as well as how it impacts the size and mass distributions of the clouds. Figure~\ref{fig:scaling_effect} shows that when we keep the \Mmol\ scaling parameter at $5\times 10^{6}~\msun$ and adjust the \Reff\ parameter between 50 and 150 pc, the total number of identified clouds changes by only about $3~\%$ compared to our default setup, where \Reff\ is set at 100 pc and \Mmol\ at $5\times10^{6}~\msun$. Even with this slight change in cloud numbers, the recovered flux and the distributions of \Reff\ and \Mmol\ remain steady and follow the same general pattern.

When we keep \Reff\ fixed at 100 pc and instead vary the \Mmol\ scaling parameter, the results stay consistent for clouds with masses above $10^{6}~\msun$, similar to what we saw when adjusting \Reff. However, setting the \Mmol\ parameter too low (below $10^{5}~\msun$) or too high (over $10^{8}~\msun$) causes clustering problems. In these cases, the algorithm either picks up too few clusters, leaving behind massive structures that stretch across large regions, for example spiral arms (as seen with the $10^{5}~\msun$ case), or it allows these large regions to stay grouped because of the high scaling parameter.

\subsection{Cloud radius prescription}\label{ss:radii}
\begin{figure*}[h]
    \centering
    \includegraphics[width=0.49\textwidth]{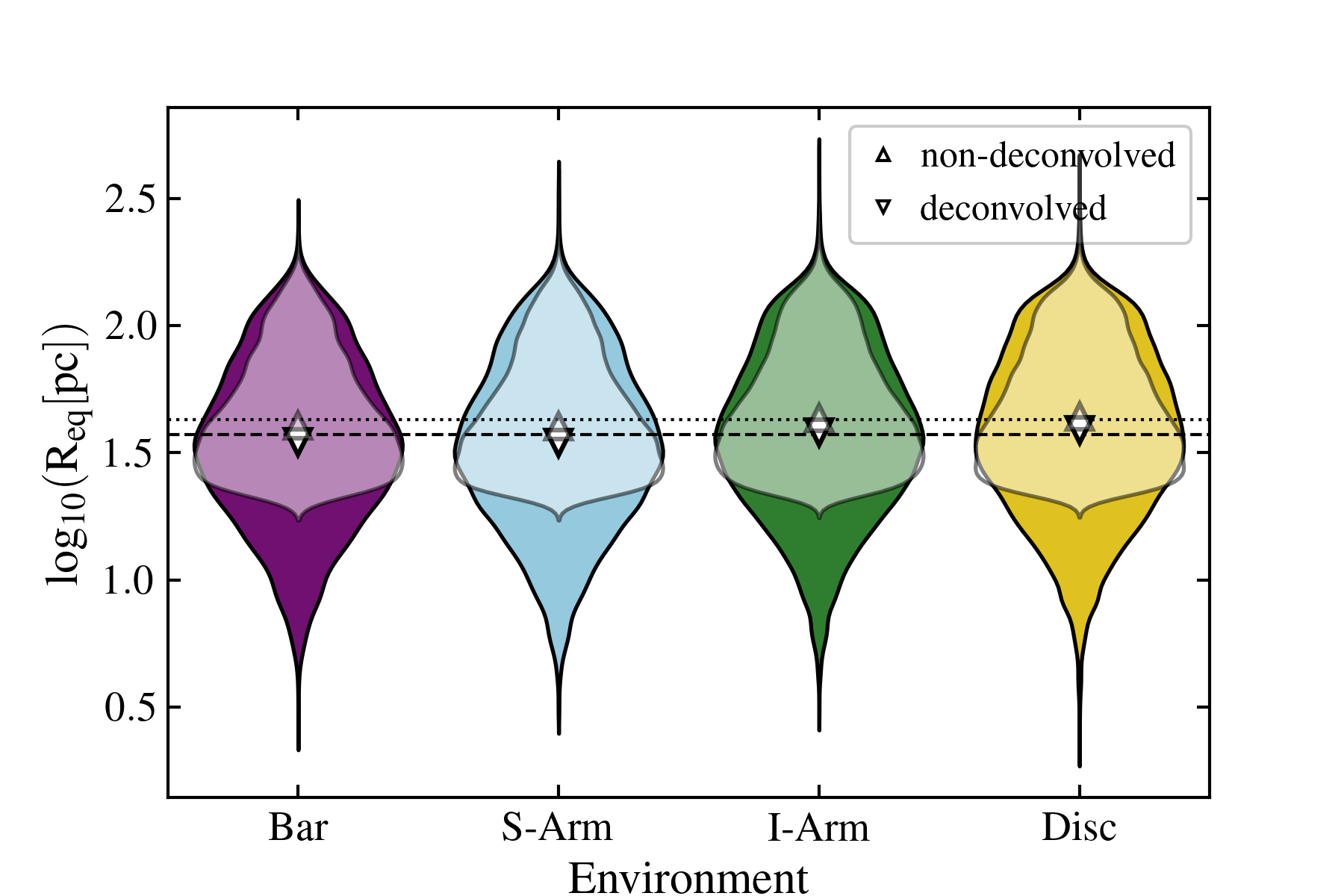}
    \includegraphics[width=0.49\textwidth]{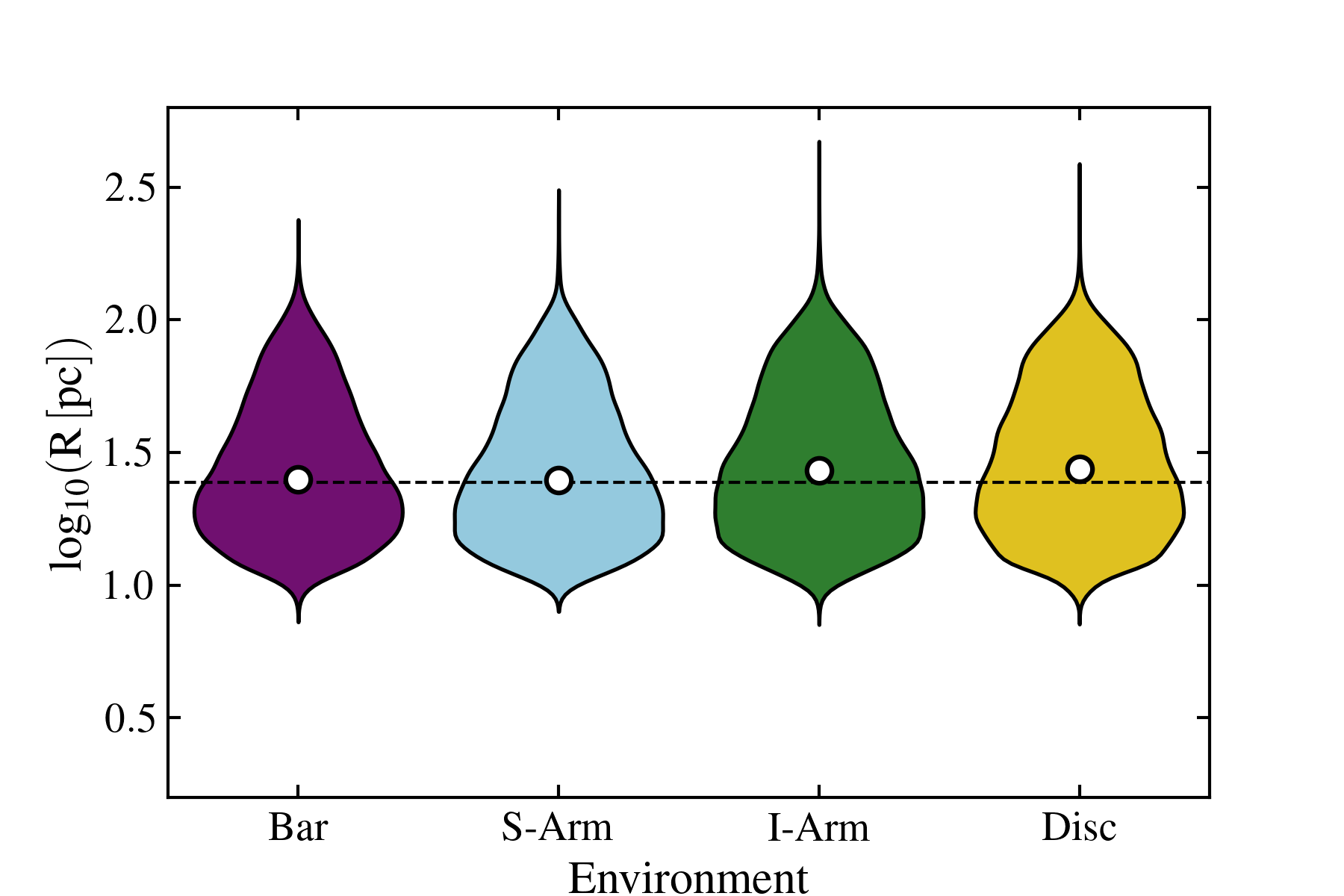}
    \caption{\textit{Left:} PAH cloud radius using the area of the cloud across different galactic environments. The beam-deconvolved radii are represented in the colored violins, and the nondeconvolved radii are shown in the transparent violins. The dashed and dotted lines are the medians of the deconvolved and nondeconvolved radii, respectively. \textit{Right:} Radii of the clouds using second-moment measurements across different galactic environments. The dashed line is the second-moment radius median.}
    \label{fig:rad_app}
\end{figure*} 

The radius of the clouds can be assessed by two different measurements (see Fig.~\ref{fig:rad_app} and Sect.~\ref{S:MC_Props}). In this paper, we recommend the usage of \Reff\, as it could be directly inferred from the number of pixels within the clouds. We applied the beam deconvolution using a Gaussian beam, which led to some inaccuracies in the measurement of \Reff . However, this is an effort to remove the beam contribution. The median \Reff\ across the sample is $37.4^{+46.3}_{-20.0}$ pc (medians with the 84th - 50th percentile and 50th - 16th percentile displayed in superscript and subscript for the full cloud distribution). For comparison, the median nondeconvolved radius is $42.6^{+43.5}_{-15.8}$ pc, and the median radius based on the second spatial moment is $24.3^{+31.1}_{-11.5}$ pc. We emphasize that our main results are robust to the choice of radius definition, and adopting any of these estimates does not alter the conclusions of our analysis.

\subsection{F1130W and F770W comparison}\label{ss:F1130W}

\begin{figure}[ht]
    \centering
    \includegraphics[width=0.49\linewidth]{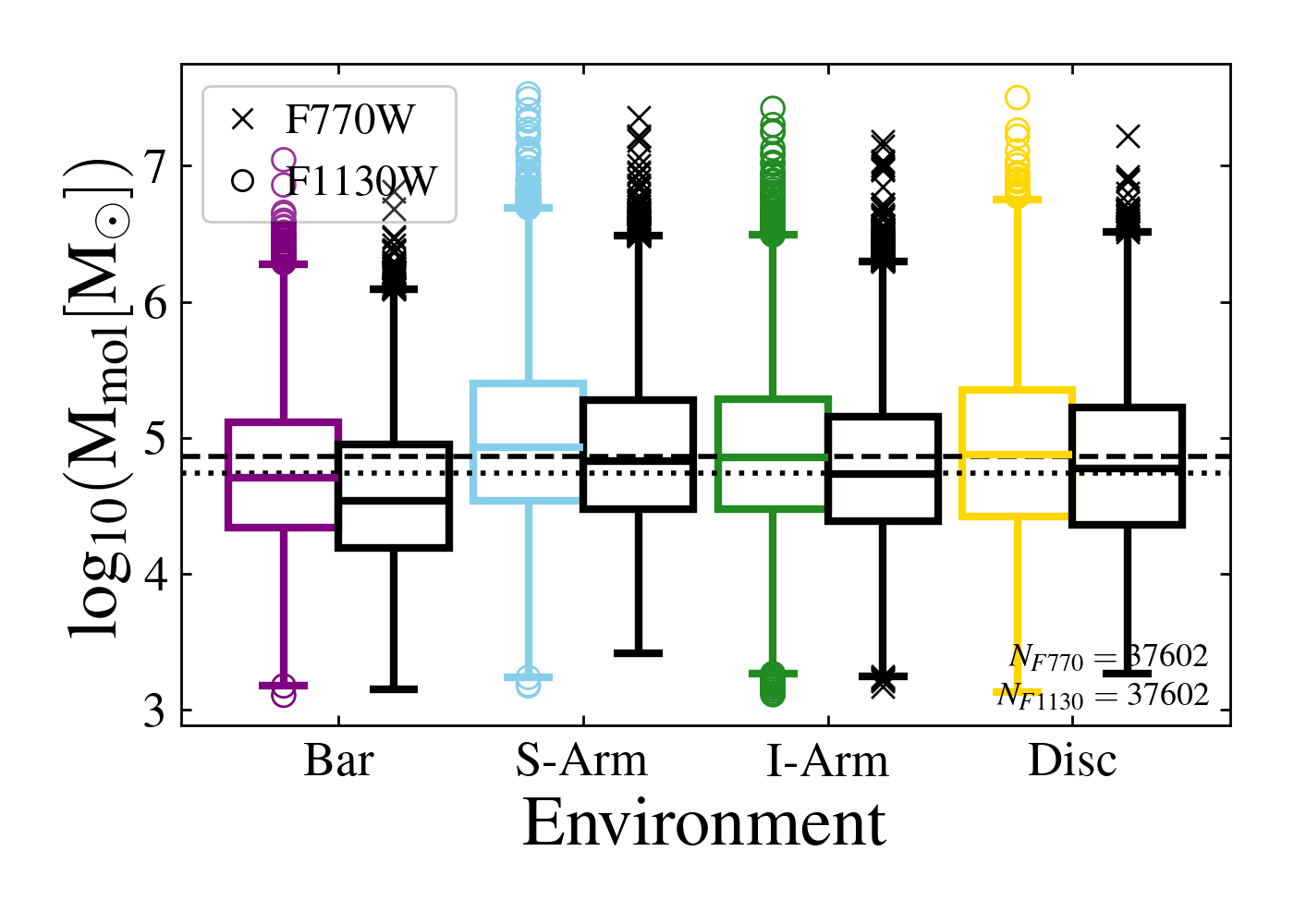}
    \includegraphics[width=0.48\linewidth]{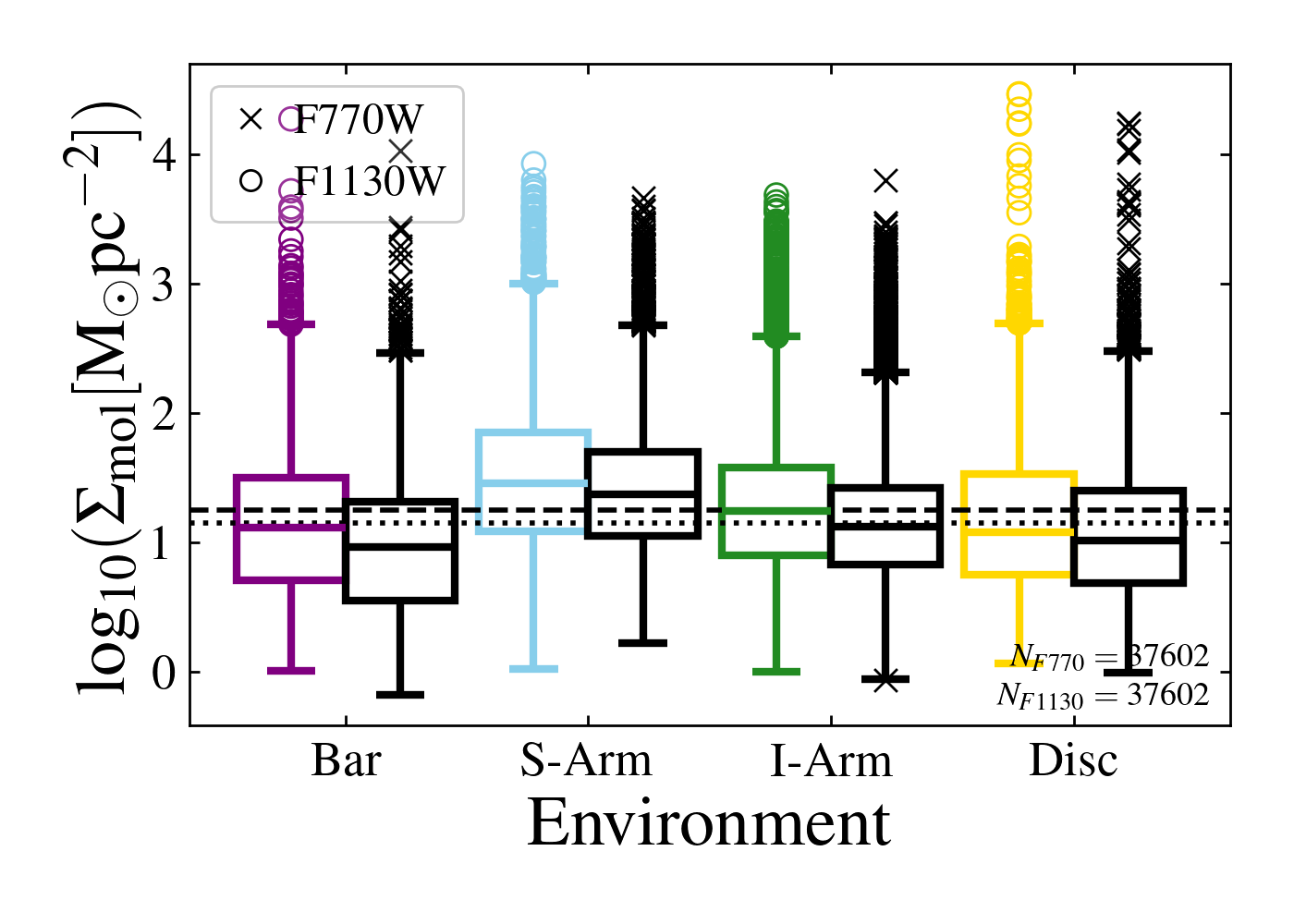}
    \caption{\Mmol and \sigmol of the PAH clouds according to  F1130W (colored) and F770W (black) in 20 galaxies. The dotted line is the median property for the F770W band, and the dashed line is the median property for the F1130W band. The total number of clouds is displayed in the lower right of the plots.}
    \label{fig:F1130W}
\end{figure}

The results presented in our analysis are consistent when considering another band. In this section, we compare cloud properties using both F770W and F1130W bands at homogenized physical resolutions for the 20 galaxies that have F1130W observations in the PHANGS-JWST sample. The difference between the molecular properties of the clouds extracted using \cite{chown2025} prescriptions between the F770W and F1130W (Eq.~C2 in \citealt{chown2025}) bands is only $\sim 0.1$ dex as seen in Fig.~\ref{fig:F1130W}. This minimal effect discrepancy between the bands could be due to stellar continuum emission that plays a minimal role in the F1130W, hence it is not subtracted. It is worth noting, however, that the F1130W band also traces more neutral PAHs than the F770W band, which traces ionized PAHs mainly. This could create further differences toward the central regions of the galaxies.

\end{appendix}

\end{document}